\documentclass[12pt]{article}
\usepackage{epsfig,amsfonts,amssymb}
\usepackage{hyperref}
\usepackage{cite}
\input epsf.sty
\usepackage{cite}

\newcommand{\hab}{}

\newcommand{\pii}{\pi}

\def\ZZZ{{\hbox{ Z\kern-1.6mm Z}}}
\def\RRR{{\hbox{ R\kern-2.4mm R}}}
\def\CCC{{\hbox{ C\kern-2.0mm C}}}
\def\zzz{{\hbox{z\kern-1mm z}}}

\newcommand{\ten}{{(10)}}
\newcommand{\bet}{{( b )}}

\newcommand{\qq}{k}
\newcommand{\pp}{l}
\newcommand{\nn}{\nonumber \\}

\newcommand{\vt}{\vartheta}

\newcommand{\vtau} {\vec \tau}
\newcommand{\vj} {\vec J}
\newcommand{\vxi} {\vec \xi}
\newcommand{\vu} {\vec u}
\newcommand{\htau} {\vec \eta}
\newcommand{\vc}{\vec\chi}
\newcommand{\vpsi} {\vec \psi}

\newcommand{\qeq}{{\hbox{=\kern-2.3mm ? \kern.5mm }}}
\renewcommand{\qeq}{=}

\newcommand{\rrho}{r}
\newcommand{\bA}{{\bf A}}
\newcommand{\tx}{\wt x}
\newcommand{\bG}{{\bf G}}
\newcommand{\bF}{{\bar F}}
\newcommand{\bbb}{{\bar b}}
\newcommand{\gam}{\tau}
\newcommand{\eps}{\epsilon}
\newcommand{\vareps}{\varepsilon}
\newcommand{\ra}{\rangle}
\newcommand{\la}{\langle}
\newcommand{\T}{\chi_{T}(k)}
\newcommand{\Tm}{\chi_{T}(k')}
\newcommand{\Cn}{{\cal C}_n}
\newcommand{\vp}{\varphi}
\newcommand{\ve}{\varepsilon}
\newcommand{\tl}{\lambda}
\newcommand{\dt}{(\vec \nabla T)^2}
\newcommand{\hp}{{\wh\Phi}}
\newcommand{\hq}{{\wh Q_B}}
\newcommand{\he}{{\wh\eta_0}}
\newcommand{\ha}{{\wh{A}}}
\newcommand{\lllb}{\Bigl\langle\Bigl\langle}
\newcommand{\rrrb}{\Bigr\rangle\Bigr\rangle}
\newcommand{\tf}{\wt f}
\newcommand{\sss}{{\cal L}_{av}}
\newcommand{\bx}{\bar x}
\newcommand{\bw}{\bar w}
\newcommand{\ws}{{\wt\sigma}}
\newcommand{\wrh}{{\wt\rho}}
\newcommand{\wv}{{\wt v}}

\newcommand{\vv} {\bar v}
\newcommand{\uu} {\bar u}

\newcommand{\K}{{\rm K_1}}
\newcommand{\Kt}{{\rm \widetilde K_1}}

\newcommand{\B}{b'}
\newcommand{\C}{c\,'}
\newcommand{\bB}{\bar b'}
\newcommand{\Bu}{B_{\vec u}}
\newcommand{\VV}{{\cal V}}
\newcommand{\BB}{{\cal B}}
\newcommand{\DD}{{\cal D}}
\newcommand{\BBB}{{\cal B}}
\newcommand{\II}{{\cal I}}
\newcommand{\AAA}{{\cal A}}
\newcommand{\GG}{{\cal G}}
\newcommand{\KK}{{\cal K}}
\newcommand{\fff}{{\bf f}}
\newcommand{\ccc}{{\bf c}}
\newcommand{\FF}{{\cal F}}
\newcommand{\JJ}{{\cal J}}
\newcommand{\HH}{{\cal H}}
\newcommand{\MM}{{\cal M}}
\newcommand{\CC}{{\cal C}}
\newcommand{\bC}{{\bf C}}
\newcommand{\OO}{{\cal O}}
\newcommand{\QQ}{{\cal Q}}
\newcommand{\PP}{{\cal P}}
\newcommand{\EE}{{\cal E}}
\newcommand{\LL}{{\cal L}}

\newcommand{\XX}{{\cal X}}

 \newcommand{\rrr}{\rangle\rangle}
\newcommand{\half}{{1\over 2}}
\newcommand{\wt}{\widetilde}
\newcommand{\wh}{\widehat}
\newcommand{\wc}{\wt}
\newcommand{\wb}{\bar}
\newcommand{\RR}{{\cal R}}
\newcommand{\NN}{{\cal N}}
\newcommand{\TT}{{\cal T}}
\newcommand{\bg}{\bar g}
\newcommand{\ba}{\bar a}
\newcommand{\bc}{\bar c}
\newcommand{\bd}{\bar d}
\newcommand{\bb}{\bar b}
\newcommand{\bT}{\bar \Theta}
\newcommand{\SSS}{{\cal S}}
\newcommand{\tlx}{\left(\wt \lambda ; X^0(0) \right)}
\newcommand{\al}{\alpha}

\newcommand{\tk}{\wt \kappa}

\newcommand{\ppp}{\prime\prime}

\newcommand{\omk}{\omega_n(\vec k)}
\newcommand{\onk}{\omega^{(N)}_{\vec k_\perp}}
\newcommand{\tI}{\wt\II}
\newcommand{\hI}{\wh\II}
\newcommand{\nI}{\II}

\newcommand{\cp}{\check\Phi}
\newcommand{\cps}{\Psi}
\newcommand{\crh}{\check\rho}
\newcommand{\cs}{\check\sigma}
\newcommand{\cv}{\check v}
\newcommand{\com}{\check\Omega}

\newcommand{\be}{\begin{equation}}
\newcommand{\ee}{\end{equation}}
\newcommand{\ben}{\begin{eqnarray}\displaystyle}
\newcommand{\een}{\end{eqnarray}}

\newcommand{\refb}[1]{(\ref{#1})}
\newcommand{\p}{\partial}
\newcommand{\sectiono}[1]{\section{#1}\setcounter{equation}{0}}
\newcommand{\subsectiono}[1]{\subsection{#1}\setcounter{equation}{0}}

\newcommand{\zet}{\zeta}

\newcommand{\gsim}{\stackrel{>}{\sim}}
\newcommand{\lsim}{\stackrel{<}{\sim}}

\newcommand{\Lamb}{\Lambda}

\def\one{{\hbox{ 1\kern-.8mm l}}}
\def\zero{{\hbox{ 0\kern-1.5mm 0}}}

\def\wa{{\wh a}}
\def\wb{{\wh b}}
\def\wc{{\wh c}}
\def\wc{\check}
\def\wdd{{\wh d}}

\newcommand{\bi}{{\bf i}}

\renewcommand{\theequation}{\thesection.\arabic{equation}}

\newcommand{\bea}[1]{\begin{eqnarray}\label{#1} }
\newcommand{\eea}{\end{eqnarray}}

\newcommand{\wJ}{\wt J}
\newcommand{\bN}{{\bf N}}

\newcommand{\aaa}{b}

\newcommand{\eqref}{\refb}

\newcommand{\un}{{\rm u}}

\newcommand{\dotalpha}{{\dot{\alpha}}}

\newcommand{\dotbeta}{{\dot{\beta}}}

\newcommand{\dotgamma}{{\dot{\gamma}}}

\newcommand{\dalpha}{\beta}

\newcommand{\Vm}{V}

\newcommand{\gb}{G}

\newcommand{\q}{e}

\def\figadded{

\def\JPicScale{0.6}
\ifx\JPicScale\undefined\def\JPicScale{1}\fi
\unitlength \JPicScale mm
\begin{picture}(105,70)(0,0)
\linethickness{0.3mm}
\multiput(20,40)(0.24,0.12){167}{\line(1,0){0.24}}
\linethickness{0.3mm}
\multiput(60,60)(0.24,-0.12){167}{\line(1,0){0.24}}
\linethickness{0.3mm}
\multiput(20,40)(0.24,-0.12){167}{\line(1,0){0.24}}
\linethickness{0.3mm}
\multiput(60,20)(0.24,0.12){167}{\line(1,0){0.24}}
\linethickness{0.3mm}
\put(20,40){\circle{10}}

\linethickness{0.3mm}
\put(60,20){\circle{10}}

\linethickness{0.3mm}
\put(60,60){\circle{10}}

\linethickness{0.3mm}
\put(100,40){\circle{10}}

\linethickness{0.3mm}
\put(40,25){\line(0,1){10}}
\linethickness{0.3mm}
\qbezier(80,0)(80,10.41)(80,17.62)
\qbezier(80,17.62)(80,24.84)(80,30)
\qbezier(80,30)(80.03,35.2)(77.62,38.81)
\qbezier(77.62,38.81)(75.22,42.42)(70,45)
\qbezier(70,45)(64.8,47.62)(61.19,47.62)
\qbezier(61.19,47.62)(57.58,47.62)(55,45)
\qbezier(55,45)(52.42,42.44)(48.81,37.62)
\qbezier(48.81,37.62)(45.2,32.81)(40,25)
\qbezier(40,25)(34.81,17.17)(30,13.56)
\qbezier(30,13.56)(25.19,9.95)(20,10)
\qbezier(20,10)(14.79,9.97)(11.78,12.38)
\qbezier(11.78,12.38)(8.77,14.78)(7.5,20)
\qbezier(7.5,20)(6.2,25.09)(5.59,37.12)
\qbezier(5.59,37.12)(4.99,49.16)(5,70)
\put(30,30){\makebox(0,0)[cc]{1}}

\put(90,30){\makebox(0,0)[cc]{2}}

\put(40,55){\makebox(0,0)[cc]{3}}

\put(80,55){\makebox(0,0)[cc]{4}}

\put(60,0){\makebox(0,0)[cc]{(a)}}

\end{picture}

}

\def\figaddedb{

\def\JPicScale{0.6}
\ifx\JPicScale\undefined\def\JPicScale{1}\fi
\unitlength \JPicScale mm
\begin{picture}(105,75)(0,0)
\linethickness{0.3mm}
\multiput(20,40)(0.24,0.12){167}{\line(1,0){0.24}}
\linethickness{0.3mm}
\multiput(60,60)(0.24,-0.12){167}{\line(1,0){0.24}}
\linethickness{0.3mm}
\multiput(60,20)(0.24,0.12){167}{\line(1,0){0.24}}
\linethickness{0.3mm}
\put(20,40){\circle{10}}

\linethickness{0.3mm}
\put(60,20){\circle{10}}

\linethickness{0.3mm}
\put(60,60){\circle{10}}

\linethickness{0.3mm}
\put(100,40){\circle{10}}

\linethickness{0.3mm}
\qbezier(80,0)(80,10.41)(80,17.62)
\qbezier(80,17.62)(80,24.84)(80,30)
\qbezier(80,30)(80.03,35.23)(77.62,37.03)
\qbezier(77.62,37.03)(75.22,38.84)(70,37.5)
\qbezier(70,37.5)(64.8,36.21)(60.59,34.41)
\qbezier(60.59,34.41)(56.38,32.6)(52.5,30)
\qbezier(52.5,30)(48.62,27.39)(43.81,26.19)
\qbezier(43.81,26.19)(39,24.98)(32.5,25)
\qbezier(32.5,25)(25.99,24.98)(21.78,26.19)
\qbezier(21.78,26.19)(17.57,27.39)(15,30)
\qbezier(15,30)(12.39,32.59)(11.19,35.59)
\qbezier(11.19,35.59)(9.98,38.6)(10,42.5)
\qbezier(10,42.5)(10,46.34)(10,54.16)
\qbezier(10,54.16)(10,61.98)(10,75)
\put(35,35){\makebox(0,0)[cc]{1}}

\put(90,30){\makebox(0,0)[cc]{2}}

\put(40,55){\makebox(0,0)[cc]{3}}

\put(80,55){\makebox(0,0)[cc]{4}}

\linethickness{0.3mm}
\put(20,40){\line(1,0){25}}
\linethickness{0.3mm}
\put(45,20){\line(1,0){15}}
\put(45,15){\makebox(0,0)[cc]{1}}

\put(60,0){\makebox(0,0)[cc]{(b)}}

\end{picture}

}

\def\figone{

\def\JPicScale{0.5}
\ifx\JPicScale\undefined\def\JPicScale{1}\fi
\unitlength \JPicScale mm

}

\def\figtwo{

\def\JPicScale{0.7}
\ifx\JPicScale\undefined\def\JPicScale{1}\fi
\unitlength \JPicScale mm
\begin{picture}(155,90)(0,0)
\linethickness{0.3mm}
\put(80,0){\line(0,1){40}}
\linethickness{0.3mm}
\put(80,40){\line(1,0){40}}
\linethickness{0.3mm}
\put(120,40){\line(0,1){20}}
\linethickness{0.3mm}
\put(80,60){\line(1,0){40}}
\linethickness{0.3mm}
\put(80,60){\line(0,1){30}}
\put(140,50){\makebox(0,0)[cc]{x}}

\put(155,50){\makebox(0,0)[cc]{x}}

\put(20,50){\makebox(0,0)[cc]{x}}

\put(100,50){\makebox(0,0)[cc]{x}}

\put(15,45){\makebox(0,0)[cc]{$Q_2$}}

\put(140,45){\makebox(0,0)[cc]{$Q_1$}}

\put(100,45){\makebox(0,0)[cc]{$Q_4$}}

\put(155,45){\makebox(0,0)[cc]{$Q_3$}}

\end{picture}

}

\def \figthree{

\def\JPicScale{0.5}
\ifx\JPicScale\undefined\def\JPicScale{1}\fi
\unitlength \JPicScale mm
\begin{picture}(150,105)(0,0)
\put(30,50){\makebox(0,0)[cc]{x}}

\put(90,50){\makebox(0,0)[cc]{x}}

\put(110,50){\makebox(0,0)[cc]{x}}

\put(150,50){\makebox(0,0)[cc]{x}}

\put(25,45){\makebox(0,0)[cc]{$Q_2$}}

\put(85,45){\makebox(0,0)[cc]{$Q_1$}}

\put(110,45){\makebox(0,0)[cc]{$Q_4$}}

\put(150,45){\makebox(0,0)[cc]{$Q_3$}}

\put(85,-15){\makebox(0,0)[cc]{(b)}}

\linethickness{0.3mm}
\qbezier(60,105)(60,102.41)(60,100)
\qbezier(60,100)(60,97.59)(60,95)
\qbezier(60,95)(60,92.44)(60,87.62)
\qbezier(60,87.62)(60,82.81)(60,75)
\qbezier(60,75)(59.95,67.22)(64.16,60)
\qbezier(64.16,60)(68.37,52.78)(77.5,45)
\qbezier(77.5,45)(86.59,37.1)(93.81,38.91)
\qbezier(93.81,38.91)(101.03,40.71)(107.5,52.5)
\qbezier(107.5,52.5)(114.02,64.31)(117.03,64.31)
\qbezier(117.03,64.31)(120.04,64.31)(120,52.5)
\qbezier(120,52.5)(120.02,40.78)(118.22,33.56)
\qbezier(118.22,33.56)(116.41,26.34)(112.5,22.5)
\qbezier(112.5,22.5)(108.62,18.58)(104.41,17.38)
\qbezier(104.41,17.38)(100.2,16.17)(95,17.5)
\qbezier(95,17.5)(89.8,18.77)(85.59,22.38)
\qbezier(85.59,22.38)(81.38,25.98)(77.5,32.5)
\qbezier(77.5,32.5)(73.6,39.03)(70.59,41.44)
\qbezier(70.59,41.44)(67.59,43.84)(65,42.5)
\qbezier(65,42.5)(62.39,41.21)(61.19,39.41)
\qbezier(61.19,39.41)(59.98,37.6)(60,35)
\qbezier(60,35)(60,32.41)(60,30)
\qbezier(60,30)(60,27.59)(60,25)
\qbezier(60,25)(60,22.42)(60,18.81)
\qbezier(60,18.81)(60,15.2)(60,10)
\end{picture}

}

\def\figfour{

\def\JPicScale{0.5}
\ifx\JPicScale\undefined\def\JPicScale{1}\fi
\unitlength \JPicScale mm
\begin{picture}(150,90)(0,0)
\put(30,50){\makebox(0,0)[cc]{x}}

\put(90,50){\makebox(0,0)[cc]{x}}

\put(110,50){\makebox(0,0)[cc]{x}}

\put(150,50){\makebox(0,0)[cc]{x}}

\put(25,45){\makebox(0,0)[cc]{$Q_2$}}

\put(85,45){\makebox(0,0)[cc]{$Q_1$}}

\put(115,45){\makebox(0,0)[cc]{$Q_4$}}

\put(150,45){\makebox(0,0)[cc]{$Q_3$}}

\put(85,-15){\makebox(0,0)[cc]{(a)}}

\linethickness{0.3mm}
\qbezier(60,90)(60,84.8)(60,81.19)
\qbezier(60,81.19)(60,77.58)(60,75)
\qbezier(60,75)(60,72.41)(60,70)
\qbezier(60,70)(60,67.59)(60,65)
\qbezier(60,65)(59.98,62.4)(61.19,60.59)
\qbezier(61.19,60.59)(62.39,58.79)(65,57.5)
\qbezier(65,57.5)(67.59,56.16)(70.59,58.56)
\qbezier(70.59,58.56)(73.6,60.97)(77.5,67.5)
\qbezier(77.5,67.5)(81.38,74.02)(85.59,77.62)
\qbezier(85.59,77.62)(89.8,81.23)(95,82.5)
\qbezier(95,82.5)(100.2,83.83)(104.41,82.62)
\qbezier(104.41,82.62)(108.62,81.42)(112.5,77.5)
\qbezier(112.5,77.5)(116.41,73.66)(118.22,66.44)
\qbezier(118.22,66.44)(120.02,59.22)(120,47.5)
\qbezier(120,47.5)(120.04,35.69)(117.03,35.69)
\qbezier(117.03,35.69)(114.02,35.69)(107.5,47.5)
\qbezier(107.5,47.5)(101.03,59.29)(93.81,61.09)

\qbezier(93.81,61.09)(86.59,62.9)(77.5,55)
\qbezier(77.5,55)(68.37,47.22)(64.16,40)
\qbezier(64.16,40)(59.95,32.78)(60,25)
\qbezier(60,25)(60,17.19)(60,12.38)
\qbezier(60,12.38)(60,7.56)(60,5)
\qbezier(60,5)(60,2.41)(60,0)
\qbezier(60,0)(60,-2.41)(60,-5)
\end{picture}

}

\def\fna{

\def\JPicScale{0.6}
\ifx\JPicScale\undefined\def\JPicScale{1}\fi
\unitlength \JPicScale mm
\begin{picture}(175,85)(0,0)
\linethickness{0.3mm}
\put(141.83,48.5){\line(0,1){0.5}}
\multiput(141.82,49.5)(0.01,-0.5){1}{\line(0,-1){0.5}}
\multiput(141.79,49.99)(0.02,-0.5){1}{\line(0,-1){0.5}}
\multiput(141.76,50.49)(0.03,-0.5){1}{\line(0,-1){0.5}}
\multiput(141.72,50.98)(0.05,-0.49){1}{\line(0,-1){0.49}}
\multiput(141.66,51.48)(0.06,-0.49){1}{\line(0,-1){0.49}}
\multiput(141.59,51.97)(0.07,-0.49){1}{\line(0,-1){0.49}}
\multiput(141.51,52.46)(0.08,-0.49){1}{\line(0,-1){0.49}}
\multiput(141.42,52.95)(0.09,-0.49){1}{\line(0,-1){0.49}}
\multiput(141.32,53.43)(0.1,-0.49){1}{\line(0,-1){0.49}}
\multiput(141.21,53.92)(0.11,-0.48){1}{\line(0,-1){0.48}}
\multiput(141.09,54.4)(0.12,-0.48){1}{\line(0,-1){0.48}}
\multiput(140.95,54.88)(0.13,-0.48){1}{\line(0,-1){0.48}}
\multiput(140.81,55.35)(0.14,-0.48){1}{\line(0,-1){0.48}}
\multiput(140.65,55.83)(0.16,-0.47){1}{\line(0,-1){0.47}}
\multiput(140.49,56.29)(0.17,-0.47){1}{\line(0,-1){0.47}}
\multiput(140.31,56.76)(0.18,-0.46){1}{\line(0,-1){0.46}}
\multiput(140.12,57.22)(0.09,-0.23){2}{\line(0,-1){0.23}}
\multiput(139.92,57.67)(0.1,-0.23){2}{\line(0,-1){0.23}}
\multiput(139.71,58.13)(0.1,-0.23){2}{\line(0,-1){0.23}}
\multiput(139.5,58.57)(0.11,-0.22){2}{\line(0,-1){0.22}}
\multiput(139.27,59.01)(0.11,-0.22){2}{\line(0,-1){0.22}}
\multiput(139.03,59.45)(0.12,-0.22){2}{\line(0,-1){0.22}}
\multiput(138.78,59.88)(0.12,-0.22){2}{\line(0,-1){0.22}}
\multiput(138.52,60.3)(0.13,-0.21){2}{\line(0,-1){0.21}}
\multiput(138.25,60.72)(0.13,-0.21){2}{\line(0,-1){0.21}}
\multiput(137.98,61.14)(0.14,-0.21){2}{\line(0,-1){0.21}}
\multiput(137.69,61.54)(0.14,-0.2){2}{\line(0,-1){0.2}}
\multiput(137.39,61.94)(0.15,-0.2){2}{\line(0,-1){0.2}}
\multiput(137.09,62.33)(0.1,-0.13){3}{\line(0,-1){0.13}}
\multiput(136.78,62.72)(0.1,-0.13){3}{\line(0,-1){0.13}}
\multiput(136.45,63.1)(0.11,-0.13){3}{\line(0,-1){0.13}}
\multiput(136.12,63.47)(0.11,-0.12){3}{\line(0,-1){0.12}}
\multiput(135.78,63.83)(0.11,-0.12){3}{\line(0,-1){0.12}}
\multiput(135.44,64.19)(0.12,-0.12){3}{\line(0,-1){0.12}}
\multiput(135.08,64.53)(0.12,-0.12){3}{\line(1,0){0.12}}
\multiput(134.72,64.87)(0.12,-0.11){3}{\line(1,0){0.12}}
\multiput(134.35,65.2)(0.12,-0.11){3}{\line(1,0){0.12}}
\multiput(133.97,65.53)(0.13,-0.11){3}{\line(1,0){0.13}}
\multiput(133.58,65.84)(0.13,-0.1){3}{\line(1,0){0.13}}
\multiput(133.19,66.14)(0.13,-0.1){3}{\line(1,0){0.13}}
\multiput(132.79,66.44)(0.2,-0.15){2}{\line(1,0){0.2}}
\multiput(132.39,66.73)(0.2,-0.14){2}{\line(1,0){0.2}}
\multiput(131.97,67)(0.21,-0.14){2}{\line(1,0){0.21}}
\multiput(131.55,67.27)(0.21,-0.13){2}{\line(1,0){0.21}}
\multiput(131.13,67.53)(0.21,-0.13){2}{\line(1,0){0.21}}
\multiput(130.7,67.78)(0.22,-0.12){2}{\line(1,0){0.22}}
\multiput(130.26,68.02)(0.22,-0.12){2}{\line(1,0){0.22}}
\multiput(129.82,68.25)(0.22,-0.11){2}{\line(1,0){0.22}}
\multiput(129.38,68.46)(0.22,-0.11){2}{\line(1,0){0.22}}
\multiput(128.92,68.67)(0.23,-0.1){2}{\line(1,0){0.23}}
\multiput(128.47,68.87)(0.23,-0.1){2}{\line(1,0){0.23}}
\multiput(128.01,69.06)(0.23,-0.09){2}{\line(1,0){0.23}}
\multiput(127.54,69.24)(0.46,-0.18){1}{\line(1,0){0.46}}
\multiput(127.08,69.4)(0.47,-0.17){1}{\line(1,0){0.47}}
\multiput(126.6,69.56)(0.47,-0.16){1}{\line(1,0){0.47}}
\multiput(126.13,69.7)(0.48,-0.14){1}{\line(1,0){0.48}}
\multiput(125.65,69.84)(0.48,-0.13){1}{\line(1,0){0.48}}
\multiput(125.17,69.96)(0.48,-0.12){1}{\line(1,0){0.48}}
\multiput(124.68,70.07)(0.48,-0.11){1}{\line(1,0){0.48}}
\multiput(124.2,70.17)(0.49,-0.1){1}{\line(1,0){0.49}}
\multiput(123.71,70.26)(0.49,-0.09){1}{\line(1,0){0.49}}
\multiput(123.22,70.34)(0.49,-0.08){1}{\line(1,0){0.49}}
\multiput(122.73,70.41)(0.49,-0.07){1}{\line(1,0){0.49}}
\multiput(122.23,70.47)(0.49,-0.06){1}{\line(1,0){0.49}}
\multiput(121.74,70.51)(0.49,-0.05){1}{\line(1,0){0.49}}
\multiput(121.24,70.54)(0.5,-0.03){1}{\line(1,0){0.5}}
\multiput(120.75,70.57)(0.5,-0.02){1}{\line(1,0){0.5}}
\multiput(120.25,70.58)(0.5,-0.01){1}{\line(1,0){0.5}}
\put(119.75,70.58){\line(1,0){0.5}}
\multiput(119.25,70.57)(0.5,0.01){1}{\line(1,0){0.5}}
\multiput(118.76,70.54)(0.5,0.02){1}{\line(1,0){0.5}}
\multiput(118.26,70.51)(0.5,0.03){1}{\line(1,0){0.5}}
\multiput(117.77,70.47)(0.49,0.05){1}{\line(1,0){0.49}}
\multiput(117.27,70.41)(0.49,0.06){1}{\line(1,0){0.49}}
\multiput(116.78,70.34)(0.49,0.07){1}{\line(1,0){0.49}}
\multiput(116.29,70.26)(0.49,0.08){1}{\line(1,0){0.49}}
\multiput(115.8,70.17)(0.49,0.09){1}{\line(1,0){0.49}}
\multiput(115.32,70.07)(0.49,0.1){1}{\line(1,0){0.49}}
\multiput(114.83,69.96)(0.48,0.11){1}{\line(1,0){0.48}}
\multiput(114.35,69.84)(0.48,0.12){1}{\line(1,0){0.48}}
\multiput(113.87,69.7)(0.48,0.13){1}{\line(1,0){0.48}}
\multiput(113.4,69.56)(0.48,0.14){1}{\line(1,0){0.48}}
\multiput(112.92,69.4)(0.47,0.16){1}{\line(1,0){0.47}}
\multiput(112.46,69.24)(0.47,0.17){1}{\line(1,0){0.47}}
\multiput(111.99,69.06)(0.46,0.18){1}{\line(1,0){0.46}}
\multiput(111.53,68.87)(0.23,0.09){2}{\line(1,0){0.23}}
\multiput(111.08,68.67)(0.23,0.1){2}{\line(1,0){0.23}}
\multiput(110.62,68.46)(0.23,0.1){2}{\line(1,0){0.23}}
\multiput(110.18,68.25)(0.22,0.11){2}{\line(1,0){0.22}}
\multiput(109.74,68.02)(0.22,0.11){2}{\line(1,0){0.22}}
\multiput(109.3,67.78)(0.22,0.12){2}{\line(1,0){0.22}}
\multiput(108.87,67.53)(0.22,0.12){2}{\line(1,0){0.22}}
\multiput(108.45,67.27)(0.21,0.13){2}{\line(1,0){0.21}}
\multiput(108.03,67)(0.21,0.13){2}{\line(1,0){0.21}}
\multiput(107.61,66.73)(0.21,0.14){2}{\line(1,0){0.21}}
\multiput(107.21,66.44)(0.2,0.14){2}{\line(1,0){0.2}}
\multiput(106.81,66.14)(0.2,0.15){2}{\line(1,0){0.2}}
\multiput(106.42,65.84)(0.13,0.1){3}{\line(1,0){0.13}}
\multiput(106.03,65.53)(0.13,0.1){3}{\line(1,0){0.13}}
\multiput(105.65,65.2)(0.13,0.11){3}{\line(1,0){0.13}}
\multiput(105.28,64.87)(0.12,0.11){3}{\line(1,0){0.12}}
\multiput(104.92,64.53)(0.12,0.11){3}{\line(1,0){0.12}}
\multiput(104.56,64.19)(0.12,0.12){3}{\line(1,0){0.12}}
\multiput(104.22,63.83)(0.12,0.12){3}{\line(0,1){0.12}}
\multiput(103.88,63.47)(0.11,0.12){3}{\line(0,1){0.12}}
\multiput(103.55,63.1)(0.11,0.12){3}{\line(0,1){0.12}}
\multiput(103.22,62.72)(0.11,0.13){3}{\line(0,1){0.13}}
\multiput(102.91,62.33)(0.1,0.13){3}{\line(0,1){0.13}}
\multiput(102.61,61.94)(0.1,0.13){3}{\line(0,1){0.13}}
\multiput(102.31,61.54)(0.15,0.2){2}{\line(0,1){0.2}}
\multiput(102.02,61.14)(0.14,0.2){2}{\line(0,1){0.2}}
\multiput(101.75,60.72)(0.14,0.21){2}{\line(0,1){0.21}}
\multiput(101.48,60.3)(0.13,0.21){2}{\line(0,1){0.21}}
\multiput(101.22,59.88)(0.13,0.21){2}{\line(0,1){0.21}}
\multiput(100.97,59.45)(0.12,0.22){2}{\line(0,1){0.22}}
\multiput(100.73,59.01)(0.12,0.22){2}{\line(0,1){0.22}}
\multiput(100.5,58.57)(0.11,0.22){2}{\line(0,1){0.22}}
\multiput(100.29,58.13)(0.11,0.22){2}{\line(0,1){0.22}}
\multiput(100.08,57.67)(0.1,0.23){2}{\line(0,1){0.23}}
\multiput(99.88,57.22)(0.1,0.23){2}{\line(0,1){0.23}}
\multiput(99.69,56.76)(0.09,0.23){2}{\line(0,1){0.23}}
\multiput(99.51,56.29)(0.18,0.46){1}{\line(0,1){0.46}}
\multiput(99.35,55.83)(0.17,0.47){1}{\line(0,1){0.47}}
\multiput(99.19,55.35)(0.16,0.47){1}{\line(0,1){0.47}}
\multiput(99.05,54.88)(0.14,0.48){1}{\line(0,1){0.48}}
\multiput(98.91,54.4)(0.13,0.48){1}{\line(0,1){0.48}}
\multiput(98.79,53.92)(0.12,0.48){1}{\line(0,1){0.48}}
\multiput(98.68,53.43)(0.11,0.48){1}{\line(0,1){0.48}}
\multiput(98.58,52.95)(0.1,0.49){1}{\line(0,1){0.49}}
\multiput(98.49,52.46)(0.09,0.49){1}{\line(0,1){0.49}}
\multiput(98.41,51.97)(0.08,0.49){1}{\line(0,1){0.49}}
\multiput(98.34,51.48)(0.07,0.49){1}{\line(0,1){0.49}}
\multiput(98.28,50.98)(0.06,0.49){1}{\line(0,1){0.49}}
\multiput(98.24,50.49)(0.05,0.49){1}{\line(0,1){0.49}}
\multiput(98.21,49.99)(0.03,0.5){1}{\line(0,1){0.5}}
\multiput(98.18,49.5)(0.02,0.5){1}{\line(0,1){0.5}}
\multiput(98.17,49)(0.01,0.5){1}{\line(0,1){0.5}}
\put(98.17,48.5){\line(0,1){0.5}}
\multiput(98.17,48.5)(0.01,-0.5){1}{\line(0,-1){0.5}}
\multiput(98.18,48)(0.02,-0.5){1}{\line(0,-1){0.5}}
\multiput(98.21,47.51)(0.03,-0.5){1}{\line(0,-1){0.5}}
\multiput(98.24,47.01)(0.05,-0.49){1}{\line(0,-1){0.49}}
\multiput(98.28,46.52)(0.06,-0.49){1}{\line(0,-1){0.49}}
\multiput(98.34,46.02)(0.07,-0.49){1}{\line(0,-1){0.49}}
\multiput(98.41,45.53)(0.08,-0.49){1}{\line(0,-1){0.49}}
\multiput(98.49,45.04)(0.09,-0.49){1}{\line(0,-1){0.49}}
\multiput(98.58,44.55)(0.1,-0.49){1}{\line(0,-1){0.49}}
\multiput(98.68,44.07)(0.11,-0.48){1}{\line(0,-1){0.48}}
\multiput(98.79,43.58)(0.12,-0.48){1}{\line(0,-1){0.48}}
\multiput(98.91,43.1)(0.13,-0.48){1}{\line(0,-1){0.48}}
\multiput(99.05,42.62)(0.14,-0.48){1}{\line(0,-1){0.48}}
\multiput(99.19,42.15)(0.16,-0.47){1}{\line(0,-1){0.47}}
\multiput(99.35,41.67)(0.17,-0.47){1}{\line(0,-1){0.47}}
\multiput(99.51,41.21)(0.18,-0.46){1}{\line(0,-1){0.46}}
\multiput(99.69,40.74)(0.09,-0.23){2}{\line(0,-1){0.23}}
\multiput(99.88,40.28)(0.1,-0.23){2}{\line(0,-1){0.23}}
\multiput(100.08,39.83)(0.1,-0.23){2}{\line(0,-1){0.23}}
\multiput(100.29,39.37)(0.11,-0.22){2}{\line(0,-1){0.22}}
\multiput(100.5,38.93)(0.11,-0.22){2}{\line(0,-1){0.22}}
\multiput(100.73,38.49)(0.12,-0.22){2}{\line(0,-1){0.22}}
\multiput(100.97,38.05)(0.12,-0.22){2}{\line(0,-1){0.22}}
\multiput(101.22,37.62)(0.13,-0.21){2}{\line(0,-1){0.21}}
\multiput(101.48,37.2)(0.13,-0.21){2}{\line(0,-1){0.21}}
\multiput(101.75,36.78)(0.14,-0.21){2}{\line(0,-1){0.21}}
\multiput(102.02,36.36)(0.14,-0.2){2}{\line(0,-1){0.2}}
\multiput(102.31,35.96)(0.15,-0.2){2}{\line(0,-1){0.2}}
\multiput(102.61,35.56)(0.1,-0.13){3}{\line(0,-1){0.13}}
\multiput(102.91,35.17)(0.1,-0.13){3}{\line(0,-1){0.13}}
\multiput(103.22,34.78)(0.11,-0.13){3}{\line(0,-1){0.13}}
\multiput(103.55,34.4)(0.11,-0.12){3}{\line(0,-1){0.12}}
\multiput(103.88,34.03)(0.11,-0.12){3}{\line(0,-1){0.12}}
\multiput(104.22,33.67)(0.12,-0.12){3}{\line(0,-1){0.12}}
\multiput(104.56,33.31)(0.12,-0.12){3}{\line(1,0){0.12}}
\multiput(104.92,32.97)(0.12,-0.11){3}{\line(1,0){0.12}}
\multiput(105.28,32.63)(0.12,-0.11){3}{\line(1,0){0.12}}
\multiput(105.65,32.3)(0.13,-0.11){3}{\line(1,0){0.13}}
\multiput(106.03,31.97)(0.13,-0.1){3}{\line(1,0){0.13}}
\multiput(106.42,31.66)(0.13,-0.1){3}{\line(1,0){0.13}}
\multiput(106.81,31.36)(0.2,-0.15){2}{\line(1,0){0.2}}
\multiput(107.21,31.06)(0.2,-0.14){2}{\line(1,0){0.2}}
\multiput(107.61,30.77)(0.21,-0.14){2}{\line(1,0){0.21}}
\multiput(108.03,30.5)(0.21,-0.13){2}{\line(1,0){0.21}}
\multiput(108.45,30.23)(0.21,-0.13){2}{\line(1,0){0.21}}
\multiput(108.87,29.97)(0.22,-0.12){2}{\line(1,0){0.22}}
\multiput(109.3,29.72)(0.22,-0.12){2}{\line(1,0){0.22}}
\multiput(109.74,29.48)(0.22,-0.11){2}{\line(1,0){0.22}}
\multiput(110.18,29.25)(0.22,-0.11){2}{\line(1,0){0.22}}
\multiput(110.62,29.04)(0.23,-0.1){2}{\line(1,0){0.23}}
\multiput(111.08,28.83)(0.23,-0.1){2}{\line(1,0){0.23}}
\multiput(111.53,28.63)(0.23,-0.09){2}{\line(1,0){0.23}}
\multiput(111.99,28.44)(0.46,-0.18){1}{\line(1,0){0.46}}
\multiput(112.46,28.26)(0.47,-0.17){1}{\line(1,0){0.47}}
\multiput(112.92,28.1)(0.47,-0.16){1}{\line(1,0){0.47}}
\multiput(113.4,27.94)(0.48,-0.14){1}{\line(1,0){0.48}}
\multiput(113.87,27.8)(0.48,-0.13){1}{\line(1,0){0.48}}
\multiput(114.35,27.66)(0.48,-0.12){1}{\line(1,0){0.48}}
\multiput(114.83,27.54)(0.48,-0.11){1}{\line(1,0){0.48}}
\multiput(115.32,27.43)(0.49,-0.1){1}{\line(1,0){0.49}}
\multiput(115.8,27.33)(0.49,-0.09){1}{\line(1,0){0.49}}
\multiput(116.29,27.24)(0.49,-0.08){1}{\line(1,0){0.49}}
\multiput(116.78,27.16)(0.49,-0.07){1}{\line(1,0){0.49}}
\multiput(117.27,27.09)(0.49,-0.06){1}{\line(1,0){0.49}}
\multiput(117.77,27.03)(0.49,-0.05){1}{\line(1,0){0.49}}
\multiput(118.26,26.99)(0.5,-0.03){1}{\line(1,0){0.5}}
\multiput(118.76,26.96)(0.5,-0.02){1}{\line(1,0){0.5}}
\multiput(119.25,26.93)(0.5,-0.01){1}{\line(1,0){0.5}}
\put(119.75,26.92){\line(1,0){0.5}}
\multiput(120.25,26.92)(0.5,0.01){1}{\line(1,0){0.5}}
\multiput(120.75,26.93)(0.5,0.02){1}{\line(1,0){0.5}}
\multiput(121.24,26.96)(0.5,0.03){1}{\line(1,0){0.5}}
\multiput(121.74,26.99)(0.49,0.05){1}{\line(1,0){0.49}}
\multiput(122.23,27.03)(0.49,0.06){1}{\line(1,0){0.49}}
\multiput(122.73,27.09)(0.49,0.07){1}{\line(1,0){0.49}}
\multiput(123.22,27.16)(0.49,0.08){1}{\line(1,0){0.49}}
\multiput(123.71,27.24)(0.49,0.09){1}{\line(1,0){0.49}}
\multiput(124.2,27.33)(0.49,0.1){1}{\line(1,0){0.49}}
\multiput(124.68,27.43)(0.48,0.11){1}{\line(1,0){0.48}}
\multiput(125.17,27.54)(0.48,0.12){1}{\line(1,0){0.48}}
\multiput(125.65,27.66)(0.48,0.13){1}{\line(1,0){0.48}}
\multiput(126.13,27.8)(0.48,0.14){1}{\line(1,0){0.48}}
\multiput(126.6,27.94)(0.47,0.16){1}{\line(1,0){0.47}}
\multiput(127.08,28.1)(0.47,0.17){1}{\line(1,0){0.47}}
\multiput(127.54,28.26)(0.46,0.18){1}{\line(1,0){0.46}}
\multiput(128.01,28.44)(0.23,0.09){2}{\line(1,0){0.23}}
\multiput(128.47,28.63)(0.23,0.1){2}{\line(1,0){0.23}}
\multiput(128.92,28.83)(0.23,0.1){2}{\line(1,0){0.23}}
\multiput(129.38,29.04)(0.22,0.11){2}{\line(1,0){0.22}}
\multiput(129.82,29.25)(0.22,0.11){2}{\line(1,0){0.22}}
\multiput(130.26,29.48)(0.22,0.12){2}{\line(1,0){0.22}}
\multiput(130.7,29.72)(0.22,0.12){2}{\line(1,0){0.22}}
\multiput(131.13,29.97)(0.21,0.13){2}{\line(1,0){0.21}}
\multiput(131.55,30.23)(0.21,0.13){2}{\line(1,0){0.21}}
\multiput(131.97,30.5)(0.21,0.14){2}{\line(1,0){0.21}}
\multiput(132.39,30.77)(0.2,0.14){2}{\line(1,0){0.2}}
\multiput(132.79,31.06)(0.2,0.15){2}{\line(1,0){0.2}}
\multiput(133.19,31.36)(0.13,0.1){3}{\line(1,0){0.13}}
\multiput(133.58,31.66)(0.13,0.1){3}{\line(1,0){0.13}}
\multiput(133.97,31.97)(0.13,0.11){3}{\line(1,0){0.13}}
\multiput(134.35,32.3)(0.12,0.11){3}{\line(1,0){0.12}}
\multiput(134.72,32.63)(0.12,0.11){3}{\line(1,0){0.12}}
\multiput(135.08,32.97)(0.12,0.12){3}{\line(1,0){0.12}}
\multiput(135.44,33.31)(0.12,0.12){3}{\line(0,1){0.12}}
\multiput(135.78,33.67)(0.11,0.12){3}{\line(0,1){0.12}}
\multiput(136.12,34.03)(0.11,0.12){3}{\line(0,1){0.12}}
\multiput(136.45,34.4)(0.11,0.13){3}{\line(0,1){0.13}}
\multiput(136.78,34.78)(0.1,0.13){3}{\line(0,1){0.13}}
\multiput(137.09,35.17)(0.1,0.13){3}{\line(0,1){0.13}}
\multiput(137.39,35.56)(0.15,0.2){2}{\line(0,1){0.2}}
\multiput(137.69,35.96)(0.14,0.2){2}{\line(0,1){0.2}}
\multiput(137.98,36.36)(0.14,0.21){2}{\line(0,1){0.21}}
\multiput(138.25,36.78)(0.13,0.21){2}{\line(0,1){0.21}}
\multiput(138.52,37.2)(0.13,0.21){2}{\line(0,1){0.21}}
\multiput(138.78,37.62)(0.12,0.22){2}{\line(0,1){0.22}}
\multiput(139.03,38.05)(0.12,0.22){2}{\line(0,1){0.22}}
\multiput(139.27,38.49)(0.11,0.22){2}{\line(0,1){0.22}}
\multiput(139.5,38.93)(0.11,0.22){2}{\line(0,1){0.22}}
\multiput(139.71,39.37)(0.1,0.23){2}{\line(0,1){0.23}}
\multiput(139.92,39.83)(0.1,0.23){2}{\line(0,1){0.23}}
\multiput(140.12,40.28)(0.09,0.23){2}{\line(0,1){0.23}}
\multiput(140.31,40.74)(0.18,0.46){1}{\line(0,1){0.46}}
\multiput(140.49,41.21)(0.17,0.47){1}{\line(0,1){0.47}}
\multiput(140.65,41.67)(0.16,0.47){1}{\line(0,1){0.47}}
\multiput(140.81,42.15)(0.14,0.48){1}{\line(0,1){0.48}}
\multiput(140.95,42.62)(0.13,0.48){1}{\line(0,1){0.48}}
\multiput(141.09,43.1)(0.12,0.48){1}{\line(0,1){0.48}}
\multiput(141.21,43.58)(0.11,0.48){1}{\line(0,1){0.48}}
\multiput(141.32,44.07)(0.1,0.49){1}{\line(0,1){0.49}}
\multiput(141.42,44.55)(0.09,0.49){1}{\line(0,1){0.49}}
\multiput(141.51,45.04)(0.08,0.49){1}{\line(0,1){0.49}}
\multiput(141.59,45.53)(0.07,0.49){1}{\line(0,1){0.49}}
\multiput(141.66,46.02)(0.06,0.49){1}{\line(0,1){0.49}}
\multiput(141.72,46.52)(0.05,0.49){1}{\line(0,1){0.49}}
\multiput(141.76,47.01)(0.03,0.5){1}{\line(0,1){0.5}}
\multiput(141.79,47.51)(0.02,0.5){1}{\line(0,1){0.5}}
\multiput(141.82,48)(0.01,0.5){1}{\line(0,1){0.5}}

\linethickness{0.3mm}
\multiput(135,65)(0.18,0.12){167}{\line(1,0){0.18}}
\linethickness{0.3mm}
\linethickness{0.3mm}
\multiput(130,30)(0.21,-0.12){167}{\line(1,0){0.21}}
\linethickness{0.3mm}
\linethickness{0.3mm}
\linethickness{0.3mm}
\linethickness{0.3mm}
\multiput(15,70)(0.15,-0.12){167}{\line(1,0){0.15}}
\linethickness{0.3mm}
\put(20,50){\line(1,0){20}}
\linethickness{0.3mm}
\multiput(25,25)(0.12,0.2){125}{\line(0,1){0.2}}
\linethickness{0.3mm}
\multiput(15,35)(0.2,0.12){125}{\line(1,0){0.2}}
\linethickness{0.3mm}
\qbezier(40,50)(50.4,60.44)(58.22,65.25)
\qbezier(58.22,65.25)(66.04,70.06)(72.5,70)
\qbezier(72.5,70)(78.99,70.01)(84.41,69.41)
\qbezier(84.41,69.41)(89.82,68.8)(95,67.5)
\qbezier(95,67.5)(100.22,66.2)(102.62,65.59)
\qbezier(102.62,65.59)(105.03,64.99)(105,65)
\linethickness{0.3mm}
\qbezier(40,50)(55.63,52.61)(64.66,53.81)
\qbezier(64.66,53.81)(73.68,55.02)(77.5,55)
\qbezier(77.5,55)(81.37,55)(86.78,55)
\qbezier(86.78,55)(92.2,55)(100,55)
\linethickness{0.3mm}
\qbezier(40,50)(50.41,39.58)(57.62,33.56)
\qbezier(57.62,33.56)(64.84,27.55)(70,25)
\qbezier(70,25)(75.2,22.39)(79.41,21.19)
\qbezier(79.41,21.19)(83.62,19.98)(87.5,20)
\qbezier(87.5,20)(91.37,19.97)(96.78,22.38)
\qbezier(96.78,22.38)(102.2,24.78)(110,30)
\linethickness{0.3mm}
\qbezier(40,50)(53.01,44.79)(62.03,41.78)
\qbezier(62.03,41.78)(71.05,38.77)(77.5,37.5)
\qbezier(77.5,37.5)(83.99,36.18)(89.41,36.78)
\qbezier(89.41,36.78)(94.82,37.38)(100,40)
\put(40,60){\makebox(0,0)[cc]{X}}

\put(75,75){\makebox(0,0)[cc]{$k_1$}}

\put(70,60){\makebox(0,0)[cc]{$k_2$}}

\put(75,15){\makebox(0,0)[cc]{$k_s$}}

\put(70,50){\makebox(0,0)[cc]{$\cdot$}}

\put(70,47){\makebox(0,0)[cc]{$\cdot$}}

\put(70,44){\makebox(0,0)[cc]{$\cdot$}}

\put(150,50){\makebox(0,0)[cc]{$\cdot$}}

\put(150,60){\makebox(0,0)[cc]{$\cdot$}}

\put(150,40){\makebox(0,0)[cc]{$\cdot$}}

\put(70,33){\makebox(0,0)[cc]{$k_{s-1}$}}

\put(120,50){\makebox(0,0)[cc]{A}}

\linethickness{0.3mm}
\put(40,50){\circle{10}}

\end{picture}

}

\def\figtwentytwo{

\def\JPicScale{0.6}
\ifx\JPicScale\undefined\def\JPicScale{1}\fi
\unitlength \JPicScale mm
\begin{picture}(120,80)(0,0)
\linethickness{0.3mm}
\multiput(10,60)(0.24,-0.12){83}{\line(1,0){0.24}}
\linethickness{0.3mm}
\multiput(10,40)(0.24,0.12){83}{\line(1,0){0.24}}
\linethickness{0.3mm}
\put(10,50){\line(1,0){20}}
\linethickness{0.3mm}
\multiput(30,50)(0.36,0.12){83}{\line(1,0){0.36}}
\linethickness{0.3mm}
\put(60,60){\line(1,0){35}}
\linethickness{0.3mm}
\multiput(60,60)(0.12,-0.12){250}{\line(1,0){0.12}}
\linethickness{0.3mm}
\multiput(30,50)(0.12,-0.12){167}{\line(1,0){0.12}}

\linethickness{0.3mm}
\put(50,30){\line(1,0){40}}

\linethickness{0.3mm}
\multiput(70,30)(0.48,0.22){10}{\line(1,0){0.48}}
\multiput(70,30)(0.48,-0.22){10}{\line(1,0){0.48}}

\linethickness{0.3mm}
\multiput(80,60)(-0.48,0.22){10}{\line(1,0){0.48}}
\multiput(80,60)(-0.48,-0.22){10}{\line(1,0){0.48}}

\linethickness{0.3mm}
\multiput(50,57)(-0.48,0.12){10}{\line(1,0){0.48}}
\multiput(51,57)(-0.48,-0.42){10}{\line(1,0){0.48}}

\linethickness{0.3mm}
\multiput(80,40)(-0.48,0.22){10}{\line(1,0){0.48}}
\multiput(80,40)(-0.28,0.62){8}{\line(1,0){0.48}}

\linethickness{0.3mm}
\put(30,50){\circle{10}}

\linethickness{0.3mm}
\put(60,60){\circle{10}}

\linethickness{0.3mm}
\put(50,30){\circle{10}}

\linethickness{0.3mm}
\put(90,30){\circle{10}}

\linethickness{0.3mm}
\put(95,60){\circle{10}}

\linethickness{0.3mm}
\multiput(40,40)(0.48,0){10}{\line(1,0){0.48}}
\put(40,40){\line(0,-1){5}}

\linethickness{0.3mm}
\multiput(50,30)(0.48,-0.12){42}{\line(1,0){0.48}}
\linethickness{0.3mm}
\put(50,15){\line(0,1){15}}
\linethickness{0.3mm}
\multiput(95,60)(0.15,0.12){167}{\line(1,0){0.15}}
\linethickness{0.3mm}
\put(95,60){\line(1,0){25}}
\linethickness{0.3mm}
\multiput(95,60)(0.15,-0.12){167}{\line(1,0){0.15}}
\linethickness{0.3mm}
\put(90,30){\line(1,0){20}}
\linethickness{0.3mm}
\put(90,10){\line(0,1){20}}
\end{picture}

}

\def\figtwenty{

\def\JPicScale{0.6}
\ifx\JPicScale\undefined\def\JPicScale{1}\fi
\unitlength \JPicScale mm
\begin{picture}(100,50)(0,0)
\linethickness{0.3mm}
\put(45,45){\circle{10}}

\linethickness{0.3mm}
\put(95,45){\circle{10}}

\linethickness{.3mm}
\put(50,45){\line(1,0){40}}

\end{picture}

}

\def\figonevr{

\def\JPicScale{0.6}
\ifx\JPicScale\undefined\def\JPicScale{1}\fi
\unitlength \JPicScale mm

}

\def\figonevrA{

\def\JPicScale{0.6}
\ifx\JPicScale\undefined\def\JPicScale{1}\fi
\unitlength \JPicScale mm

}

\def\figonevrB{

\def\JPicScale{0.6}
\ifx\JPicScale\undefined\def\JPicScale{1}\fi
\unitlength \JPicScale mm

}

\def\figonevrr{

\def\JPicScale{0.6}
\ifx\JPicScale\undefined\def\JPicScale{1}\fi
\unitlength \JPicScale mm
\begin{picture}(100,70)(0,0)
\linethickness{0.3mm}
\put(45,65){\circle{10}}

\linethickness{0.3mm}
\put(70,5){\circle{10}}

\linethickness{0.3mm}
\put(95,25){\circle{10}}

\linethickness{0.3mm}
\put(45,25){\circle{10}}

\linethickness{0.3mm}
\put(70,45){\circle{10}}

\linethickness{0.3mm}
\put(95,65){\circle{10}}

\linethickness{0.3mm}
\linethickness{0.3mm}

\linethickness{0.3mm}
\multiput(70,5)(0.3,0.25){83}{\line(1,0){0.3}}
\linethickness{0.3mm}
\multiput(45,25)(0.3,-0.25){83}{\line(1,0){0.3}}

\linethickness{0.3mm}

\linethickness{0.3mm}
\multiput(70,45)(-0.3,-0.25){82}{\line(1,0){0.3}} %

\linethickness{0.3mm}
\multiput(70,45)(0.3,-0.25){82}{\line(1,0){0.3}} %

\linethickness{0.3mm}
\multiput(70,45)(-0.3,0.25){82}{\line(1,0){0.3}} %

\linethickness{0.3mm}
\multiput(70,45)(0.3,0.25){82}{\line(1,0){0.3}} %

\linethickness{0.3mm}
\put(45,65){\line(1,0){50}} %
\end{picture}

}

\def\figdis{

\def\JPicScale{0.6}
\ifx\JPicScale\undefined\def\JPicScale{1}\fi
\unitlength \JPicScale mm

}

\def\figupdpwn{

\def\JPicScale{0.6}
\ifx\JPicScale\undefined\def\JPicScale{1}\fi
\unitlength \JPicScale mm
\begin{picture}(236.25,70)(0,0)
\linethickness{0.3mm}
\put(45,40){\circle{12.5}}

\linethickness{0.3mm}
\put(120,40){\circle{12.5}}

\linethickness{0.3mm}
\put(85,25){\circle{12.5}}

\linethickness{0.3mm}
\put(85,55){\circle{12.5}}

\linethickness{0.3mm}
\put(195,25){\circle{12.5}}

\linethickness{0.3mm}
\put(230,40){\circle{12.5}}

\linethickness{0.3mm}
\put(195,55){\circle{12.5}}

\linethickness{0.3mm}
\put(155,40){\circle{12.5}}

\linethickness{0.3mm}
\multiput(45,40)(0.32,0.12){125}{\line(1,0){0.32}}
\linethickness{0.3mm}
\multiput(85,55)(0.28,-0.12){125}{\line(1,0){0.28}}
\linethickness{0.3mm}
\multiput(85,25)(0.28,0.12){125}{\line(1,0){0.28}}
\linethickness{0.3mm}
\multiput(45,40)(0.32,-0.12){125}{\line(1,0){0.32}}
\linethickness{0.3mm}
\multiput(155,40)(0.32,0.12){125}{\line(1,0){0.32}}
\linethickness{0.3mm}
\multiput(195,55)(0.28,-0.12){125}{\line(1,0){0.28}}
\linethickness{0.3mm}
\multiput(195,25)(0.28,0.12){125}{\line(1,0){0.28}}
\linethickness{0.3mm}
\multiput(155,40)(0.32,-0.12){125}{\line(1,0){0.32}}
\put(80,10){\makebox(0,0)[cc]{(a)}}

\put(195,10){\makebox(0,0)[cc]{(b)}}

\put(65,25){\makebox(0,0)[cc]{$P_1$}}

\put(105,25){\makebox(0,0)[cc]{$P_2$}}

\put(175,25){\makebox(0,0)[cc]{$P_1$}}

\put(215,25){\makebox(0,0)[cc]{$P_2$}}

\linethickness{0.3mm}
\put(105,30){\line(0,1){10}}
\linethickness{0.3mm}
\put(60,10){\line(0,1){60}}
\linethickness{0.3mm}
\put(170,10){\line(0,1){60}}
\linethickness{0.3mm}
\put(180,25){\line(0,1){10}}
\linethickness{0.3mm}
\put(220,30){\line(0,1){10}}
\end{picture}

}

\begin{document}

\baselineskip 24pt

\begin{center}
{\Large \bf  Cutkosky Rules for Superstring Field Theory}

\end{center}

\vskip .6cm
\medskip

\vspace*{4.0ex}

\baselineskip=18pt

\centerline{\large \rm Roji Pius$^a$ and Ashoke Sen$^b$}

\vspace*{4.0ex}

\centerline{\large \it $^a$Perimeter Institute for Theoretical Physics} 
\centerline{\large \it  Waterloo, 
ON N2L 2Y5, Canada}

\centerline{\large \it $^b$Harish-Chandra Research Institute}
\centerline{\large \it  Chhatnag Road, Jhusi,
Allahabad 211019, India}

\vspace*{1.0ex}
\centerline{\small E-mail:  rpius@perimeterinstitute.ca, sen@mri.ernet.in}

\vspace*{5.0ex}

\centerline{\bf Abstract} \bigskip

Superstring field theory expresses the perturbative 
S-matrix of superstring theory as a sum of Feynman diagrams
each of which 
is manifestly free from ultraviolet divergences. The interaction vertices 
fall off exponentially for large space-like external momenta making the ultraviolet finiteness
property manifest, but blow up exponentially for large time-like external momenta making it
impossible to take the integration contours for loop energies to lie along the
real axis. This forces us to carry out the integrals over the loop energies by 
choosing appropriate contours in the complex plane whose ends go to infinity along
the imaginary axis but which take complicated form in the interior navigating 
around
the various poles of the propagators. We consider the general class of quantum 
field theories with this property and prove Cutkosky rules for the amplitudes to all orders
in perturbation theory. Besides having applications to string field theory, these results
also give an alternative derivation of Cutkosky rules in ordinary quantum
field theories.

\vfill \eject

\tableofcontents

\baselineskip=18pt

\sectiono{Introduction} \label{s0}

Unitarity is a necessary property of any theory that aims at describing the fundamental constituents
of matter and their interactions. Since superstring theory is, at present, the leading candidate for
such a theory, it is necessary to ensure that the scattering matrix computed from superstring theory
is unitary. The goal of this paper will be to address this issue in superstring perturbation theory.

Our strategy will be to make use of superstring field theory\footnote{Our analysis
will not require using any specific version of superstring field theory. For
definiteness we can consider the version of quantum superstring field theory
considered in \cite{1508.05387}. Most of the other recent work has been towards the
construction of classical  open and/or closed superstring field 
theory\cite{wittenssft,9202087,
9503099,0109100,0406212,0409018,1312.2948,1312.7197,
1403.0940,1407.8485,1412.5281,1505.01659,1506.05774,1506.06657,1507.08250,
1508.00366,1512.03379,1602.02582,1602.02583}. 
If they can be elevated to consistent
quantum theory, they may provide equally good candidates for our analysis. 
One may also be able to use
non-local versions of closed superstring field theory of the kind suggested in 
\cite{1303.2323}.} 
-- a quantum field theory whose
Feynman rules reproduce the perturbative amplitudes  computed using 
the conventional 
Polyakov approach. 
The advantage of using superstring field theory is that we can use the well 
known techniques of 
quantum field theory to address various issues. 
In particular one might expect that the conventional approach to proving unitarity
of quantum field theories using Cutkosky 
rules\cite{Cutkosky, fowler,veltman,diagrammar,1512.01705}
may be used to give a proof of unitarity of
superstring perturbation theory, since these rules
encode the  perturbation expansion
of the relation $S^\dagger S=1$
satisfied by the S-matrix $S$.

It turns out however that
there is one way in which superstring field theory differs from conventional
quantum field theories. The interaction vertices of superstring field theory 
have the property that
they fall off exponentially when the external states carry large space-like momenta. 
This property
is what makes the superstring perturbation expansion manifestly free from ultraviolet divergences.
However there is a flip side to this story -- for large time-like momenta the interaction vertices
diverge exponentially. For this reason, the only way to make sense of integration over loop energies
is to let the energy integration contours reach infinity along the imaginary axis. If we 
consider the Wick rotated 
Green's function in which all the external states carry imaginary energy,
this is straightforward. We simply take all the loop energy integrals to lie along the imaginary
axis so that all the propagators and vertices carry imaginary energy. This leads to
non-singular integrand with exponential fall-off at infinity and the integral is well defined.
In a conventional quantum field theory, we could inverse Wick rotate\footnote{In our
notation, Wick rotation will denote taking the energies from the real axis 
to the imaginary axis,
while inverse Wick rotation will correspond to taking them from the imaginary axis
to the real axis.}
all the external 
energies towards the real axis and at the same time rotate the energy integration contours 
clockwise
from the imaginary axis to the real axis, eventually arriving at the formalism where the 
energies of external states are real, and the loop energy
integrals run along the real axis with $i\eps$ prescription for dealing with the poles of the
propagator. However such an integral will be ill defined in string field theory, since the
vertex factors will blow up exponentially as the loop 
energy integrals approach infinity along the real
axis. For this reason, even when we inverse Wick rotate the external energies from the
imaginary axis back to the real axis, we must continue to let the loop
energy integration 
contours reach infinity along the imaginary axis. However we can no longer ensure that 
these integrals run all along the imaginary axis since during the inverse
Wick rotation of 
external energies, some of the
poles of the propagator will approach the imaginary energy axis and we have to deform the
integration contour 
away from these poles in order to ensure that we get the analytic continuation of
the Wick rotated 
result. As a result, when the external energies reach the real axis, we typically will
have a complicated integration contour over the loop energies with their ends tied at
$\pm i\infty$. 
For example for the one loop amplitude shown in Fig.~\ref{f1} in page
\pageref{f1}, a possible integration
contour over the loop energy is shown in Fig.~\ref{f3} in page \pageref{f3}.

Since the proof of unitarity involves identifying the anti-hermitian part of the amplitude,
we now have to identify the anti-hermitian part of this Feynman integral. {\it A priori}
the result looks complicated due to the fact that the choice of integration contour does not
have simple reality properties. One can in fact 
show that the prescription for computing the hermitian
conjugate of the T-matrix reduces to the 
computation of a Feynman integral similar to the original integral,
with all the external energies replaced by their complex conjugates and
the integration contour over the loop energies  
related to the original contour by complex conjugation.
The main result of this paper involves proving that to all orders in perturbation theory,
this difference between the two integrals is given by Cutkosky rules in the limit when
the external energies approach the real axis.

If we denote by $T$ the T-matrix related to the S-matrix via the relation $S=1-iT$, then
Cutkosky rules express the difference between $T$ and its hermitian
conjugate $T^\dagger$ as a sum of cut Feynman diagrams in which we draw an
oriented line through
the Feynman diagrams contributing to the original T-matrix, dividing the diagram into
two pieces. In every cut propagator, the original propagator is replaced by the product
of a  delta function that sets the momentum 
along the propagator on-shell and a step function that forces the energy of the propagator
to flow from the left to the
right of the cut. The contribution from part of the Feynman diagram to
the left of the cut is computed using the usual
Feynman rules and the contribution from part of the Feynman diagram to the 
right of the cut is given by the hermitian
conjugate of the corresponding Feynman diagram. 
The contributions from the cut diagrams have
the interpretation of the matrix elements of $T^\dagger T$, computed by inserting a complete set
of states between $T^\dagger$ and $T$ represented by the cut propagators.
After taking into account the factors of $i$
we arrive at the relation $T-T^\dagger = -i T^\dagger T$, which is precisely the statement
of unitarity of the S-matrix.

In quantum field theories where all the fields represent fundamental particles, the
Cutkosky rules establish the unitarity of the S-matrix. For theories with local
gauge symmetry, including string field theory,
Cutkosky rules are necessary ingredients for the proof of unitarity, but they are not
sufficient. These theories contain many unphysical and pure gauge states
besides physical states, and we must show that only the physical states 
contribute
to the sum over intermediate states. In conventional gauge theories this is proved
using Ward identities (see {\it e.g.} \cite{diagrammar}). 
Since gauge invariance of string theory leads to similar Ward
identities\cite{1508.02481}, we expect that they can be used to complete the proof of unitarity.
We leave this for future work.

The paper is organized as follows. In \S\ref{s1} we introduce a toy scalar field theory that 
captures all the essential properties of string field theory that goes into the proof of
Cutkosky rules. 
In order to define an amplitude in this theory 
with Lorentzian external momenta, with the $s$-th external particle carrying 
spatial momenta $\vec p_s$
and 
energy $E_s$, we begin with an amplitude where 
the $s$-th particle has  spatial momenta $\vec p_s$
and energy $\lambda E_s$, where $\lambda$ is a complex parameter. 
For purely imaginary $\lambda$ the amplitude is defined by taking all the loop
energy integrals along the imaginary axis. We then define the physical amplitude, corresponding
to $\lambda=1$, by analytic continuation of the result on the imaginary $\lambda$-axis
to the real $\lambda$-axis {\it via the first quadrant of the complex $\lambda$-plane.}
In \S\ref{s3} we prove that this analytic continuation procedure 
is well defined by showing that the amplitude does not have any singularity 
in the first quadrant of the $\lambda$-plane.
In \S\ref{s1}
we also derive an algorithm for computing 
the hermitian conjugate of an amplitude. 

In \S\ref{s2} we consider
a simple one loop amplitude in this theory and show how Cutkosky rules hold for this
amplitude. The complete proof to all orders in perturbation theory is carried out
in \S\ref{scut}. This is done in several steps. First we show that for fixed values of 
the spatial components of
loop momenta the contribution to the anti-hermitian part of a
connected amplitude is non-vanishing 
only when some of the integration contours over the loop 
energy integrals are pinched, i.e. 
two poles approach each other from opposite sides of a contour
so that we cannot deform the contour away
from the poles without passing through a pole. Then we divide the contribution from the
pinch singularities into two classes, one vertex irreducible (1VI) diagram and one vertex 
reducible (1VR) diagrams, and show that Cutkosky rules hold for the 1VR diagrams as
long as they hold for the 1VI diagrams. Next we prove the 
Cutkosky rules for 1VI diagrams. Finally we prove
that the Cutkosky rules for disconnected diagrams follow as a consequence of the
Cutkosky rules for connected diagrams. Our proof uses the method of induction in the
number of loops, and holds to all orders in perturbation theory.

In \S\ref{s6} we describe how the analysis of the toy model
 in the previous sections captures most, but not all, of the
ingredients needed to prove unitarity of superstring field theory.
We
discuss 
what else needs to be done to prove the unitarity of superstring perturbation theory. Some
of these are common to ordinary quantum field theories, {\it e.g.} we need to work in
sufficiently high dimensions so that we avoid the usual infrared divergence problems 
that plague quantum field theories in dimensions $\le 4$, and we need to prove the
cancellation of the contributions from intermediate unphysical and pure gauge states using
Ward identities. However some of them are
purely technical problems in string field theory -- {\it e.g.} proving the reality of the superstring field
theory action -- which we believe can be proven with some effort but has not been done
so far.

We conclude this introductory 
section by reviewing some of the previous work on this subject.  A complete proof
of unitarity of superstring perturbation theory was attempted in \cite{dhoker}
by showing the equivalence of the perturbative amplitudes in superstring theory
and the amplitudes in light-cone string field theory. 
Since the latter is manifestly
unitary, this would imply unitarity of the S-matrix computed in the covariant formulation.
In view of recent understanding of the subtleties of superstring perturbation 
theory\cite{1209.5461,1304.7798,1404.6257,1504.00609}
one should reinvestigate this correspondence.  Nevertheless it seems quite likely
that this will lead to a concrete formulation of light-cone string field theory which will
still be manifestly unitary and at the same time generate the usual amplitudes
of perturbative superstring theory. 
This would establish the unitarity of perturbative superstring
amplitudes. However the main advantage of using a covariant superstring field theory
for our analysis is that this theory can be used to analyze unitarity and other properties
of string theory not only in the perturbative vacuum, but also in situations where loop
corrections require us to shift the vacuum expectation values of the fields away from
that in the perturbative vacuum\cite{1508.02481}. 
The shift in the field will change the vertices, but not
their general properties on which we shall base our analysis as long as the 
string field theory action in the shifted background continues to be real.

One could also try to prove the unitarity of superstring perturbation theory directly
by using the $i\eps$ prescription for defining the perturbative amplitudes as given
in \cite{berera,1307.5124}. 
At this stage it is not known how this can be done, but it is conceivable
that one can translate this $i\eps$ prescription into a direct proof of unitarity
of the 
perturbative superstring amplitudes. However this will still suffer from the 
fact that the proof will not extend in a straightforward manner to the cases where
the true vacuum is related to the perturbative vacuum by a shift in the fields.

Finally we would like to add one word about convention. Throughout this paper
we shall use the notion of a pinch singularity to denote that the integration contours
over some loop energies
encounter poles approaching each other from opposite sides of the contours
so that 
by deforming the contours into the complex loop energy plane
we cannot avoid these poles. However
we shall always keep the integration over the spatial components of loop momenta
along the real axes. This notion differs from that used in the standard
literature {\it e.g.} in \cite{Cutkosky, fowler}, where a contour is declared to be pinched
only if it cannot be deformed away from the pole by deforming the integration contour
into the complex energy and / or complex spatial momentum plane. Due to this, some of
our results, {\it e.g.} that the anti-hermitian part of the amplitude comes only from pinch
singularities, may look unfamiliar to the experts. On the other hand, our approach leads
to a proof of the Cutkosky rules at fixed values of the spatial components of the
loop momenta and for general off-shell external states. This is close in spirit
to the results of \cite{sterman}, although the analysis of \cite{sterman} cannot be
applied directly to the class of field theories we consider due to essential singularities
of the interaction vertices at infinite momenta.

\sectiono{The field theory model} \label{s1}

In this section we shall introduce a toy quantum field theory that captures all
the essential features of the subtleties of string field theory action. Our model 
will involve a single scalar field. But the analysis we shall perform can be easily generalized 
to the case of multiple fields including fields of higher spin, since Lorentz invariance will
not play any significant role in our analysis. In \S\ref{s6} we shall discuss what additional
subtleties we need to address in order to translate the result of this paper to a complete
proof of unitarity of superstring field theory.

\subsection{The model} \label{s1.1}

We consider a scalar field theory in $(d+1)$-dimensions containing a single real scalar field
$\phi$, with the following 
action written in momentum space:
\ben \label{e2.1ff}
S &=& -{1\over 2} \int {d^{d+1} k\over (2\pi)^{d+1}} \phi(-k) (k^2 + m^2) \phi(k)  
\nonumber \\
&& -\sum_n {1\over n!}
(2\pi)^{-(n-1)(d+1)} \int d^{d+1} k_1 \cdots d^{d+1} k_n \, 
\delta^{(d+1)}(k_1+\cdots +k_n)  \nonumber \\ && 
\qquad \qquad \qquad \qquad  \qquad \qquad \times
V^{(n)} (k_1, \cdots k_n) \, \phi(k_1)\cdots \phi(k_n)\, . \een
Here $k^2\equiv -(k^0)^2 + (k^1)^2 +\cdots + (k^d)^2$,
$d^{d+1} k \equiv dk^0 dk^1 \cdots dk^{d}$ and the vertices $V^{(n)}$
satisfy the reality condition
\be  \label{ereal}
(V^{(n)}(k_1, \cdots k_n))^* = V^{(n)}(-k_1^*, \cdots -k_n^*)\, ,
\ee
where $*$ denotes complex conjugation. We take the  
$V^{(n)}$'s to be invariant under arbitrary permutation
of the arguments,
and assume that they
have no singularities in the
$k_s^\mu$ planes at finite values.
Furthermore, they vanish exponentially as  one or more $k_s^0$ approach $\pm i\infty$ 
along the
imaginary axis and/or  one or more 
$k_s^i$ for $1\le i\le d$ approach $\pm\infty$ along the real axis,
keeping the other $k_r$'s
fixed.  On the other hand, $V^{(n)}$ may
blow up exponentially as $k_s^0$ and/or
$k_s^i$ 
approach infinity in certain other directions, {\it e.g.}
along the  real $k_s^0$ axis or the imaginary $k_s^i$ axis. 
Therefore $V^{(n)}$'s have essential 
singularities at infinity.

Note that due to the exponential growth of $V^{(n)}$ for large time-like momentum,
a classical field configurations $\phi(k)$ with real argument $\{k^\mu\}$ 
will have finite action only
if it falls off sufficiently fast for large  $|k^0|$ so as to compensate for the exponential
growth of the $V^{(n)}$'s. Once this condition is satisfied and the
action is finite, then \refb{e2.1ff} is real for real field configuration satisfying $\phi(k)^*
=\phi(-k)$. We shall of course not be interested in classical field configirations -- for
us the significance of \refb{e2.1ff} lies in the fact that this is the property that we expect
the superstring field theory action to possess.

The Feynman rules 
for computing the T-matrix, related to
the S-matrix via $S=I - i\, T$, are as follows:
\ben \label{eifactorpre}
\hbox{propagator of momentum $k$} &:& -i \, (k^2+m^2)^{-1}\nonumber \\
\hbox{$n$-point vertex with incoming momenta $k_1,\cdots k_n$} &:&
-i \,V^{(n)}(k_1,\cdots k_n) \nonumber \\
\hbox{each loop momentum integration} &:& {d^{d+1} \ell\over (2\pi)^{d+1}}
\nonumber \\
\hbox{overall factor} &:& i \, (2\pi)^{d+1} \delta^{(d+1)} \left(\sum_s p_s\right)\, ,
\een
where in the last equation the sum over $s$ runs over all the external
momenta $p_s$ in the convention that $p_s$ denotes the momentum entering
the diagram from outside. If the diagram has disconnected components then there
will be separate momentum conserving delta function for each component.
These rules are derived from path integral expressions for the Green's
functions with weight factors $e^{iS}$.

We can simplify the Feynman rules somewhat by extracting a factor of $i$ from
each propagator, a factor of $-i$ from each vertex and the overall factor of $i$ given 
in the last line of 
\refb{eifactorpre}. This gives a total factor of $(i)^{n_p - n_v+1}$ where $n_p$ is the
number of propagators and $n_v$ is the number of vertices.
If $n_\ell$ is the
number of loops, then using the relation
\be \label{eloopver}
n_\ell = n_p - n_v + 1\, ,
\ee
that holds for a connected diagram,
we get a net factor of
\be \label{einl}
(i)^{n_p-n_v+1}=(i)^{n_\ell}\, .
\ee
If the diagram has $n_c$ disconnected components then \refb{eloopver}
will be replaced by $n_\ell = n_p - n_v + n_c$, and hence the net factor given
in the right hand side of \refb{einl} will be
\be \label{e2.5a}
(i)^{n_\ell - n_c+1}\, .
\ee
For now we shall proceed by assuming that the
diagram is connected so that \refb{einl} holds.
Therefore each loop integral is accompanied by a factor of $i$.
The modified Feynman rules now involve
\ben \label{eifactornew}
\hbox{propagator of momentum $k$} &:& P(k) = - (k^2+m^2)^{-1} 
\nonumber \\ &&
= \left(k^0 - \sqrt{\vec k^2+m^2}\right)^{-1}
 \left(k^0 + \sqrt{\vec k^2+m^2}\right)^{-1}\nonumber \\
\hbox{vertex with incoming momenta $k_1,\cdots k_n$} &:&
V^{(n)}(k_1,\cdots k_n) \nonumber \\
\hbox{each loop momentum integration} &:& i\, {d^{d+1} \ell\over (2\pi)^{d+1}}
\nonumber \\
\hbox{overall factor} &:&  (2\pi)^{d+1} \delta^{(d+1)} \left(\sum_k p_k\right)\, .
\een

Since the vertices diverge exponentially for large time-like external momenta,
individual Feynman diagrams in this theory have somewhat strange properties.
For example the s-channel diagram for a tree level 4-point function, in which a pair of
3-point functions are joined by a single internal propagator, will blow up 
exponentially in the limit of large center of mass energy of the incoming particles. 
In string field theory this effect is cancelled by the contribution from the 4-point vertex.
On the other hand, this property of the vertices makes the individual 
Feynman diagrams manifestly free from ultraviolet divergences once we choose 
the loop momentum integration
contours appropriately. This will be described in the next subsection.

\subsection{Loop momentum integration contours} \label{s2.2}

Due to the peculiar behavior of the interaction vertices at large momenta, the 
integration over loop momenta has to be defined somewhat carefully.
The integrals over $\vec\ell_k\equiv (\ell_k^1,\cdots \ell_k^d)$  -- the spatial
components of the loop momenta -- 
are taken to be along the real axis, but
the integration contours for the $\ell_k^0$'s  -- the zeroth
components of all the loop momenta -- are chosen as follows.
Let us denote collectively by $\{p_s^0\}$ the zero components of the external
momenta $\{p_s\}$. We shall introduce a set of numbers $\{E_s\}$ which denote
the actual real values of the $\{p_s^0\}$ for which we want to compute the 
Green's function,
and consider a more general set of external momenta where $p_s^0=\lambda
E_s$ with $\lambda$ an arbitrary complex number. 
When $\lambda$ is purely imaginary then we can get a well-defined expression
for the Green's function by taking the integration contour for $\ell_k^0$'s 
to run along the imaginary axis. In this case all the internal propagators carry imaginary
energy and real spatial momenta and hence are free from singularities. Furthermore
since the $\ell_k^0$ integration contour approaches infinity along
the imaginary axis and the $\ell_k^1,\cdots \ell_k^d$  integration contours approach 
infinity along
the real axis, we get a convergent loop momentum integral due to the convergence
property of $V^{(n)}$ discussed above.
We now define the off-shell Green's function for general $\lambda$ as the
analytic continuation of this expression in the complex $\lambda$ plane. 
Operationally this means that
as we deform $\lambda$ away from the imaginary
axis, we continue to define the Green's function by taking the integration contour over
$\ell_k^0$'s to run from $-i\infty$ to $i\infty$ till 
some pole of the integrand approaches the imaginary $\ell_k^0$ axis.
When a pole approaches the imaginary $\ell_k^0$ axis, we
deform the integration contour away from the imaginary axis to avoid the poles,
keeping its ends fixed at $\pm i\infty$. The integrations over the spatial components of
loop momenta are always taken to be along the real axis.\footnote{This prescription is 
not manifestly
Lorentz invariant, since the standard proof of Lorentz invariance requires us
to transform the loop momenta by the same Lorentz transformation that acts on the
external states, and this does not leave the end points of the
contour invariant. However the new
contour obtained by Lorentz transformation 
will also have the property that the integrand falls off 
exponentially at the two ends since the integrand is manifestly Lorentz invariant. 
As a result we can prove the equality of the new 
integral with the old integral by deforming the new contour to the old
one near the end points by successive infinitesimal Lorentz transformations.
}
We shall show in \S\ref{s3} that the off-shell
Green's function defined this way 
is an analytic function of
$\lambda$ in the first quadrant of the complex plane,
i.e. for Re($\lambda)\ge 0$, Im($\lambda) > 0$. 
This is simply the statement that as long as $\lambda$
remains in the first quadrant, a deformation of the integration contour of the kind 
mentioned above
is always possible.  For any amplitude, we shall denote by $C$ the collective 
prescription for all the $\ell_k^0$ integration contours.

As is well known, if the integrands had sufficiently rapid fall off as $\ell_k^0\to\infty$
in any direction in the complex plane,
the above prescription is equivalent to the usual $i\eps$ prescription
for computing Green's functions. To see this let us replace $m^2$ by $m^2-i\eps$
in the propagators. Since the euclidean path integral has no divergences, 
in the $\eps\to 0^+$ limit this
replacement has no effect on the Euclidean Green's functions with 
$p_s^0=i E_s$. Now we rotate 
each external $p_s^0$ from imaginary to real axis by taking $p_s^0=E_s e^{i\theta}$
and letting $\theta$ vary from $\pi/2$ to 0. We can accompany this by a deformation
of the integration contour over $\ell_k^0$'s by replacing $\ell_k^0$ by 
$e^{i\theta}u_i$ with real  $u_i$. As long as the integrand 
falls off sufficiently fast as
$\ell_k^0\to \pm \, e^{i\theta}\times \infty$, this is an 
allowed deformation of the contour. During this deformation
the momentum $k_j\equiv (k_j^0,\vec k_j)$ flowing through the $j$-th 
internal propagator 
will take the form $(e^{i\theta} \kappa_j, \vec k_j)$ with real 
$\kappa_j$. Therefore we have
\be
k_j^2 + m^2-i\eps \equiv -(k_j^0)^2 + \vec k_j^2 + m^2-i\eps =
-\kappa_j^2 e^{2i\theta} + \vec k_j^2 + m^2 -i\eps\, .
\ee
For $\eps>0$ and $0\le\theta\le\pi/2$ this has strictly negative imaginary
part and hence does not vanish. Therefore the deformed contour
does not cross any pole as we vary $\theta$ from $\pi/2$ to 0. 
For $\theta=\pi/2$ this gives the euclidean expression
whereas for $\theta=0$ we get the usual Feynman rules with Lorentzian
momentum integration with the $i\eps$ prescription.

Of course for the kind of vertices we are using here this rotation of the integration
contours is not 
allowed due the essential singularity that the integrand has at infinity. Therefore 
we
have to work with the integration contour with the end-points of $\ell_k^0$
contour integrals fixed at $\pm i\infty$, as mentioned above. 
Nevertheless, replacing $m^2$ by $m^2-i\eps$ serves a useful purpose
of determining which side of the integration contour a given singularity lies.
For this let us express the propagator $-(k^2+m^2)^{-1}$ as
$\left(k^0 - \sqrt{\vec k^2+m^2}\right)^{-1}
 \left(k^0 + \sqrt{\vec k^2+m^2}\right)^{-1}$. In this case if we replace
 $m^2$ by $m^2-i\eps$ and pretend that the $k^0$ integral runs along the real
 axis towards $+\infty$, then the first pole
 lies to the right of the $k^0$ 
integration contour whereas the second pole lies to the 
left of the integration contour. 
This property is inherited from the original definition of the integral for
purely imaginary $\lambda$ where the $k^0$ integral runs
along the imaginary axis from $-i\infty$ to $i\infty$, and must be satisfied by the $k^0$ integration
contour for any $\lambda$ in the first quadrant.
Therefore the $i\eps$ prescription may be regarded as a way of keeping track of
on which side of the integration contour a pole lies, -- we simply have to pretend that the
$k^0$ integration runs along the real axis towards $+\infty$, 
and read off which side of the contour the pole
is on when we replace $m^2$ by $m^2-i\eps$.

\subsection{Hermitian conjugate of the T-matrix} \label{s2.3}

\begin{figure}
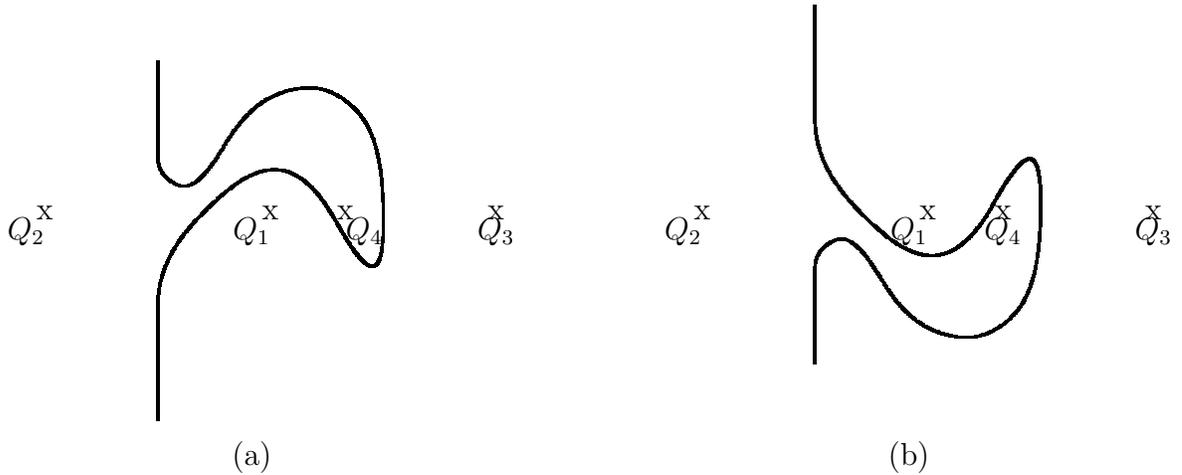

\begin{center}
\hbox{\figfour \quad  \quad \figthree}
\end{center}
\caption{(a)~The integration contour over $\ell^0$ with the locations of the poles marked by
x. (b)~The complex conjugate contour.
\label{f3}}
\end{figure}

The Feynman rules described above directly compute the T-matrix. Our goal will
be to compute the difference $\langle a| (T-T^\dagger)|b\rangle$ for incoming
states $|b\rangle$ and outgoing states $\langle a|$. For this we use the 
relation\footnote{We use the shorthand notation $\langle a|S|b\rangle
\equiv \langle a, \hbox{out}|b, \hbox{in}\rangle$ and 
$\langle a|S^\dagger|b\rangle
\equiv \langle a, \hbox{in}|b, \hbox{out}\rangle$.}
\be \label{etdag}
\langle a| T^\dagger | b\rangle = \langle b| T |a\rangle^*\, ,
\ee
and proceed as
follows:
\begin{enumerate}
\item We use the Feynman rules to compute the right hand side
of \refb{etdag}. Now in our convention
where all external states have their momenta entering the Feynman diagram,
an external line of momentum $p_i$
with positive $p_i^0$ is to be interpreted as an incoming state of 
$(d+1)$-momentum $p_i$  whereas an external state of momentum $p_i$
with negative $p_i^0$ is to be interpreted as an outgoing state of 
$(d+1)$-momentum $-p_i$. Therefore $\langle b| T |a\rangle$ can be obtained from
$\langle a| T | b\rangle$  by
simply switching the signs of all the external momenta. 
\item Due to the change in sign of the external momenta, the $\ell_k^0$ integration
contours for computation of $\langle b| T |a\rangle$
will have to be deformed in a way that is different from what we have for 
$\langle a|T|b\rangle$, since the poles
are at different places. However if we 
make a change of variables in which each loop momentum
$\ell_k$ is replaced by $-\ell_k$ 
then all the momenta carried by the internal vertices and propagators\footnote{The
change in sign of the momentum does not affect the propagator, but we have included
it to facilitate generalizations in \S\ref{s6}.} in the expression for $\langle b|T|a\rangle$
will have
their signs reversed compared to the integrand appearing in the
computation of $\langle a|T|b\rangle$. 
Since this does not change the positions of the poles, the contours for $\langle b|T|a\rangle$
can now be defined
in the same way as for $\langle a|T|b\rangle$.
The $(-1)^{d+1}$ factor picked up by the measure $d^{d+1}\ell_k$ during the change of
variables is
compensated by an orientation reversal of the integration contours, so that each
$\ell_k^i$ integral for $1\le i\le d$ still runs from $-\infty$ to $\infty$ along the real
axis and each $\ell_k^0$ integration still runs from $-i\infty$ to $i\infty$ along the
original contour.
This shows that  $\langle b|T|a\rangle$ can be computed by
taking the expression for $\langle a|T|b\rangle$ 
and changing the sign of the arguments of each 
vertex factor $V^{(n)}$ and internal propagator 
that appears in the amplitude, keeping the integration contours
unchanged.
\item 
Next we study the effect of the complex conjugation appearing
on the right hand side of \refb{etdag}.
This
changes the factor of
$i$ accompanying each $\ell_k^0$ integral to $-i$, and complex conjugates all
the vertices and propagators. 
We can now use \refb{ereal} and the minus signs in the arguments
of $V^{(n)}$ introduced at the previous step to bring the factors of $V^{(n)}$
back to the form in which they appeared in the expression for
$\langle a|T|b\rangle$, except that their arguments are replaced by 
their complex conjugates.
On the other hand in each propagator factor the momentum gets complex
conjugated.
Therefore the net difference between the expressions for $\langle a|T|b\rangle$ and
$\langle b|T|a\rangle^*$ is that all the  momentum factors
in the integrand are replaced by their complex conjugates and the factor of $i$
accompanying each loop integral is replaced by $-i$.
\item
If we make a further change in the variables
$\ell_k^0\to (\ell_k^0)^*$, it sends the integrand for $\langle b|T|a\rangle^*$
to its original form that appears
in the computation of $\langle a|T|b\rangle$ except that all the external momenta
are replaced by their complex conjugates. The new $\ell_k^0$
contour would run from $i\infty$ to $-i\infty$, but we compensate for this by changing
its orientation by
absorbing the $-$ sign from the factor of $-i$ mentioned at the end of the last
paragraph. However the new contours now are related
to the original contours by complex conjugation.
We denote the new choice of contours collectively by $C^*$.  An example of how 
$C^*$ is constructed from $C$ can be found in \S\ref{s2} (see Fig.~\ref{f3})
and a systematic procedure for constructing $C$ and $C^*$
will be described in \S\ref{schoice}.
\item As long as the contours can be kept away from the poles, we can take the
limit in which the external energies approach real values. In that case the integrands
in the expressions for $\langle a|T|b\rangle$ and $\langle b|T|a\rangle^*$ become
identical. We shall see however that this is not always possible since the contours
may encounter pinch singularities in this limit. In such cases we have to take
the limit after carrying out the integration.
\end{enumerate}

To summarize, we have shown that the expression for $\langle a|T^\dagger|b\rangle$ takes
a form similar to that for $\langle a|T|b\rangle$, 
except that in the integrand the external momenta are replaced by their
complex conjugates and 
the choice of integration contours over $\ell_k^0$, denoted collectively
by $C$, is replaced by $C^*$. Therefore the difference between $T$ and
$T^\dagger$ can be computed by calculating the difference between these two contour
integrals.
Our goal will be to prove that this difference  is given by the Cutkosky rules.  
In carrying out this analysis we shall make use of the freedom of
deforming the $\ell_k^0$ contours in the complex $\ell_k^0$ plane, 
possibly picking up
residues from the poles that the contour crosses, but keep the
ends of the integration contour always tied at $\pm i\infty$ to ensure convergence
of the integral. 

\sectiono{One loop four point function} \label{s2}

\begin{figure}
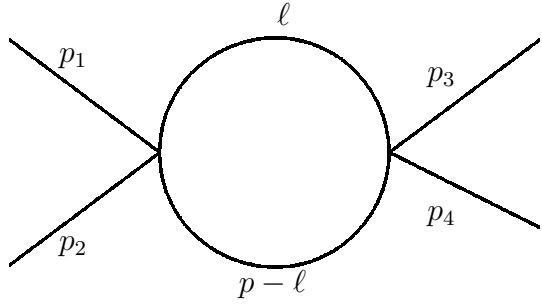

\begin{center}
\figone
\end{center}
\caption{A one loop contribution to the four point function. All external momenta flow
inwards, the internal momenta $\ell$ and $p-\ell$ flow
from left to right and $p=p_1+p_2=-(p_3+p_4)$.
\label{f1}}
\end{figure}

In this section we shall analyze in detail 
the simple example of a contribution to the four point
function shown in Fig.~\ref{f1}. Using \refb{eifactornew}
the contribution to the Green's function is given by
\ben \label{e4point}
A(p_1,p_2,p_3,p_4) &=& 
i \int_C {d^{d+1} \ell\over (2\pi)^{d+1}} \{(\ell^0)^2-\vec \ell^2 -m^2\}^{-1} \{(p^0 - \ell^0)^2 
- (\vec p -\vec \ell)^2 - m^2\}^{-1} \nonumber \\
&& \hskip 1in \times \, V^{(4)}(p_1, p_2, -\ell, \ell-p) \, V^{(4)}(\ell, p-\ell, p_3, p_4)\, .
\een
We have dropped the overall factor of $(2\pi)^{d+1}\delta^{(d+1)}(p_1+p_2+p_3+p_4)$ 
to avoid cluttering, and $C$ denotes the integration contour as described in \S\ref{s1}.
Following the logic described at the end of \S\ref{s1} we also get, after some change
of variables,
\ben \label{e4pointstar1}
A(-p_1,-p_2,-p_3,-p_4)^* 
&=& i \int_{C^*} {d^{d+1} \ell\over (2\pi)^{d+1}} \{(\ell^0)^2-\vec \ell^2 -m^2\}^{-1} \{((p^0)^*
- \ell^0)^2 
- (\vec p -\vec \ell)^2 - m^2\}^{-1} \nonumber \\
&& \hskip .5in \times \, V^{(4)}(p_1^*, p_2^*, -\ell, \ell-p^*) \, V^{(4)}(\ell, p^*-\ell, p_3^*, p_4^*)\, ,
\een
where $C^*$ is the contour obtained from $C$ after complex conjugation and an 
orientation reversal so that it still runs from $-i\infty$ to $i\infty$. 
Therefore $A(p_1,p_2,p_3,p_4)$ and $A(-p_1,-p_2,-p_3,-p_4)^*$ differ
from each other only in the choice of the integration contour, and complex conjugation of
the external momenta.

Let us return to \refb{e4point}.
The poles in the $\ell^0$ plane are at
\be  \label{eqpos}
Q_1 \equiv \sqrt {\vec \ell^2 + m^2}, \quad Q_2 \equiv -\sqrt{\vec \ell^2 + m^2}, \quad
Q_3 \equiv p^0 + \sqrt{(\vec p - \vec \ell)^2 + m^2} , \quad Q_4 \equiv
p^0 - \sqrt{(\vec p - \vec \ell)^2 + m^2}\, .
\ee
Let $E_s$'s be the physical real values of $p_s^0$ that we are interested in, 
and define $E\equiv E_1+E_2=-(E_3+E_4)$. 
Let us for definiteness take $E_1$, $E_2$ to be positive and $E_3$, $E_4$
to be negative so that
$E$ is positive. This corresponds to
choosing $p_1$ and $p_2$ as incoming momenta and $-p_3$ and $-p_4$ as 
outgoing momenta. 
In order that
this amplitude can be defined via analytic continuation from the Euclidean 
result as suggested in \S\ref{s1}, we have to ensure that for $p_s=\lambda E_s$
the amplitude defined
above has no singularity for $\lambda$ in the first quadrant. This in the
present circumstances correspond to $p^0=\lambda(E_1+E_2)$ 
lying in the first quadrant. 
Now our prescription for defining the Green's function for $p^0$ on the imaginary axis
is to take the integration contour
of $\ell^0$ from $-i\infty$ to $i\infty$ along the imaginary axis. In this case 
the poles $Q_2$ and $Q_4$ are to the left
of the integration contour and the poles $Q_1$ and $Q_3$ are to the right of the 
integration contour.  As $p^0$ moves into the first quadrant, the positions of the
poles shift. If they come towards the imaginary $\ell^0$ axis, then analytic
continuation of the original results is obtained by deforming the $\ell^0$
integration contour into the complex plane to avoid the pole, keeping its ends 
fixed at $\pm i\infty$. We shall hit a singularity in the $p^0$-plane if the
singularity is pinched, i.e.\ two poles approach the same point on the integration
contour from opposite sides so that we cannot deform the contour away from
the pole without passing through one of the poles. In the present context this
would happen if $Q_2$ approaches $Q_1$ or $Q_3$, or $Q_4$ approaches
$Q_1$ or $Q_3$. Now from the expressions given in \refb{eqpos} it is clear
that for real $\vec \ell$, $Q_1$ cannot approach $Q_2$ and $Q_3$ cannot
approach $Q_4$. Therefore the only possibilities are $Q_2$ approaching $Q_3$
or $Q_4$ approaching $Q_1$. The conditions for these to happen 
can be written as
\be \label{epinch}
p^0 = \pm \left( \sqrt {\vec \ell^2 + m^2} + \sqrt{(\vec p - \vec \ell)^2 + m^2}\right)
\, .
\ee
Since the right hand side of \refb{epinch} is real, this can be avoided as
long as $p^0$ is away from the real axis. This shows that the amplitude is free
from singularities as long as $p^0$ lies in the first quadrant,
and we can define the amplitude for
real positive $p^0$ by taking the Im$(p^0)\to 0$ limit from above. 
An alternative but equivalent approach will be to replace $m^2$ by $m^2-i\eps$.
In this case
the right hand sides of \refb{epinch} 
will lie in the fourth and the second quadrants.
Therefore as $p^0$ approaches a positive real value from the first quadrant,
the analyticity property of the integral extends all the
way up to real $p^0$ axis, and we can
define the amplitude for real $p^0$ by taking
the $\eps\to 0^+$ limit after setting $p^0$
to be real.

The story can be repeated even in the case $E<0$. In this case $p^0=\lambda E$
lies in the third quadrant and the possible solution to \refb{epinch} comes from
the choice of minus sign on the right hand side. 
This can be avoided as long as Im$(p^0)<0$, 
i.e. Im$(\lambda)>0$, and we define the amplitude for real negative
$p^0$ by taking 
Im$(p^0)\to 0$ from below. Alternatively, 
replacing $m^2$ by $m^2-i\eps$ shifts the right hand side of
\refb{epinch} with the choice of minus sign to the second quadrant. Therefore
the analyticity property of the integral also holds when we consider real negative
$p^0$, i.e.\ real positive $\lambda$. This allows us to
take $p^0$ to be real keeping
$\eps>0$
and then take $\eps\to 0^+$ limit.

\begin{figure}
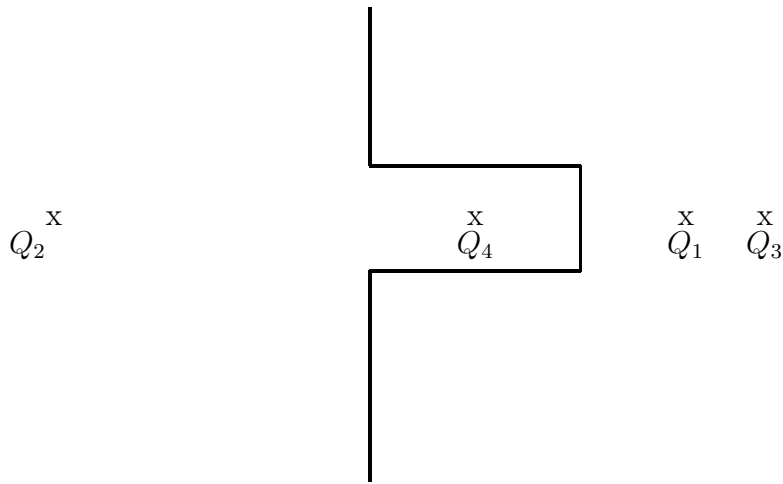

\begin{center}
\figtwo
\end{center}
\caption{The integration contour over $\ell^0$ and the locations of the poles marked by
x.
\label{f2}}
\end{figure}

This proves the desired analyticity property of the Green's function that allows
us to define the amplitudes for real $p^0$ via analytic continuation of the amplitude
for imaginary $p^0$. Let us now
focus on deriving the Cutkosky rules for this amplitude. For this we need
to compute the difference between \refb{e4point} and \refb{e4pointstar1}
in the limit 
$p^0\to E$ from the first quadrant.
Eq.\refb{eqpos} shows that in this limit all the poles approach the real axis. The original
contour needs to be deformed when $Q_4$ crosses the imaginary axis
so that
the poles $Q_2$ and $Q_4$ will
continue to lie to the left of the integration contour and the poles $Q_1$ and
$Q_3$ will lie to the right of the integration contour. 
This has been
shown in Fig.~\ref{f2}. If we include the $i\eps$ term, then the poles $Q_2$
and $Q_4$ get lifted slightly above the real axis, while the poles $Q_1$ and
$Q_3$ get shifted slightly below the real axis.

Now as long as $p^0< \sqrt{\vec \ell^2+m^2} + 
\sqrt{(\vec p-\vec \ell)^2+m^2}$,  
$Q_4$ lies to the left of $Q_1$ and the contour can be taken to be invariant under
complex conjugation as shown in Fig.~\ref{f2}. 
As a result $C$ and $C^*$ are identical.\footnote{A more 
general statement is that the new contour obtained after complex
conjugation can be deformed to the original
contour without passing through a pole.} Also all external momenta are real
so that the integrands in \refb{e4point} and 
\refb{e4pointstar1} become equal.
In this case using \refb{e4point} and 
\refb{e4pointstar1} we see that for the range of values of $\vec\ell$ satisfying the
above inequality, 
the contributions to
$A(-p_1,-p_2,-p_3, -p_4)^*$ and 
$A(p_1, p_2, p_3, p_4)$ are equal.

Let us now consider the case when $Q_4$ approaches $Q_1$. If we
use the $i\eps$ prescription, $Q_4$ always lies above the real axis and
$Q_1$ lies below the real axis, and we can continue to choose the contour
so that $Q_2$ and $Q_4$ are to the left and $Q_1$ and $Q_3$ are to the
right. However in this case we can no longer ignore the $i\eps$ term.
Equivalently we can set $\eps=0$ but take $p^0$ to have a small
positive imaginary
part as its real part approaches $\sqrt{\vec \ell^2+m^2} + 
\sqrt{(\vec p-\vec \ell)^2+m^2}$. In either case, $Q_4$ lies above $Q_1$ in
the complex plane when their real parts approach each other.
We shall return to this contribution later.

For $p^0 > \sqrt{\vec \ell^2+m^2} + 
\sqrt{(\vec p-\vec \ell)^2+m^2}$, $Q_4$ is to the right of $Q_1$ and the 
deformed contour takes the form shown in Fig.~\ref{f3}(a). In drawing this
we have used the fact that $Q_4$ remains above $Q_1$ as it 
passes $Q_1$ and that during this process the contour needs to be
deformed continuously without passing through a pole. The complex
conjugate contour of Fig.~\ref{f3}(a) 
has been shown in Fig.~\ref{f3}(b). 
Since in both cases the contours can be taken far away from the poles,
we can set the external energies to be real and $\eps$ to zero
so that the integrands in
\refb{e4point} and \refb{e4pointstar1} become identical. 
Now even though
the contours in Fig.~\ref{f3}(a) and \ref{f3}(b) are topologically distinct, 
each of them can be split into two contours -- an anti-clockwise contour around
$Q_4$ and a contour from $-i\infty$ to $i\infty$ keeping $Q_1$, $Q_3$ and $Q_4$
to the right. Therefore their contributions are equal. 
This shows that  integration over the contours in Fig.~\ref{f3}(a) and Fig.~\ref{f3}(b)
give the same result, and hence 
the contribution to $A(p_1, p_2, p_3, p_4)-A(-p_1,-p_2,-p_3, -p_4)^*$ from the
region of integration where $\vec\ell$ satisfies the above inequality also vanishes.

Therefore we see that the contribution to the imaginary part of \refb{e4point}
comes from the region around $p^0 \simeq \sqrt{\vec \ell^2+m^2} + 
\sqrt{(\vec p-\vec \ell)^2+m^2}$ when the poles $Q_4$ and $Q_1$
approach each other. We shall now evaluate this contribution.
For both $C$ and $C^*$, we proceed  by 
deforming the $\ell^0$ contour through the pole $Q_4$ so that 
the poles $Q_4$, $Q_1$ and $Q_3$ now lie to the right of the
integration contour. This new contour is far away from all poles and can be
chosen to be
invariant under complex 
conjugation followed by a reversal of orientation. As a result the integral along this
contour gives equal contribution to 
$A(p_1, p_2, p_3, p_4)$ and $A(-p_1,-p_2,-p_3, -p_4)^*$
 according to our previous argument.
In the process of passing the contour through $Q_4$ we also 
pick up the residue from $Q_4$.  Let us denote by $A_r$ the contribution 
from the
residue at $Q_4$. 
This can be
expressed as
\ben \label{eAr}
A_r &=& \int {d^{d} \ell\over (2\pi)^{d}} 
\left\{ 2\sqrt{(\vec p -\vec \ell)^2 + m^2}\right\}^{-1} \nonumber \\ &&
\times\, 
\left\{ p^0 - \sqrt{(\vec p -\vec \ell)^2 + m^2} - \sqrt{\vec \ell^2 + m^2}\right\}^{-1}
\left\{ p^0 - \sqrt{(\vec p -\vec \ell)^2 + m^2} + \sqrt{\vec \ell^2 + m^2}\right\}^{-1}
\nonumber \\
&& \times \, \,
  V^{(4)}(p_1, p_2, -\ell, \ell-p) \, V^{(4)}(\ell, p-\ell, p_3, p_4)\, ,
\een
where  it is understood that
$\ell^0$ in the argument of $V^{(4)}$
is given by its pole value $p^0 - \sqrt{(\vec p -\vec \ell)^2+m^2}$ and
that this contribution is being evaluated only for those values of $\vec\ell$ for which
$p^0$ is close to 
$\sqrt{(\vec p -\vec \ell)^2 + m^2} + \sqrt{\vec \ell^2 + m^2}$ so that $Q_4$
is close to $Q_1$.
In this case the second term in the second line
remains finite over the entire range of integration of $\vec \ell$.  However the
first term in the second line can encounter a divergence. To regulate this we 
either replace
$m^2$ by $m^2-i\eps$ or take $p^0$ in the first quadrant. 
On the other hand for the
hermitian conjugate amplitude \refb{e4pointstar1}
the situation is opposite and we have to either replace
$m^2$ by $m^2+i\eps$ or take $(p^0)^*$ in the fourth quadrant. 
Using the result 
\be 
(x + i\eps)^{-1} - (x-i\eps)^{-1} = -2 i \pi \delta (x)\, ,
\ee
we see that the contribution to 
$A(p_1, p_2, p_3, p_4)-A(-p_1,-p_2,-p_3, -p_4)^*$ is
given by
\ben \label{ecut}
&& -2 \pi \, i \, \int {d^{d} \ell\over (2\pi)^{d}} \, 
\delta \left(E - \sqrt{(\vec p -\vec \ell)^2 + m^2} - \sqrt{\vec \ell^2 + m^2}\right)
\left\{ 2\sqrt{(\vec p -\vec \ell)^2 + m^2}\right\}^{-1}  
\nonumber \\
&& \times \, \,
\left\{ 2\sqrt{\vec \ell^2 + m^2}\right\}^{-1} 
\,  V^{(4)}(p_1, p_2, -\ell, \ell-p) \, V^{(4)}(\ell, p-\ell, p_3, p_4)\, ,
\een
where now all external momenta are taken to be real.
Interpreting $V^{(4)}(p_1, p_2, -\ell, \ell-p)$ as the matrix element of $T$ with
initial state carrying momentum $(p_1,p_2)$ and final state carrying momentum
$(\ell, p-\ell)$ and $V^{(4)}(\ell, p-\ell, p_3, p_4)
= V^{(4)}(-\ell, \ell-p, -p_3, -p_4)^*$ as the matrix element of 
$T^\dagger$ with the initial state carrying momentum $(\ell, p-\ell)$ and the final
state carrying momentum $(-p_3, -p_4)$ we see that \refb{ecut} is precisely the
statement of the relation
\be 
T - T^\dagger = -i \, T^\dagger T\, .
\ee
In order to check the precise normalization we must also put back the momentum
conserving $\delta$-functions in the expressions for $T$ and $T^\dagger$.

\sectiono{Analytic property of general Green's functions} \label{s3}

In this section we shall prove the 
analyticity of the general off-shell Green's function
in the first quadrant of the complex $\lambda$ plane 
as stated in \S\ref{s1}. More specifically, we shall show that if we restrict
the external momenta so that the spatial components are real, and
the time components have the form $\lambda$ times real numbers 
for a complex parameter $\lambda$, 
then the
amplitudes, defined via analytic continuation from imaginary $\lambda$ axis, are
free from any singularities for Re($\lambda)\ge 0$, Im($\lambda)>0$.
We shall also describe explicitly the procedure for
choosing the
integration contour that implements the analytic continuation.

\subsection{Analyticity in the first quadrant}

Our strategy for proving analyticity of the Green's function in the first quadrant
of the complex $\lambda$-plane will be as follows. We shall show that for any
fixed real values of the spatial components $\{\vec \ell_k\}$ of the loop momenta,
the integral over the $\{\ell_k^0\}$'s can always be deformed away from 
all singularities of the integrand, i.e.\ the integration contour is not pinched. 
As a result the contribution to the integral over
$\{\ell_k^0\}$
is non-singular. Since this is true at every $\{\vec \ell_k\}$, the result remains
non-singular even after integration over $\{\vec\ell_k\}$.

We shall prove the result by assuming the contrary and then showing
that there
is a contradiction. Therefore
let us  suppose that there is a subspace $R$ of the space spanned by $\{\vec\ell_k\}$
where there is a pinch singularity. This means that on this subspace 
the integrand becomes singular at 
some points on the $\ell_k^0$ integration contours, 
and we cannot deform the $\ell_k^0$ contours away from
these points without passing through a singularity. 
We can classify the regions $R$ of this type using `reduced diagram' which
is obtained from the original Feynman diagram by collapsing
all propagators
whose energies can be deformed away from the poles. 
This means that we remove each of these propagators and join the pair of
vertices that were originally connected
by the propagator into a single vertex with larger number of external legs.
The vertices of the reduced diagram 
will be called reduced vertices.  A given reduced vertex may receive contribution
from many different Feynman diagrams. We shall assume that the loop energies flowing
through the propagators inside the reduced vertex -- i.e.\ the propagators
which have been collapsed to points -- have been deformed if needed
to keep
these propagators finite distance away from their poles. 
Henceforth the propagators of a reduced diagram will refer to
only those propagators which have not been collapsed to points.

\begin{figure}
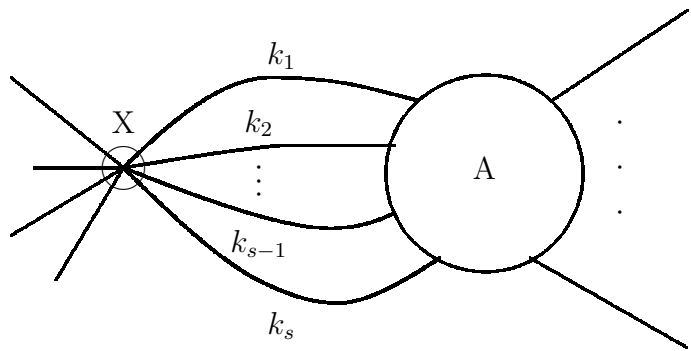

\begin{center}
\fna
\end{center}
\caption{The reduced diagram displaying only the on-shell propagators
of a potentially singular region of integration. $A$ denotes a blob 
containing arbitrary number of internal lines and reduced vertices, 
and X denotes a specific reduced vertex 
of this reduced diagram where some external lines are connected.\label{fna}}
\end{figure}

Let us take any reduced vertex $X$ 
of the reduced diagram and label by $k_1,\cdots k_s$ 
the momenta carried away by internal propagators emerging from the vertex.
This has been shown in Fig.~\ref{fna}
with $X$ marking a particular reduced vertex, and
A denoting a blob containing arbitrary number of internal lines 
and reduced vertices.
 Since each of the internal propagators of the reduced diagram
is on-shell at the pinch 
(otherwise they would have been collapsed to points in the reduced
diagram) we have
\be \label{er1}
(k_i^0) = \pm \sqrt{\vec k_i^2 + m^2} \quad \hbox{for $1\le i\le s$}\, .
\ee
On the other hand if $p$ denotes the total momentum entering the vertex X from
the external lines, we have, by momentum conservation,
\be \label{er2}
p^0 = \sum_{i=1}^s k_i^0 \, .
\ee
Now \refb{er1} shows that $k_i^0$ is real. Therefore it follows from \refb{er2} that
$p^0$ must be  real.
On the
other hand we have taken $p^0$ to be of the form $\lambda E$ for some real number
$E$, with $\lambda$ lying in the first quadrant. This shows that the only way
to satisfy \refb{er1} and \refb{er2} for finite Im$(\lambda)$ 
is to take $E=0$.
Repeating this analysis for every reduced vertex we see that the 
total energy entering externally into every reduced 
vertex of the reduced diagram must
vanish.

Let us now denote by the set $\{k_\alpha\}$ the momenta carried by all the
propagators of the reduced diagram -- not only the ones that leave a given 
reduced vertex $X$. Using the $i\eps$ convention to label the side of the
contour on which a pole lies, we see that the relevant poles at the pinched
singularity are at
\be\label{erelevant}
k_\alpha^0 = \pm \sqrt{\vec k_\alpha^2+m^2-i\eps}\, .
\ee
Near the singularities \refb{erelevant}, we now
deform all the loop energy integration contours of the reduced diagram  
by
multiplying them by $\wt\lambda$ where $\wt\lambda$ is a complex number
close to 1, lying in the first quadrant. Since the deformation is small, it does
not lead to any new singularity from the propagators that are inside the reduced
vertices. On the other hand since the external energies entering each reduced
vertex vanishes, it multiplies each $k_\alpha^0$ by $\wt\lambda$. It is easy to
see that this deforms the contours away from each of the poles given in
\refb{erelevant}. Therefore the loop energy integration contours are not pinched
at the poles \refb{erelevant}, showing that our initial assumption was incorrect.

This proves the
desired result.

\subsection{Choice of integration contour} \label{schoice}

For future use, we shall now describe a specific operational procedure for 
choosing the integration contour.
As before we denote by
$\{p_i\}$ the external momenta, by
$\{\ell_k\}$ the loop momenta and by $\{k_i\}$ the momenta carried by the propagators.
We express the $n_p$ propagator factors $(-k_i^2- m^2)^{-1}$ as 
$\left(k_i^0+\sqrt{\vec k_i^2 + m^2}
\right)^{-1} \left(k_i^0-\sqrt{\vec k_i^2 + m^2}
\right)^{-1}$ 
and assign fixed labels $1,\cdots 2n_p$ to the $2n_p$ poles obtained from the
$n_p$ propagators.
Our analytic continuation involves  
choosing the external energies $\{p_i^0\}$ to be $\{\lambda E_i\}$ with
real $\{E_i\}$. 
When $\lambda$ is on the imaginary axis, 
each of the $\ell_k^0$ integration contours can be taken to run along the imaginary
axis from $-i\infty$ to $i\infty$.
Furthermore, for each $\ell_k^0$ integration contour, there is a definite notion 
of whether a given pole that depends of $\ell_k^0$ 
lies to the left or right of the integration contour. We make these into permanent
assignments in what follows below. As mentioned in \S\ref{s1} we could keep
track of this information using the $i\eps$ prescription even when the contours
are deformed.

Consider now a general value of $\lambda$ in the first quadrant, and 
choose a specific order in which we carry
out the integration over  $\{\ell_k^0\}$, for fixed values of the spatial components
of all loop momenta. Without any loss of generality we can take this order to be
$\ell_1^0, \ell_2^0,\ell_3^0,\cdots$.  Now let us regard the integrand as a general complex
function of $\{\ell_k^0\}$, and for fixed complex values of $\ell_2^0,\ell_3^0,\cdots$,
carry out the $\ell_1^0$ integration along a contour from $-i\infty$ to $i\infty$ that
keeps the $\ell_1^0$ dependent poles on the same side of the integration contour as
the original contour defined for purely imaginary $\lambda$. 
Note that there may be more than one contour satisfying this condition that are
not deformable to each other, {\it e.g.} as shown in Fig.~\ref{f3}(a) and (b). 
However, the result of
integration does not depend on the choice of contour. To see this we note that given any
two contours satisfying the above condition, one
can be deformed to the other by allowing it to pass through the poles and
picking up residues,
and during any such deformation a given pole will have to be crossed an 
even number of times in opposite directions since every time a pole is 
crossed it moves from the right of the contour to the left or vice versa. 
Therefore all the residues cancel and the result of integration becomes independent of the
choice of contour.
This gives a function of $\ell_2^0,\ell_3^0,
\cdots$. The resulting function can develop new poles as a function of these
variables from the $\ell_1^0$ integration. For example a pole in the $\ell_k^0$ plane
can arise when an $\ell_k^0$ dependent pole $A$ in $\ell_1^0$ plane collides with an
$\ell_k^0$ independent pole $B$ in the $\ell_1^0$ plane from opposite sides
of the contour.\footnote{To see that this generates a pole in the $\ell_k^0$ plane, we note
that the singular part of the integrand has the form $(\ell_1^0 \pm \ell_k^0 - R_A)^{-1}
(\ell_1^0 - R_B)^{-1}$ for some $R_A$ and $R_B$ that are independent of $\ell_1^0$ and
$\ell_k^0$.
We can deform the $\ell_1^0$
contour through the pole $B$ picking up the residue. The deformed
contour integral has no singularity from $A$ or $B$, while the residue at $B$ produces
a pole in $\ell_k^0$ of the form $(\pm \ell_k^0 -R_A+R_B)^{-1}$.}
We assign this new pole in the
$\ell_k^0$ plane to be on the same side of the $\ell_k^0$ contour  that the pole $A$ was
before $\ell_1^0$ integration. We now carry out the integration over $\ell_2^0$ along
a contour from $-i\infty$ to $i\infty$ keeping all the $\ell_2^0$ dependent poles
on the `correct side' of the contour. Repeating the same argument as before, we
get a function of $\ell_3^0,\ell_4^0,\cdots$ with definite assignment of which side 
of the contours in the  $\ell_3^0,\ell_4^0,\cdots$ plane a
given pole should lie. This
way we can successively 
carry out the integration over all the $\ell_k^0$'s and get a finite result as
a function of $\lambda$, the spatial components of the loop momenta, and the external
momenta. The set of rules defined
above for constructing the $\{\ell_k^0\}$ integration contours will be collectively denoted
by $C$. 

The complex conjugate contour $C^*$ introduced in \S\ref{s1}, needed for computing the matrix
elements of $T^\dagger$, is defined as follows. 
Let us suppose that the original integration over $\ell_s^0$
was done along a contour
\be \label{edeffs}
\ell_s^0 = f_s(t; \ell_{s+1}^0, \ell_{s+2}^0,\cdots; \{p_i\}, \{\vec\ell_k\})\, ,
\ee
for some function $f_s$ of a real variable $t$ labelling the contour, the other 
$\ell_k^0$'s for $k>s$, all the external momenta $\{p_i\}$ and the spatial components
of all the loop momenta $\{\vec \ell_k\}$.
The set of functions $\{f_s\}$ is what we collectively call the choice of the contour $C$.
Now the contour in terms of the variables $\{(\ell_k^0)^*\}$ will be
\be \label{edeftfs}
(\ell_s^0)^* = \left(f_s(t; \ell_{s+1}^0, \ell_{s+2}^0,\cdots; \{p_i\}, \{\vec\ell_k\})\right)^* 
\equiv \wt f_s\left(t;(\ell_{s+1}^0)^*, (\ell_{s+2}^0)^*,\cdots; \{p_i^*\}, \{\vec\ell_k\}\right)\, ,
\ee
where we have used the fact that the spatial components of loop momenta are always kept
real.
After renaming the variables $(\ell_k^0)^*$ as $\ell_k^0$,
the function $\wt f_s$ defined this way gives the new $\ell_s^0$ integration
contour. Operationally $\wt f_s$ is obtained from $f_s$ by replacing all explicit factors
of $i$ by $-i$.
We shall denote collectively by $C^*$ the information on the
integration contours encoded in the functions  $\wt f_1, \wt f_2,\cdots$.

\sectiono{Cutkosky rules} \label{scut}

Our next task is to compute the matrix elements of $T-T^\dagger$ and show that
the result is given by Cutkosky rules.
The matrix element of $T$ is given by the Green's function $A(\{p_i\}$ for
on-shell external momenta $\{p_i\}$ and the matrix element of $T^\dagger$ 
between the same external states is given by $A(\{-p_i\})^*$.
As mentioned before, throughout our analysis we shall keep
the spatial components of loop momenta real and allow only the 0-components
of the loop momenta to be deformed so as to avoid the poles. 
It was shown in \S\ref{s1} that in this case the contributions to 
$A(\{p_i\}$ and $A(\{-p_i\})^*$
are given by similar integrals with the integrands related by the replacement
of $p_i$ by $p_i^*$, and integration contours $C$ and $C^*$ related by complex
conjugation. In absence of pinch singularity we can set the external momenta
$p_i$'s to be real and the integrands become identical.

Consider now
a pinch singularity where the 0-components of
$N$ of the loop momenta are constrained. In this
case, in order that each of these $N$ loop momenta are pinched, we need at
least $N+1$ of the denominator factors to vanish. This means that there will
be at least one constraint among the spatial components of these $N$ loop momenta.
More generally we can say that pinch singularities will arise in subspaces
of codimension $\ge 1$ in the space spanned by the spatial components
of the loop momenta. We shall call such subspaces pinched 
subspaces.\footnote{Since we shall eventually look for functions with
$\delta$-function support on the pinched subspaces, it is more appropriate to
consider subspaces of small thickness around the pinched subspaces.
\label{fo4}}

We shall now prove the Cutkosky rules in three steps.
\begin{enumerate}
\item We shall begin our analysis with connected diagrams.
First we shall show that when 
the spatial components of loop momenta are
away from the pinched subspaces, and the spatial components of the external
momenta are away from the subspaces on which some single particle
intermediate state is
on-shell, the result
of carrying out integration over the 0-components of all loop momenta
gives the same contribution to $A(\{p_i\}$ and $A(\{-p_i\})^*$.
Therefore there is no contribution
to   $A(\{p_i\}-A(\{-p_i\})^*$ from this region.
\item Then we shall show that for connected diagrams,
the contribution to $A(\{p_i\})-A(\{-p_i\})^*$ from
the pinched subspaces and/or from on-shell single particle intermediate states
is given by the Cutkosky rules.
\item Finally we shall prove the Cutkosky rules for disconnected diagrams.
\end{enumerate}
Throughout this analysis we shall be using the method of induction, i.e.\ while proving
any of these results for an $N$-loop amplitude, we shall assume that all the results are
valid for any $(N-1)$ loop amplitude. Also during this analysis we ignore the effect of
mass renormalization. This is discussed separately in \S\ref{smass}.

Due to the iterative nature of our proof, and given that the full analysis is somewhat long,
some subtle points may be overlooked
if we are not careful. We shall give some examples below:
\begin{enumerate}
\item 
Cutkosky rules, as explained in \S\ref{s0}, 
require that the part of the diagram on the right of
the cut is conjugated. Much of our analysis that follows will go through even
if we do not take the hermitian conjugate of the amplitude to the right of the cut.
For example in the analysis of the class of diagrams considered in \S\ref{s5.2.2} we do
not need to use explicitly the fact that the part of the diagram to the right of the cut needs
to be hermitian conjugated.
This may give the reader the impression that for this class of diagrams, 
Cutkosky rules will hold even if we do not take the hermitian conjugate 
of the
amplitude to the right of the cut.
We shall now argue that this is not the case.
In \S\ref{s4.2.1}
there is a crucial minus sign on the right hand side of the fourth
line of eq.\refb{esecond} that is there due to the hermitian conjugation, and without it
the analysis following this equation will not hold. Hermitian conjugation of the amplitude
to the right of the cut
also plays a crucial role in the analysis of disconnected diagrams in \S\ref{s5.3}.
Now  while applying recursive methods to
the diagrams of \S\ref{s5.2.2} we often end up with lower order diagrams of the type 
analyzed in \S\ref{s4.2.1} and \S\ref{s5.3}, and assume that Cutkosky rules hold for
these diagrams.  For these we must take the hermitian conjugate of the diagram to the
right of the cut. As a result even for the diagrams analyzed in \S\ref{s5.2.2}, Cutkosky
rules hold only if we take the hermitian conjugate of the diagram to the right of the
cut.
\item In our analysis we give an iterative proof that for reduced diagrams, Cutkosky rules 
require us to sum over only those cut diagrams for
which the cut does not pass through a reduced vertex.
As usual we assume this to be true to a given order and then prove
the result to the next order. The reader may feel somewhat uneasy at the lack of a direct
proof, and wonder if the iterative proof 
would have gone through even if we had relaxed the
constraint that the cut does not pass through a reduced vertex. 
However, if we examine the iterative proof carefully we shall find that during the
course of iteration we often end up with diagrams where the whole diagram is a single
reduced vertex. The result of \S\ref{s4.1} shows that this has
no anti-hermitian part. This would have been in conflict with the Cutkosky rules if the
cuts were allowed to pass through the reduced vertex leading to a non-vanishing result
for the anti-hermitian part of the amplitude. Therefore we again see that different
parts of the analysis are intimately tied together, and relaxing 
any ansatz made during one  part of the
analysis also affects the results of all other parts.
\end{enumerate}

\subsection{Hermiticity of the connected diagrams 
in absence of pinch singularity} \label{s4.1}

In this subsection we shall prove that for connected diagrams, $A(\{p_i\})-
A(\{-p_i\})^*$ vanishes in the absence of pinch singularities and on-shell single
particle intermediate states.
We follow the algorithm described at the end of \S\ref{s3} to define the 
analytically continued amplitude
as a function of $\lambda$ in the first quadrant and the amplitude at
$\lambda=1$ as the limit from the first quadrant. As long as there is no pinch
singularity at $\lambda=1$, we can systematically choose the integration contours $C$ 
over
$\ell_1^0,\ell_2^0,\cdots$ appearing in $A(\{p_i\})$, 
and compute the integrals following the procedure described
in \S\ref{s3}. 
The contribution to $A(\{-p_i\})^*$ can be computed by evaluating the same integral
over the integration contours $C^*$. Since the external momenta are real at $\lambda=1$
the integrands in the expressions for $A(\{p_i\})$ and $A(\{-p_i\})^*$ are identical.
Therefore
if we can show that the choice of
the contours $C$ and $C^*$ are identical,  or deformable to each other without passing through
a singularity, we would have proved that the integrals
are the same. Actually the same arguments as in \S\ref{schoice}
shows that we need less -- all we need to show is
that for each $s$, the
choice of contour in the $\ell_s^0$ plane encoded in the functions $f_s$ and
$\wt f_s$ introduced in \refb{edeffs} and \refb{edeftfs}
have all the poles lying on the same side, i.e.\ if a given pole lies on the
left (right) of the first contour then it must lie on the left (right) of the second 
contour.\footnote{This includes the case where the contours are not necessarily deformable to each other,
as in Fig.~\ref{f3}(a) and (b).}
This can be proved by considering the special case where 
$\ell_{s+1}^0,\ell_{s+2}^0,\cdots$ are real since the side of the contour on which a 
pole in the $\ell_s^0$-plane
lies is by construction independent of $\ell_{s+1}^0,\ell_{s+2}^0,\cdots$. For real
$\ell_{s+1}^0,\ell_{s+2}^0,\cdots$ the poles in the $\ell_s^0$-plane are along the
real axis, whereas the $\ell_s^0$ integration contours in $C$ and $C^*$ are
related by a reflection about the real axis together with a change in orientation.
Under this operation the different segments of the real axis lie
on the same side of the contours in $C$ and $C^*$ independent of how many times
the contours cross the real axis, and hence all the poles on the
real axis also lie on the same side of the contours in $C$ and $C^*$.
This establishes the desired
result, that the contribution to $A(\{p_i\})-A(\{-p_i\})^*$ vanishes
as long as there is no
pinch singularity at $\lambda=1$. 

There is one exception to the above result, and this occurs when the external
momenta are such that some intermediate one particle state
goes on-shell. In this case there are Feynman
diagrams in which some propagator carrying momentum $p$, given by some
linear combination of external momenta, blows up. 
In order to compute the
contribution to $A(\{p_i\})- A(\{-p_i\})^*$ from such Feynman diagrams we 
again work with a general complex $\lambda$ and define the amplitude
by analytic continuation from imaginary $\lambda$-axis to $\lambda=1$ along
the first quadrant.
As mentioned before, in $A(\{p_i\})$ 
this is equivalent to replacing $m^2$ by $m^2-i\eps$
in the propagator. After going through
the manipulations described at the end of \S\ref{s1} we can bring the
expression for $A(\{-p_i\})^*$ to an 
identical form, except that due to the operation of complex conjugation,
in this amplitude $m^2$ is replaced by $m^2+i\eps$. Therefore in the difference 
between $A(\{p_i\})$ and $A(\{-p_i\})^*$, the propagator $((p^0)^2-\vec p^{\, 2} - m^2)^{-1}$
will be replaced by
\be \label{e5.1}
((p^0)^2-\vec p^{\, 2} - m^2+i\eps)^{-1} - ((p^0)^2-\vec p^{\, 2} - m^2-i\eps)^{-1}
= -2\pi i \, \delta((p^0)^2-\vec p^{\, 2} - m^2)  \, .
\ee
This shows that 
in this case we can get a non-vanishing imaginary part of the amplitude even in
the absence of pinch singularity. We shall take into account contributions of this type in our
analysis below.

\subsection{Anti-hermitian part of connected amplitude} \label{s4.2}

We now turn to the second problem, i.e.\ the computation of the 
anti-hermitian
part of a connected amplitude 
when the spatial components of the loop momentum integrals lie on
-- or more precisely around as stated in footnote \ref{fo4} -- a pinched
subspace, or the external momenta lie on a subspace on which some
intermediate single particle state goes on-shell. 
In carrying out the analysis we shall again 
use the notion of reduced diagram in which we collapse  to points
all lines which are not put on-shell at the pinch singularity of the energy
integration contours. In one particle reducible diagrams we also have internal 
propagators which are not part of any loop and carries 
momenta given by linear combinations of the external momenta 
only. For these lines, we collapse to points those lines 
which are not on-shell for the specific values of the external momenta
we work with.
On such a reduced diagram we shall
draw an arrow on each of the propagators to indicate the direction of energy flow
at the pinch singularity.

\begin{figure}
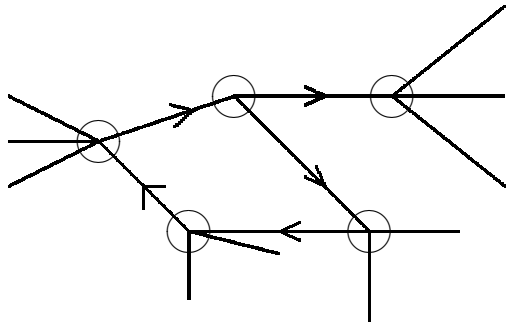


\begin{center}
\figtwentytwo
\end{center}

\vskip -.5in

\caption{A reduced diagram containing an oriented loop. Such a diagram is not
allowed.
\label{f22}
}
\end{figure}

We shall now show that the reduced diagram defined this way cannot have a directed
closed loop  -- i.e. a closed loop with the property that we can traverse the loop
by following the directions of the arrows. Such a diagram has been shown in Fig.~\ref{f22}.
If there is such a loop, then we can find a loop momentum
$\ell$ that appears only in each propagator in the loop,
and the direction of $\ell$ is along the direction of energy flow for each of the propagators.
As a result these propagators will carry momenta $K_i+\ell$ where $K_i$ is linear
combination of other loop momenta and external momenta, and at the pinch we have
\be
K_i^0 + \ell^0 = \sqrt{(\vec K_i+\vec \ell)^2+m^2-i\eps}\, .
\ee
Note the + sign on the right hand side, reflecting the fact that $\ell$ is  directed
along the energy flow. 
The $-i\eps$ is a formal way of stating the fact all the poles are to the right of the
$\ell^0$ integration contour from $-i\infty$ to $i\infty$.
Furthermore there is no other propagator that involves $\ell$.
It is now easy to see that the $\ell^0$ contour is not pinched and can be deformed away
keeping all the poles to the right. This proves the desired result.

In what follows, we shall use an even more minimal representation of a reduced
diagram in which we suppress all external legs and represent reduced vertices
by circles. Furthermore, the absence of an oriented loop in the diagram allows us
to do a partial ordering of the vertices in the diagram 
so that all arrows are directed from the left to the right.  With this understanding we
can also drop the arrows from the diagram.

\subsubsection{Statement of Cutkosky rules for reduced diagrams} \label{sstate}

Our task will be to compute the contribution to $A(\{p_i\})- A(\{-p_i\})^*$ from
such a reduced diagram and show that the result is consistent with unitarity.
For this let us first examine what we need for unitarity.
Using $S=1-iT$ and the unitarity relation $S^\dagger S=1$ we get
\be \label{ett}
T - T^\dagger = -i \, T^\dagger T\, .
\ee
The
computation of $T^\dagger T$ is done by inserting a complete set of states between
$T$ and $T^\dagger$. 
For a multi-particle intermediate state we need to integrate over the spatial
components $\vec k_i$ of momenta of each particle subject to an overall
energy and momentum conserving 
delta function, and a measure factor
\be \label{epcut}
\left( 2\sqrt{\vec k_i^2+m^2}\right)^{-1} \, .
\ee
We can formally express this as
\be \label{enewk0}
i \, \int{dk_i^0\over 2\pi} \, P_c(k_i)\, ,
\ee
where
\be \label{eppcut}
P_c(k_i) \equiv - 2\pi i \, \delta \left((k_i^0)^2  - (\vec k_i^2 + m^2)\right) \theta(k_i^0)\, ,
\ee
and  the $k_i^0$ integral in \refb{enewk0} is taken 
to run along the real axis near the support of the 
$\delta$-function.
The factor \refb{eppcut} 
is precisely what we would get from the residue at the pole of the propagator 
$((k_i^0)^2-\vec k_i^2 - m^2)^{-1}$ if we take the difference between two contour
integrals in the
complex $k_i^0$ plane, one keeping 
the pole at $k_i^0=\sqrt{\vec k_i^2+m^2}$ to the right and the other keeping the same
pole to the left. 
We shall denote by a {\em cut propagator}, with the momentum $k$
flowing from the left to the right of the cut, the effect of replacing a 
propagator by \refb{eppcut}.
A {\em cut diagram}, obtained by drawing a line that divides the diagram into a left half and
a right half, will involve replacing each cut internal 
propagator by \refb{eppcut} and in addition
replacing the amplitude on the right of the cut by its hermitian conjugate. 
A cut across an external line has no effect.
With this
convention \refb{ett} is equivalent to the statement 
 that the amplitude $A(\{p_i\}) - A(\{-p_i\})^*$
will be given by sum over all cut diagrams of the amplitude $A(\{p_i\})$ up to some
phases. We shall now describe the origin of these phases and compute them. 
\begin{enumerate}
\item First of all the $-i$ factor on the right hand side of \refb{ett} will give an explicit
factor of $-i$ multiplying each cut diagram.
\item Replacing each cut propagator by \refb{eppcut} will produce the measure factor
\refb{epcut} if for each cut propagator there is an integral $i d k_i^0/2\pi$ in the 
original Feynman diagram for $A(\{p_i\})$. 
If each $k_i^0$ had represented an independent loop momentum then such a factor will
indeed be present according to \refb{eifactornew}.
However
typically there are energy conserving constraints relating the $k_i^0$'s which
reduce the
number of $k_i^0$ integrals, and hence also 
the number of $i$'s.
If there are $n_L$ disconnected components of the diagram 
to the left of the cut and $n_R$
disconnected components to the right of the cut, then the total number of constraints
is $n_L+n_R$. Of these one represents overall energy conservation instead of
imposing relations between $k_i^0$'s but the other $n_L+n_R-1$ constraints reduce
the number of independent  $k_i^0$'s and hence the number of factors of $i$. Therefore we
need to supply the missing $i$'s by
multiplying the cut diagram by a factor of $(i)^{n_L + n_R-1}$ so that we get back
the correct number of $i$'s that is needed to get the correct expression for 
$T^\dagger T$.
(The factors of $2\pi$ work out automatically since each momentum conserving delta
function is accompanied by a factor of $2\pi$.)
\item Eq.~\refb{e2.5a} shows that if the diagram on the left of the cut
has $n_L$ disconnected components then the expression for the matrix elements of
$T$ should contain an extra factor of $(i)^{-n_L+1}$. Similarly if the diagram
on the right of the cut has $n_R$ disconnected components then the matrix element
of $T^\dagger$ will contain a net extra factor of $(-i)^{-n_R+1}$ where the
replacement of $i$ by $-i$ is due to hermitian conjugation.  
There is no such factor in the original diagram without
cut, since we have assumed that to be connected.
Therefore this factor is also
absent in the cut diagram, and we need
to multiply the cut diagram by a
net factor of $(i)^{-n_L+n_R}$.
\end{enumerate}
Combining all these factors we see that we need to weigh a cut diagram by a factor
of 
\be \label{esignrule1}
(-i) (i)^{n_L + n_R -1} (i)^{-n_L+n_R} = (-1)^{n_R-1}\, ,
\ee
to reproduce the right hand side of \refb{ett}.  

A further simplification of cutting rules is possible for reduced diagrams. Let us
consider a cut Feynman diagram in which $n$ propagators carrying momenta
$k_1,\cdots k_n$ from left to right are cut. Then we have the relation
\be 
k_i^0 = \sqrt{\vec k_i^2 + m^2}\, .
\ee
Furthermore the $k_i$'s satisfy a momentum conservation law
\be 
p = \sum_{i=1}^n k_i\, ,
\ee
where $p$ is some linear combination of external momenta. This 
imposes constraint on the spatial components $\vec k_i$ of the momenta.
Now consider the same Feynman diagram without a cut but with the same spatial 
components of momenta along the propagators that were cut earlier. It is easy to
see that the integration contour over $k_i^0$ for $1\le i \le (n-1)$
are now pinched at
\be 
k_i^0 = \sqrt{\vec k_i^2 + m^2-i\eps}\, , \quad p^0 - \sum_{i=1}^{n-1} k_i^0
= \sqrt{(\vec p - \vec k)^2+m^2-i\eps}\, .
\ee
Reversing this result we see that for fixed spatial momenta flowing along
the loops, a Feynman diagram allows a cut passing through propagators
$P_1,\cdots P_n$ only if in the original diagram
the energy integration contour has a pinch where all the propagators 
$P_1,\cdots P_n$ are on-shell. This is turn means that in a reduced diagram
a cut cannot intersect the propagators inside a reduced vertex. This allows
us to state the required Cutkosky rule for a reduced diagram as 
follows:

\noindent{\it The contribution to $A(\{p_i\})-A(\{-p_i\})^*$ from a reduced diagram 
is given by the sum over all cut diagrams with the cuts 
avoiding the reduced vertices, weighted by the
factor given in \refb{esignrule1}.}

\begin{figure}
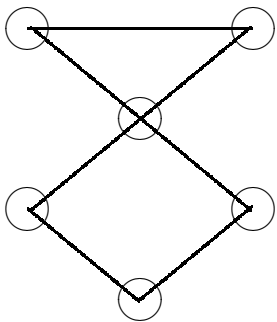

\begin{center}
\figonevrr
\end{center}

\vskip -.2in

\caption{Example of a 1VR diagram. The external lines are suppressed,
the reduced vertices are denoted by circles, and the arrows on all
lines are understood to be directed towards the right.
\label{figonevrr}
}
\end{figure}

We shall in fact prove a slightly more general result. Consider an amplitude in
which we have replaced the integration contour $C$ 
required to compute $A(\{p_i\})$
by a different contour $\wt C$ leaving the integrand unchanged. 
Let us call this contribution $\wt A(\{p_i\})$.
We again work at fixed values of the spatial
components of loop momenta, and define 
$R$ as the original
reduced diagram 
associated with integration along the contour $C$ and $\wt R$ as the reduced
diagram obtained
by shrinking to
points all propagators which are not pinched in $\wt C$. 
We shall show that the version of the Cutkosky rules for
reduced diagrams, as stated above, holds for 
$\wt R$ as 
long as the poles coming from the surviving propagators in $\wt R$ 
lie on the same side
of the integration contour $\wt C$ as they were for $C$. However there is no restriction on
how the poles associated with the propagators inside the reduced vertices
of $\wt R$ are situated relative
to $\wt C$ as long as there is no pinch singularity that prevents us from deforming $\wt C$
away from these poles. In particular 
even if the original contour $C$
was pinched at the poles of some of the propagators inside a reduced vertex
of $\wt R$, Cutkosky rules for $\wt R$ will not include sum over cuts passing through this 
reduced vertex.

We shall prove this in two steps.
\begin{enumerate}
\item First we shall introduce the notion of one vertex irreducible (1VI) and 
one vertex reducible (1VR) reduced diagrams and show that the Cutkosky rules for 
1VR diagrams hold as long as they hold for 1VI diagrams.
\item Then we shall prove the Cutkosky rules for 1VI reduced diagram.
\end{enumerate}

\subsubsection{One vertex reducible reduced diagrams} \label{s4.2.1}

We shall define a reduced diagram to be  1VR if it can be
regarded as two reduced diagrams joined at a single reduced vertex. 
An example of such a diagram has been shown in Fig.~\ref{figonevrr}.
Reduced
diagrams which are not 1VR will be called  1VI. We shall
now show that for a 1VR  reduced diagram, the Cutkosky rules follow if they hold for the
individual components that are joined at a single reduced vertex to produce the 1VR
diagram. By repeated application of this result, one can then show that the Cutkosky
rules will hold for a general reduced diagram as long as they hold for 1VI diagrams.

\begin{figure}
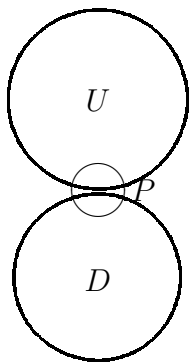


\begin{center}
\figonevr
\end{center}

\vskip -.5in

\caption{Schematic representation of a
1VR reduced diagram consisting of two components $U$ and $D$
joined at a single reduced vertex $P$. All external lines have been 
suppressed.
\label{figonevr}
}
\end{figure}

\begin{figure}
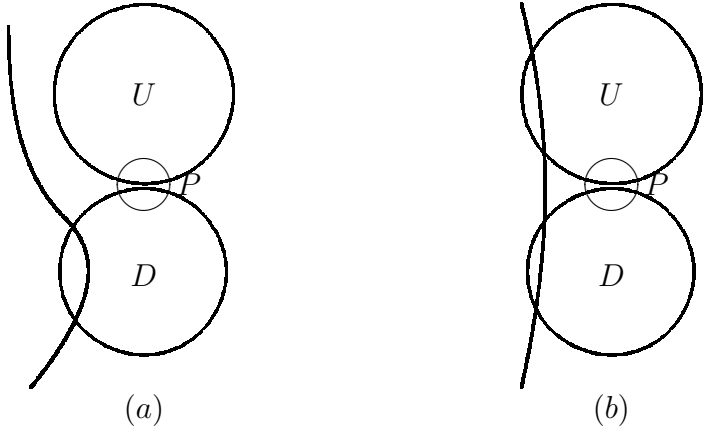

\begin{center}
\hbox{\figonevrB \quad \figonevrA}
\end{center}

\vskip -.3in

\caption{Cut diagrams of the reduced diagram of Fig.\ref{figonevr} corresponding to the first
and the second line of \refb{esecond}.
\label{figtwovr}
}
\end{figure}

Let us consider a 1VR diagram shown in Fig.~\ref{figonevr} consisting of
two pieces $U$ and $D$ connected at a single reduced vertex $P$. 
$U$ and $D$ may be either 1VI or 1VR -- our analysis holds in all cases.
In general the contribution from the reduced vertex $P$ will depend on the
momenta entering it from the blobs $U$ and $D$ and the amplitude will not be
factorized. First let us assume that the dependence on these momenta are
factorized so that the full amplitude can be regarded as a product of the amplitudes
associated with the two blobs -- we shall deal with the general case later. 
We denote by $A_U$
and $A_D$ the amplitudes associated with the reduced diagrams $U$ and $D$.
Then the full amplitude is given by $A_U A_D$.
Our goal will be to show that 
$A_U A_D - A_U^* A_D^*$ is given by the sum over cut diagrams of the full diagram 
weighted by the phase factor \refb{esignrule1}, if we assume that similar result
holds for $A_U - A_U^*$ and $A_D-A_D^*$.   Now since the cuts do not pass through the 
reduced vertex $P$, the cut diagrams of $A_U$ and $A_D$
can be divided into two parts, those with the cut
on the left of the vertex $P$ and those with the cut on the
right of the vertex $P$.  
We denote by $\Delta_{UL}$, $\Delta_{UR}$, $\Delta_{DL}$ and $\Delta_{DR}$ respectively
the sum over
all cut diagrams,  weighted by \refb{esignrule1}, 
(a) of $U$ with the cut on the left of $P$,  
(b) of $U$  with the cut on the right of $P$, 
(c) of $D$  with the cut on the left of $P$ and
(d) of $D$  with the cut on the right of $P$.
Then the assumption that the diagrams $U$ and $D$ satisfy Cutkosky rules imply that
\be\label{efirst}
A_U - A_U^* = \Delta_{UL} + \Delta_{UR}, \qquad A_D -A_D^* = \Delta_{DL} + \Delta_{DR}\, .
\ee
We shall now compute the sum over all cut diagrams of the full diagram shown in
Fig.~\ref{figonevr}. These diagrams can be divided into six classes. 
Two of them, described by the first two lines of \refb{esecond}, are
shown in Figs.~\ref{figtwovr}(a) and (b)
respectively; 
the rest can be
drawn in a similar fashion. 
Below we describe
these six classes of cut diagrams and their contribution:
\ben \label{esecond}
\hbox{cuts of $D$ on the left of $P$, passing on the left of $U$}
&:& A_U^* \Delta_{DL}\nonumber \\
\hbox{cuts of $D$ on the left of $P$, cuts of $U$ on the left of $P$}
&:& \Delta_{UL} \Delta_{DL}\nonumber \\
\hbox{cuts of $U$ on the left of $P$, passing on the left of $D$}
&:& \Delta_{UL} A_D^* \nonumber \\
\hbox{cuts of $D$ on the right of $P$, cuts of $U$ on the right of $P$}
&:& - \Delta_{UR} \Delta_{DR}\nonumber \\
\hbox{cuts of $D$ on the right of $P$, passing on the right of $U$}
&:& A_U \Delta_{DR}\nonumber \\
\hbox{cuts of $U$ on the right of $P$, passing on the right of $D$}
&:&  \Delta_{UR} A_D\, .
\een
The minus sign on the right hand side of the fourth line is a consequence of
\refb{esignrule1} and 
the fact that $n_R-1$ for
the corresponding cut 
is given by $(n_{UR}-1)+(n_{DR}-1)+1$.
The sum of these, using \refb{efirst}, can be easily seen to be
given by
\be 
A_U A_D - A_U^* A_D^*\, .
\ee
This is precisely the Cutkosky rule for the full diagram. 
This proves the desired relation. 

Let us now turn to the general case where the reduced vertex $P$ depends on the
momenta entering it from both $U$ and $D$, and the contribution is not factorized.
Let us denote by $\{\ell_{U,i}\}$ the momenta entering $P$ from $U$
and by $\{\ell_{D,i}\}$  the momenta entering $P$ from $D$. 
Each set satisfies an overall momentum conservation constraint that sets 
$\sum_i \ell_{U,i}=-\sum_i \ell_{D,i}$
to some linear combination of external momenta giving the 
total momentum flowing across the reduced vertex $P$.
Our starting
assumption will be that for fixed $\{\ell_{D,i}\}$ the sub-diagram $U$,
including the contribution from the reduced vertex $P$, satisfies the
Cutkosky rules and that for fixed $\{\ell_{U,i}\}$ the subdiagram $D$,
including the contribution from the reduced vertex $P$,  satisfies 
the Cutkosky rules. Let us denote the corresponding amplitudes by $A_U$ and $A_D$
respectively, and the sum over cuts as described above \refb{efirst} by $\Delta_{UR}$,
$\Delta_{UL}$, $\Delta_{DR}$ and $\Delta_{DL}$ so that \refb{efirst} holds.
We also denote by $V_P(\{\ell_{U,i}\}, \{\ell_{D,i}\})$ the contribution from the reduced
vertex $P$, and
 by $a_U$, $a_U^*$, $\delta_{UL}$ and $\delta_{UR}$ the quantities
 appearing in the expressions for 
 $A_U$, $A_U^*$, $\Delta_{UL}$, $\Delta_{UR}$ introduced
above \refb{efirst}, {\it before
doing integration over $\{\ell_{U,i}\}$ and without including the contribution 
$V_P$ from the reduced vertex}. 
$a_D$, $a_D^*$, $\delta_{DL}$ and $\delta_{DR}$ will
denote similar contributions that would
enter the computation of $A_D$, $A_D^*$, $\Delta_{DL}$ and $\Delta_{DR}$.
In that case $a_U$, $a_U^*$, $\delta_{UL}$ and $\delta_{UR}$ depend on 
$\{\ell_{U,i}\}$ but not on $\{\ell_{D,i}\}$ and 
$a_D$, $a_D^*$, $\delta_{DL}$ and $\delta_{DR}$ depend on 
$\{\ell_{D,i}\}$ but not on $\{\ell_{U,i}\}$.
We now have
\ben \label{eff1}
&& A_U =  \int_{\{\ell_{U,i}^0\}} a_U V_P, \quad A_U^* =  \int_{\{\ell_{U,i}^0\}} a_U^* V_P,
\quad \Delta_{UL}= \int_{\{\ell_{U,i}^0\}} \delta_{UL} V_P,
\quad \Delta_{UR}= \int_{\{\ell_{U,i}^0\}} \delta_{UR} V_P\, ,
\nonumber \\
&& A_D =  \int_{\{\ell_{D,i}^0\}} a_D V_P, \quad A_D^* =  \int_{\{\ell_{D,i}^0\}} a_D^* V_P,
\quad \Delta_{DL}= \int_{\{\ell_{D,i}^0\}} \delta_{DL} V_P,
\quad \Delta_{DR}= \int_{\{\ell_{D,i}^0\}} \delta_{DR} V_P\, .
\nonumber \\
\een
In these equations it is understood that while doing the integration over 
$\{\ell_{U,i}^0\}$ and $\{\ell_{D,i}^0\}$, the choice of integration contour may
depend on the integrand. For example the integration contours for 
$\{\ell_{U,i}^0\}$ for integral over $a_U$ and $a_U^*$ may not be the same.
Also note that we have not included integration over the spatial components of
$\{\ell_{U,i}\}$ and $\{\ell_{D,i}\}$ 
since we have been working at fixed values of the spatial components
of loop momenta.
Eq.\refb{efirst} now takes the form
\be \label{ethird}
\int_{\{\ell_{U,i}^0\}} (a_U - a_U^*)  V_P= \int_{\{\ell_{U,i}^0\}} 
(\delta_{UL}+\delta_{UR}) V_P, \quad
\int_{\{\ell_{D,i}^0\}} (a_D - a_D^*) V_P = \int_{\{\ell_{D,i}^0\}} 
(\delta_{DL}+\delta_{DR}) V_P\, .
\ee
Eq.\refb{esecond} can be similarly generalized, leading to the following contribution
to the sum over cut diagrams of Fig~\ref{figonevr}:
\be \label{efourth}
\int_{\{\ell_{U,i}^0\}}  \int_{\{\ell_{D,i}^0\}}
\Big( a_U^* \delta_{DL} + \delta_{UL} \delta_{DL} + \delta_{UL}a_D^*
- \delta_{UR} \delta_{DR} + a_U \delta_{DR} +\delta_{UR}a_D\Big)\,  V_P 
\, .
\ee
Again we should keep in mind that for different integrands we have to integrate
over different contours. 
After some algebra using \refb{ethird}, the expression \refb{efourth} can be brought to
the form
\be \label{efifth}
\int_{\{\ell_{U,i}^0\}}  \int_{\{\ell_{D,i}^0\}}  (a_U a_D - a_U^* a_D^*)\, V_P \, .
\ee
This is precisely the difference between the original amplitude shown in 
Fig.~\ref{figonevr} and its hermitian conjugate. 
This gives the desired result.

\subsubsection{One vertex irreducible reduced diagrams} \label{s5.2.2}

We now turn to the task of proving Cutkosky rules for 1VI diagrams.
As mentioned before,
we shall carry out the proof recursively, i.e.\ assume that the result holds for
all reduced diagrams with $(N-1)$ loops and then prove that the result holds
for 1VI reduced diagrams with $N$ loops. 
To this end let us consider a 1VI
reduced diagram with $N$ loops and label the independent loop momenta by
$\ell_1,\cdots \ell_N$. We now consider the particular loop $S$ that carries loop
momentum $\ell_1$ and analyze the integral over $\ell_1^0$ at fixed values of
other loop momenta. Let us suppose that as we traverse this loop along the
direction of $\ell_1$, $n$ of the propagators in the
loop -- which we denote by
$P_1,\cdots P_n$ -- have their 
arrows directed along $\ell_1$ while the others have their arrows directed opposite
to $\ell_1$. In that case near the pinch the relevant pole in the $\ell_1^0$ plane 
from  the propagator $P_i$ has the form
\be \label{epoles}
\left( -(\ell_1 + K_i)^2 - m^2 + i\eps\right)^{-1} \theta(\ell_1^0+K_i^0)\, ,
\ee
for $1\le i\le n$.
Here $K_i$'s are linear combinations of the external momenta and other
loop momenta in the reduced diagram. The $\theta(\ell_1^0+K_i^0)$ is a formal
expression that tells us that at the pinch the relevant pole is the one that
appears at positive value of $\ell_1^0+K_i^0$.
The $i\eps$ prescription reflects that for the $\ell_1^0$ integration 
contour beginning at
$-i\infty$ and ending at $i\infty$, the pole of \refb{epoles}
lies to the right of the integration contour. On the other hand the poles from all other
 propagators in the loop $S$ that are not in the set $P_1,\cdots P_n$ lie to the
left of the $\ell_1^0$ integration contour.
 We denote by $C$ the original integration
contour, and by $C^*$ the integration contour needed to compute the hermitian conjugate
amplitude. Below we follow the convention that
for any contour $\CC$ required for computing an amplitude,
$\CC^*$ will denote
the integration contour required to compute the hermitian
conjugate of the amplitude.

We shall now deform the integration contours of $\ell_1^0$ in $C$
through each of the poles
given in \refb{epoles} to the other side, at the expense of picking up residues at the
poles. Let us denote the deformed integration contour 
by $\wh C$ and the amplitude obtained by integrating over this deformed
contour by $\wh A$. This contour will have all the relevant poles in the
$\ell_1^0$ plane
to the left of the integration contour and hence the contour is not pinched.
Therefore by deforming the $\ell_1^0$
integration contours we can ensure that the momenta along 
all the  propagators in the loop $S$ can be deformed away from the on-shell
values. On the other hand the difference between the original amplitude $A$
and the new amplitude $\wh A$ is given by the sum of residues at the poles
\refb{epoles} through which we deform the contour. This
may be computed using the
relation
\ben \label{efactss}
&& \prod_{i=1}^n \left\{
\left(- (\ell_1 + K_i)^2  - m^2 - i\eps\right)^{-1} \theta(\ell_1^0+K_i^0)\right\}
\\
&=& \prod_{i=1}^n \left\{ \left( -(\ell_1 + K_i)^2  - m^2 + i\eps\right)^{-1} \theta(\ell_1^0+K_i^0)
 + 2\pi i \, \delta \left( (\ell_1 + K_i)^2  + m^2 \right) \theta(\ell_1^0+K_i^0)\right\}\, ,
\nonumber
\een
which gives
\ben \label{eyy1}
&& \prod_{i=1}^n \left( -(\ell_1 + K_i)^2  - m^2 +i\eps\right)^{-1} \theta(\ell_1^0+K_i^0)
\nonumber \\
&=& \prod_{i=1}^n \left( -(\ell_1 + K_i)^2  - m^2 -
i\eps\right)^{-1} \theta(\ell_1^0+K_i^0)
\nonumber \\  && 
+ \sum_{j=1}^n  \left\{
- 2\pi i \, \delta \left( (\ell_1 + K_j)^2  + m^2 \right) \theta(\ell_1^0+K_j^0)
\right\}
\prod_{i=1\atop i\ne j }^n \left( -(\ell_1 + K_i)^2  - m^2 +
i\eps\right)^{-1} \theta(\ell_1^0+K_i^0)
\nonumber \\ &&
- \sum_{j,k=1\atop j<k }^n  \left\{
- 2\pi i \, \delta \left( (\ell_1 + K_j)^2  + m^2 \right) \theta(\ell_1^0+K_j^0)
\right\}
 \left\{
- 2\pi i \, \delta \left( (\ell_1 + K_k)^2  + m^2 \right) \theta(\ell_1^0+K_k^0)
\right\} \nonumber \\ &&\qquad  \times
\prod_{i=1\atop i\ne j,k }^n \left( -(\ell_1 + K_i)^2  - m^2 +
i\eps\right)^{-1} \theta(\ell_1^0+K_i^0)
\nonumber \\
&& + \cdots 
\nonumber \\
&& + (-1)^{n-1} \prod_{j=1}^n 
\left\{
- 2\pi i \, \delta \left( (\ell_1 + K_j)^2  + m^2 \right) \theta(\ell_1^0+K_j^0)
\right\} \, .
\een
The $(-2\pi i)\, \delta \left( (\ell_1 + K_j)^2  + m^2 \right) \theta(\ell_1^0+K_j^0)$ factor
should again be regarded as a formal expression that 
has to be made sense of by regarding the $\ell_1^0+K_j^0$ integration to be along
the real axis near the pinch singularity.
The product over the propagator factors given in the left hand side of
\refb{eyy1} appears in the integrand needed for computing the amplitude $A$.
When we replace this by the right hand side of \refb{eyy1} inside the integral,
the first term on the right hand side represents 
integration over the deformed contour $\wh C$ generating
the amplitude $\wh A$ and the other terms on the
right hand side represent the residues at various poles from the propagators
$P_1,\cdots P_n$ picked up during the
deformation from $C$ to $\wh C$.
Comparison with the right hand side of \refb{eppcut} shows that the effect of
replacing the  propagator $P_j$ by the 
$(-2\pi i) \delta \left( (\ell_1 + K_j)^2  + m^2 \right) \theta(\ell_1^0+K_j^0)$
factor may be represented by a cut on the $j$-th  propagator.
Let us denote by $A^{(j)}$ the amplitude obtained by replacing 
the propagator $P_j$ by the cut 
propagator  in the original
amplitude. More generally we denote by 
$A^{(i_1\cdots i_s)}$ the amplitude obtained by replacing the  propagators 
$P_{i_1},\cdots P_{i_s}$ by cut  propagators. 
Then \refb{eyy1} inside the integral translates to
\be \label{expC}
A = \wh A + \sum_{j=1}^n A^{(j)} - \sum_{j,k=1\atop j<k}^n A^{(jk)}
+\cdots + (-1)^{n-1} A^{(12\cdots n)}\, .
\ee

Even though $A^{(i_1\cdots i_s)}$ is obtained from the original amplitude $A$ by
replacing some of its internal propagators by cut propagators,
it is important to recognize that $A^{(i_1\cdots i_s)}$ is not a cut diagram. There is no
cut separating the graph into two parts and there is no part of the diagram that
is to be replaced by its hermitian conjugate. 
Therefore it is more appropriate to interpret 
$A^{(i_1\cdots i_s)}$  as an amplitude where the
propagators $P_{i_1},\cdots P_{i_s}$ have been replaced by on-shell
external states. Furthermore, since all the propagators factors on the right hand
side of \refb{eyy1} except the first term have the correct $i\eps$ prescription, 
$A^{(i_1\cdots i_s)}$ is defined in the same way as the original amplitude $A$, i.e.\
by taking all the external state energies to be $\lambda E_s$ for real $E_s$, and then
taking the $\lambda\to 1$ limit from the first quadrant.

We can also carry out a similar manipulation for the hermitian conjugate 
amplitude $A^*$.\footnote{Naively one might expect that the effect of 
hermitian conjugation will 
change the $i$'s to $-i$ in the expression for the cut propagators, and hence 
give an extra minus sign for each cut propagator. However the way we have defined
the contour $C^*$ involves a complex conjugation together with orientation reversal,
and this ensures that any given pole lies on the same side of $C$ and $C^*$.
Therefore during the deformation from $C$ to $\wh C$ and $C^*$ to $\wh C^*$ we
cross various poles in the same direction, and 
there is no minus sign in the expression for the cut propagators of the
hermitian conjugate amplitude $A^*$.}
Manipulations similar to the one given above, applied to $A^*$,
give
\be \label{expCstar}
A^* = \wh A^* + \sum_{j=1}^n A^{(j)*} - \sum_{j,k=1\atop j<k}^n A^{(jk)*}
+\cdots + (-1)^{n-1} A^{(12\cdots n)*}\, .
\ee

\begin{figure}
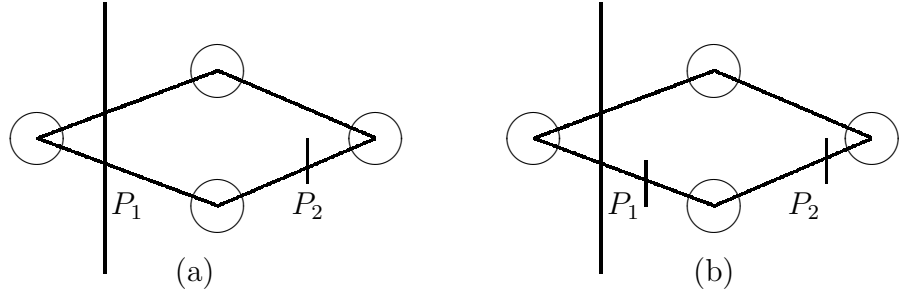


\begin{center}
\figupdpwn
\end{center}

\vskip -.2in

\caption{Fig.~(a) shows a cut reduced diagram in which a
propagator $P_2$ is replaced by a cut propagator in the original diagram and
the cut passes through the propagator
$P_1$. In our notation this
will be labelled as $A^{(2)}_1$. Fig.~(b) 
shows a cut reduced diagram in which 
propagators $P_1$ and $P_2$ are replaced by cut propagators in the original diagram and
the cut passes through the propagator
$P_1$. In our notation this
will be labelled as $A^{(12)}_1$. These two contributions are identical. In this example there
is no cut diagram in which the cut passes through both the propagators $P_1$ and $P_2$, and
hence $A^{(12)}_{12}=0$. But this is not always the case.
\label{fupdown}
}
\end{figure}

Since $A^{(i_1\cdots i_s)}$
has less number of loops than the original diagram contributing to the
amplitude $A$,  
the Cutkosky rules hold for  $A^{(i_1\cdots i_s)}$.
Therefore the anti-hermitian part
of $A^{(i_1\cdots i_s)}$ is given by the sum over all its cut diagrams. We denote by 
$A^{(i_1\cdots i_s)}_{j_1\cdots j_r}$ the sum over all cut diagrams of the amplitude
$A^{(i_1\cdots i_s)}$ which can be considered as cuts of the original amplitude,
and for which the cut passes through
$P_{j_1},\cdots P_{j_r}$ and possibly other propagators,
but not any of the other $P_i$'s in the set $\{P_1,\cdots P_n\}$.
Some examples of this have been shown in Fig.~\ref{fupdown}.
$A^{(i_1\cdots i_s)}_{\emptyset}$ will denote the sum over all the cut diagrams of 
$A^{(i_1\cdots i_s)}$ for which the cut does not pass through any of the  propagators
in the set $\{P_1,\cdots P_n\}$. Then we may express the Cutkosky rule
applied to the amplitude associated with $A^{(i_1\cdots i_s)}$ as
\be\label{ethisth}
A^{(i_1\cdots i_s)} - A^{(i_1\cdots i_s)*} 
= A_\emptyset ^{(i_1\cdots i_s)} + \sum_{j_1=1}^n A_{j_1} ^{(i_1\cdots i_s)}
+ \sum_{j_1, j_2=1\atop j_1<j_2}^n A_{j_1j_2}^{(i_1\cdots i_s)}
+ \cdots + A_{1\cdots n}^{(i_1\cdots i_s)} + R^{(i_1\cdots i_n)}\, ,
\ee
where $R^{(i_1\cdots i_n)}$ denotes sum over cuts of 
$A^{(i_1\cdots i_n)}$ which cannot be considered
as cuts of the original amplitude $A$. Some examples of such cuts
can be found in Fig.~\ref{fadded} below, but we shall postpone discussion on them now
and return to them below \refb{elhs}.
Using \refb{ethisth} and \refb{expC}, \refb{expCstar} we get
\ben  \label{eccstar}
A - A^* &=& \wh A - \wh A^* 
+ \sum_{i=1}^n \left[A^{(i)}_\emptyset + \sum_{j_1=1}^n A^{(i)}_{j_1} 
+ \sum_{j_1,j_2=1\atop j_1<j_2}^n A^{(i)}_{j_1 j_2} + \cdots
+ A^{(i)}_{1\cdots n} 
\right] \nonumber \\ &&
- \sum_{i,j=1\atop i<j}^n \left[A_\emptyset^{(ij)}
+ \sum_{j_1=1}^n A^{(ij)}_{j_1} 
+ \sum_{j_1,j_2=1\atop j_1<j_2}^n A^{(ij)}_{j_1 j_2} + \cdots
+ A^{(ij)}_{1\cdots n} \right]
\nonumber \\ &&
+\cdots \nonumber \\ && \hskip -1in 
+ (-1)^{n-1} \left[A^{(12\cdots n)}_\emptyset + \sum_{j_1=1}^n 
A^{(12\cdots n)}_{j_1} 
+ \sum_{j_1,j_2=1\atop j_1<j_2}^n A^{(12\cdots n)}_{j_1 j_2} + \cdots
+ A^{(12\cdots n)}_{1\cdots n} 
\right] + R\, ,
\een
where $R$ is the sum over the contributions from $R^{(i_1\cdots i_n)}$.

We now note the following relations. First of all, since we have seen that in $\wh A$
the $\ell_1^0$ contour is not pinched, the loop $S$ can be shrunk to a reduced vertex.
The resulting reduced diagram has one less loop than the original diagram, and hence
the Cutkosky rules should hold for this diagram. Furthermore in none of the cut diagrams of this
diagram the cut will pass through any of the  propagators $P_i$ since they have all
been shrunk to a reduced vertex. This gives, in our previous notation,\footnote{Note
that \refb{ewhcdiff} requires the generalization of the Cutkosky rules for the reduced
diagrams mentioned at the end of \S\ref{sstate}.}
\be \label{ewhcdiff}
\wh A - \wh A^* =\wh A_\emptyset\, .
\ee
Second we note that in $A^{(i_1\cdots i_s)}_{j_1\cdots j_r}$ the cut passes
through the 
propagators $P_{j_1},\cdots P_{j_r}$ putting them on-shell, and the propagators 
$P_{i_1},\cdots P_{i_s}$ are replaced by cut propagators, putting them on-shell  
from the beginning. Therefore
the result remains
the same if we append to the set $i_1,\cdots i_s$ appearing in the superscript one or more
elements of the set $\{j_1,\cdots j_r\}$
that are not already part of $\{i_1,\cdots i_s\}$. This has been illustrated in Fig.~\ref{fupdown}.
This gives 
\be \label{e5.25}
A^{(i_1\cdots i_s)}_{j_1\cdots j_r} 
= A^{(\{i_1,\cdots i_s\}\cup \{ j_1,\cdots j_r\})}_{j_1\cdots j_r} \, .
\ee
Using this we can compute the coefficient of
$A^{(i_1\cdots i_s)}_{j_1\cdots j_r}$ on the right hand side of
\refb{eccstar} as follows. 
Due to \refb{e5.25} we can choose the independent $A$'s to be of the form
$A^{(i_1\cdots i_s j_1\cdots j_r)}_{j_1\cdots j_r}$ with 
$\{i_1,\cdots i_s\}\cap\{j_1,\cdots j_r\}=\emptyset$. In this case
for $s\ne 0$, $r\ne 0$, the coefficient of
$A^{(i_1\cdots i_s j_1\cdots j_r)}_{j_1\cdots j_r}$ comes from the following terms
in \refb{eccstar}:
\ben\label{eline}
A^{(i_1\cdots i_s)}_{j_1\cdots j_r} &:&  (-1)^{s-1}\nonumber \\
A^{(i_1\cdots i_s j_m)}_{j_1\cdots j_r} &:& (-1)^{s} \quad \hbox{for $1\le m\le r$}
\nonumber \\
A^{(i_1\cdots i_s j_m j_p)}_{j_1\cdots j_r} &:& (-1)^{s+1} \quad \hbox{for $1\le m<p\le r$}
\nonumber \\
\cdots &:& \cdots \nonumber \\
A^{(i_1\cdots i_s j_1\cdots j_r )}_{j_1\cdots j_r} &:& (-1)^{s+r-1}
\een
The 
net contribution to the coefficient from all the terms is given by
\be 
(-1)^{s-1} \left[ 1 - r +{r \choose 2} - \cdots + (-1)^r {r \choose r}
\right] = (-1)^{s-1} (1-1)^r = 0\, .
\ee
For $s=0$, i.e.\ for $A^{(j_1\cdots j_r )}_{j_1\cdots j_r}$, the first line of
\refb{eline} will be missing. As a result the contribution is given by
\be
- \left[- r +{r \choose 2} - \cdots + (-1)^r {r \choose r}
\right] = 1 - (1-1)^r = 1\, .
\ee
Finally
for $r=0$, i.e.\ for $A^{(i_1\cdots i_s)}_\emptyset$, only the term in the first line
of \refb{eline} is present and the contribution is given by
\be 
(-1)^{s-1}\, .
\ee
This, together with \refb{ewhcdiff} can be used to rewrite \refb{eccstar} as
\ben \label{elhs}
A - A^* &=& \wh A_\emptyset  + \sum_{i=1}^n A^{(i)}_i + \sum_{i,j=1\atop i<j}^n
A^{(ij)}_{ij} + \cdots + A^{(1\cdots n)}_{1\cdots n} \nonumber \\ &&
+ \sum_{i=1}^n A^{(i)}_\emptyset - \sum_{i,j=1\atop i<j}^n
A^{(ij)}_{\emptyset} + \cdots + (-1)^{n-1} A^{(1\cdots n)}_{\emptyset} + R
\, .
\een

\begin{figure}
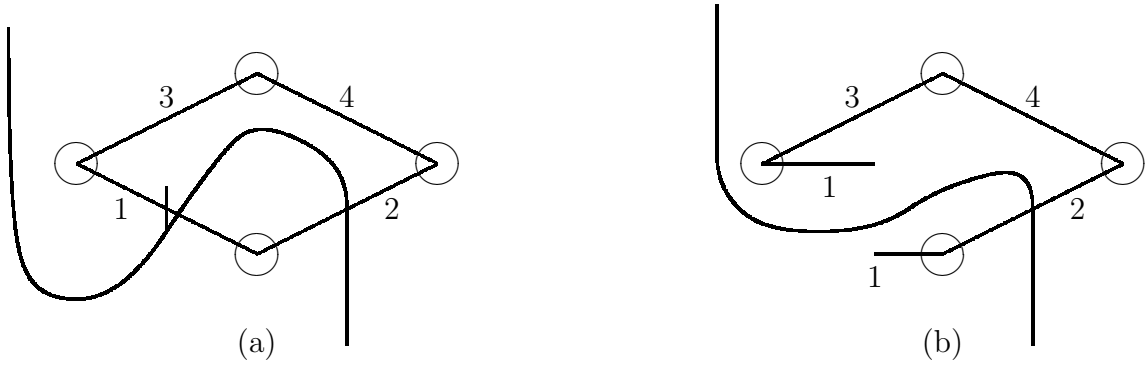


\begin{center}

\hbox{\figadded \hskip 1in \figaddedb}

\end{center}

\vskip -.2in

\caption{Fig.~(a) shows a cut reduced diagram in which a
propagator $P_1$ is replaced by a cut propagator in the original diagram and
the cut passes through the propagator
$P_1$ in the reverse direction and the propagator $P_2$ in the correct direction. 
In our notation this
will be labelled as $A^{(1)}_{\underline{1} 2}$. Fig.~(b) 
shows a more conventional depiction of the same diagram in which the cut propagator 
$P_1$ is depicted as one outgoing and one incoming particle carrying 
identical quantum numbers. This figure makes it clear that this is an allowed cut of
$A^{(1)}$ even though it is not an allowed cut of $A$.
\label{fadded}
}
\end{figure}

We now turn to the additional contribution $R$.
By definition, $A^{(i_1\cdots i_n)}_{j_1\cdots j_m}$
includes sum over only the cuts of the original diagram. However since in 
$A^{(i_1\cdots i_n)}$ the propagators $P_{i_1},\cdots P_{i_n}$ are put on-shell, there are
other possible cuts which, in the original diagram, will appear as if the cut passes through 
one or more of the propagators $P_{i_1},\cdots P_{i_n}$ in the reverse direction. 
$R$ denotes the contributions from all such cuts. An example of such a cut 
has been
shown in Fig.\ref{fadded}(a). 
While it may seem strange to include such cuts, as explained in
Fig.~\ref{fadded}(b), this is a regular cut of $A^{(i_1\cdots i_n)}$ 
if we regard each of the propagators $P_{i_1},\cdots P_{i_s}$
as a pair of incoming and outgoing external particles carrying identical momentum. Effectively
putting a propagator on-shell creates a gap in the propagator through which the cut can pass in
either direction.

We shall now evaluate the contribution from these cut diagrams. As is clear from Fig.~\ref{fadded}
if we have a cut passing through an on-shell propagator in the reverse order, there must be at least
one other propagator in the set $\{P_1,\cdots P_n\}$
through which the cut passes through in the correct order -- in Fig.~\ref{fadded}(a)
it is the propagator $P_2$. Without any loss of generality we can assume that among the
on-shell propagators $P_{i_1},\cdots P_{i_s}$, the propagators
$P_{i_1},\cdots P_{i_m}$ for $m\le s$ are traversed by the cut in the reverse direction. 
We can define 
the contributions from these cut diagrams to $A^{(i_1\cdots i_n)}-A^{(i_1\cdots i_n)*}$ as
\be
A^{(i_1\cdots i_s)}_{\underline{i_1\cdots i_m}\, j_1\cdots j_t}\, ,
\ee
where $P_{j_1},\cdots P_{j_t}$ are the propagators in the set $\{P_1,\cdots P_n\}$ 
that are traversed by the
cut in the right direction.
The set $\{j_1,\cdots j_t\}$ may or may not have overlap with the set
$\{i_1,\cdots i_s\}$. We now note that as before, we can append to the set 
$\{i_1\cdots i_s\}$ in the superscript 
one or more members of the set $\{j_1,\cdots j_t\}$ that is not present
there, without changing the result. Therefore we can begin with the term where $\{i_1,\cdots i_s\}$
has no overlap with $\{j_1,\cdots j_t\}$ and then add to it the result of appending one or more
members of $\{j_1,\cdots j_t\}$.
The total coefficient of such a term in $A-A^*$ will be given by
\be
(-1)^s \left(1 - t + {t\choose 2} - \cdots + (-1)^t\right) = (-1)^s (1-1)^t\, .
\ee
Since we have already argued that $t\ge 1$, we see that this contribution vanishes.
Therefore \refb{elhs}, with $R=0$, gives the complete result for $A-A^*$.

In order to show that the Cutkosky rules hold for the amplitude associated
with $A$, we have to show that the right hand side of \refb{elhs} agrees with the 
sum of all the cut diagrams of this amplitude. Let us denote by $A_\emptyset$ 
the sum of cut diagrams of $A$ in
which the cut does not pass through any of the  propagator $P_{1},\cdots P_{n}$,
and by $A_{i_1\cdots i_s}$ the sum of
cut diagrams of $A$ in which the cut passes through the 
propagators $P_{i_1},\cdots P_{i_s}$ and possibly other propagators
but not any of the other  propagators
in the set $\{P_{1},\cdots P_{n}\}$. Then the 
sum over all the cut diagrams of $A$ is given by
\be \label{esumcut}
A_\emptyset + \sum_{i=1}^n A_i + \sum_{1\le i<j\le n} A_{ij} +\cdots + A_{1\cdots n}
\, .
\ee
Since in $A_{i_1\cdots i_s}$ the  propagators $P_{i_1}, \cdots P_{i_s}$
are put on-shell, we have
\be \label{erhs1}
A_{i_1\cdots i_s} = A_{i_1\cdots i_s}^{(i_1\cdots i_s)}\, .
\ee
On the other hand since in $A_\emptyset$ none of the  propagators $P_1,\cdots P_n$
are cut, the cut does not enter the loop $S$. As a result the entire loop lies on one
side of the cut. We can now repeat the analysis that led to \refb{expC},
\refb{expCstar} on the
sub-diagram of $A_\emptyset$ that contains the loop $S$, leading to
\be\label{erhs2}
A_\emptyset 
= \wh A_\emptyset + \sum_{i=1}^n A_\emptyset^{(i)} - \sum_{i,j=1\atop i<j}^n A_\emptyset^{(ij)}
+\cdots + (-1)^{n-1} A_\emptyset^{(12\cdots n)}\, .
\ee
Substituting 
\refb{erhs1} and \refb{erhs2} into \refb{esumcut} we get the following expression for
the sum over all the cut diagrams of $A$:
\ben \label{eeachred}
&& \wh A_\emptyset  + \sum_{i=1}^n A^{(i)}_\emptyset - \sum_{i,j=1\atop i<j}^n
A^{(ij)}_{\emptyset} + \cdots + (-1)^{n-1} A^{(1\cdots n)}_{\emptyset}
\nonumber \\ &&
+ \sum_{i=1}^n A^{(i)}_i + \sum_{i,j=1\atop i<j}^n
A^{(ij)}_{ij} + \cdots + A^{(1\cdots n)}_{1\cdots n} \, .
\een
This
precisely agrees with the right hand side of \refb{elhs} with $R=0$. This shows that
the Cutkosky rules hold for the 1VI reduced diagrams 
with $N$ loops if it holds for 
amplitudes with $\le (N-1)$ loops.

\begin{figure}
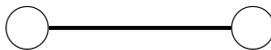


~

\vskip .5in

\begin{center}
\figtwenty
\end{center}

\vskip -1in

\caption{A reduced diagram containing a single propagator. 
\label{f20}
}
\end{figure}

In order to complete the proof we need to verify that the result holds for 
1VI reduced diagrams with zero loops. This corresponds to two reduced vertices
connected by a single propagator as shown in Fig~\ref{f20}. If $p$ denotes the
momentum flowing from the left to the right then $p^0$ is positive by convention, and the 
contribution to the diagram is given by
\be 
A(p) = {1\over (p^0)^2 - \vec p^{\, 2} - m^2+i\eps} F(p)\, ,
\ee
where $F(p)$ is the contribution from the reduced vertices. The $i\eps$ factor is
equivalent to defining this amplitude via analytic continuation of the amplitude
with $p^0$ replaced by $\lambda p^0$, and taking the limit $\lambda\to 1$ 
from the first quadrant
of the complex $\lambda$-plane. Now since the reduced vertices are not
pinched we have $F(p)^*= F(-p)$ for real $p$ and hence
\ben \label{elowest}
A(p) - A(-p)^* &=&  \left[ {1\over (p^0)^2 - \vec p^{\, 2} - m^2+i\eps} 
- {1\over (p^0)^2 - \vec p^{\, 2} - m^2-i\eps}\right] F(p) \nonumber \\
&=& - 2\pi i \, \delta \left( (p^0)^2 - \vec p^{\, 2} - m^2 \right) \theta(p^0) F(p)\, ,
\een
where the $\theta(p^0)$ factor has been included since we are considering positive
$p^0$ anyway.
Comparison with \refb{eppcut} shows that this precisely corresponds to
replacing the propagator in Fig.~\ref{f20} by the cut propagator, in accordance
with the Cutkosky rules.

\subsection{Amplitudes with disconnected components} \label{s5.3}

Finally we turn to the proof of Cutkosky rules for amplitudes with disconnected
components. We shall prove the result by showing that if an amplitude has two 
disconnected components $A$ and $B$, each of which may be connected or disconnected,
then as long as the Cutkosky rules hold for $A$ and $B$, they also hold for the diagram
with components $A$ and $B$.  Repeated use of this result, and the fact that
Cutkosky rules hold for connected diagrams, then proves that 
Cutkosky rules hold for diagrams with arbitrary number of disconnected components.

\begin{figure}
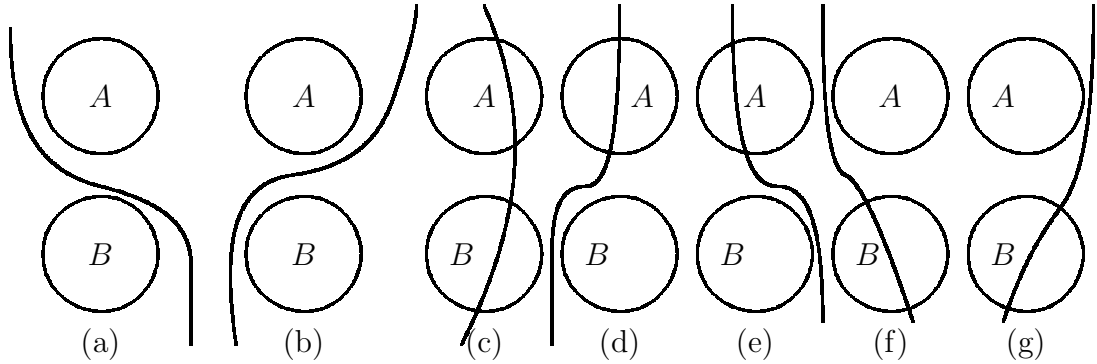

\begin{center}
\figdis
\end{center}
\caption{The cut diagrams of a disconnected diagram.
\label{fdis}}
\end{figure}

We begin by analyzing the left hand side of the Cutkosky rules.
Let $T_A$ and $T_B$ denote the T-matrix
associated with individual blobs and  $T_{AB}$ denote 
the T-matrix associated with
the combined diagram.
If $A$ has $n_A$ disconnected components and $B$ has $n_B$
disconnected components, then according to
\refb{e2.5a}  $A$ carries an extra factor of $i^{1-n_A}$ and 
$B$ carries an extra 
factor of $i^{1-n_B}$, while the combined amplitude carries an extra factor
of $i^{1-n_A-n_B}$. Therefore we need to remove a factor of $i$ from the 
product $T_A\otimes T_B$ to get the combined amplitude $T_{AB}$.
This gives
\be \label{epr0}
T_{AB}= -i \, T_A\otimes T_B\, .
\ee
We define the amplitudes $A$ and $B$ associated with the two blobs
as the matrix elements of $T_A$ and
$T_B$ between external states.  We now see using \refb{epr0} that 
the combined amplitude is given by $-i AB$.  
Using the shorthand notation $A^*$ and $B^*$ for
hermitian conjugates of $A$ and $B$ we get the left hand side of the Cutkosky
rules, encoding the anti-hermitian part of the full amplitude, to be
\be \label{eleft}
-i A B - (i A^* B^*) = -i (AB + A^* B^*)\, .
\ee

The right hand side of the Cutkosky rules is given by the sum over all the cut diagrams
of the original diagram. These are shown in Fig.~\ref{fdis}. 
The phases of different cut diagrams must be chosen such that 
the sum over cut diagrams represent matrix elements of $-i T^\dagger T$. 
However now in $T$ we also need to include the 
terms $T_A\otimes I_B$ and $I_A\otimes T_B$ besides $T_{AB}$ given in
\refb{epr0}. Therefore the total relevant contribution
to $T$ is given by
\be \label{edeftab}
T = T_A\otimes I_B + I_A \otimes T_B-i\, T_A\otimes T_B \, .
\ee
We can now compare different terms in Fig.~\ref{fdis} with the corresponding
terms in $-i T^\dagger T$ to determine their phases. For example
Fig.~\ref{fdis}(a) represents the matrix element of
$-i  (T_A\otimes I_B)^\dagger ( I_A \otimes T_B) = -i T_A^\dagger\otimes T_B$,
and hence gives the contribution 
 $-i A^* B$. Similarly  
Fig.~\ref{fdis}(b) gives the contribution $-i A B^*$. 
In order to evaluate the 
contribution from Fig.~\ref{fdis}(c) including its phase we note that 
this
diagram should represent the matrix element of
\be \label{epr1}
-i \, (-i \,T_A\otimes T_B)^\dagger (-i\, T_A\otimes T_B)
= -i \, (T_A^\dagger T_A )\otimes (T_B^\dagger T_B) = i \, (T_A-T_A^\dagger) \otimes
(T_B-T_B^\dagger)\, ,
\ee
where in the last step we have used the fact that the blobs $A$ and $B$ 
individually satisfy the Cutkosky rules and hence $-i T_A^\dagger T_A$ and
$-i T_B^\dagger T_B $ are given respectively by $(T_A-T_A^\dagger)$ and 
$(T_B-T_B^\dagger)$. The matrix element of \refb{epr1} gives 
$i (A-A^*) (B-B^*)$. 
Following similar logic we get the contributions from Fig.~\ref{fdis}(d), (e), (f) and
(g) to be, respectively, $i(A-A^*) B^*$, $-i (A-A^*) B$, $i A^* (B-B^*)$ and
$-i A (B-B^*)$. The last four terms add up to $-2i(A-A^*) (B - B^*)$. 
Therefore the total contribution to the
right hand side of the Cutkosky rules is given by\footnote{Note that two
other contributions given by $-i (T_A^\dagger T_A)\otimes I_B$ and
$I_A\otimes (-i T_B^\dagger T_B )$ are present in the expressions for
$-iT^\dagger  T$, but are not included in Fig.~\ref{fdis}. They represent
diagrams where either the blob $B$ or the blob $A$ are replaced by forward
scattering amplitudes. They will reproduce the anti-hermitian parts of the first
two terms on the right hand side of \refb{edeftab}.}
\be \label{eright}
-i A^* B - i A B^* + i (A-A^*) (B-B^*)
-2 i  (A-A^*) (B-B^*)
= - i (AB + A^*B^*)\, . 
\ee
This is in perfect agreement with the left hand side of the Cutkosky rules
given in \refb{eleft}.

\subsection{Mass and wave-function renormalization} \label{smass}

In the derivation of the Cutkosky rules we have given, the momentum $k$ carried by
a cut propagator is forced to satisfy $k^2+m^2=0$ where $m$ is the tree level mass
of the scalar field. However in general a theory of the kind we have analyzed will have
finite mass renormalization and hence the constraint on the cut propagator should have
been that $k^2+m_p^2=0$ where $m_p$ is the renormalized physical mass. In our analysis
this issue shows up in the fact that if we have self energy insertions on a cut
propagator on either side of the cut, we get extra propagator factors proportional
to $(k^2+m^2)^{-1}$ which diverge. Therefore the Cutkosky rules 
become only formal
relations.

This problem can be avoided by the usual trick of reorganizing Feynman diagrams
at each order in perturbation theory. If $m_p$ is the physical mass computed to a given
order in perturbation theory, and $Z$ is the wave-function renormalization factor so that
the two point function has a pole at $k^2=-m_p^2$ with residue $-i\, Z$,
then for computing any amplitude at higher order, we
change $k^2+m^2$ to $Z^{-1}(k^2+m_p^2)$ in \refb{e2.1ff} and compensate for it 
by adding to the two point vertex $V^{(2)}$ a new term proportional to $(m^2-m_p^2)
+(1-Z^{-1})(k^2+m_p^2)$.
This makes the propagator $-i\, Z\, (k^2+m_p^2)^{-1}$.
Now the cut propagator will set $k^2+m_p^2$ to zero, and the self energy insertions
on the cut propagators, after including the contribution from the new two point vertex, 
 will vanish at $k^2+m_p^2=0$.  This makes
the contribution from the cut diagrams manifestly finite. 

Note that the new contribution to $V^{(2)}$ does not carry the exponential suppression
factor that was assumed to be present for all $V^{(n)}$. However whenever this new
vertex is inserted into an internal propagator of a loop diagram carrying momentum $k$,
there will be some other vertex whose external line carries the same momentum $k$ and
hence exponentially
suppresses the integrand in the large $k^2$ region. Therefore the new two point
vertex does not affect the ultraviolet finiteness property of individual Feynman diagrams.

\sectiono{Field theory model to superstring field theory} \label{s6}

We shall now  discuss what is involved in going from the toy model we have analyzed
to the full string field theory. This discussion will be divided into two parts. In
the first part we shall describe generalizations of our analysis to more
general quantum field theories, and in the second part we shall turn to the
specific case of string field theory.

\subsection{More general quantum field theories} \label{e6.1}

In this subsection we shall describe the extension of our analysis to more
general class of quantum field theories.

\begin{enumerate}

\item {\bf Multiple fields of higher spin:}
The toy model of \S\ref{s1} has only one scalar field. This 
can be easily generalized to the case of multiple fields including
those carrying higher spins and also complex fields. If we denote 
the complex conjugate of a field $\phi_\alpha$ by $\phi_{\bar\alpha}$
then the reality condition on the vertices $V^{(n)}_{\alpha_1\cdots \alpha_n}$
and the propagator $P_{\alpha\beta}$ appearing in \refb{eifactornew} take the form
\be \label{erexx}
(V^{(n)}_{\alpha_1\cdots \alpha_n}(p_1,\cdots p_n))^* = V^{(n)}_{\bar\alpha_1\cdots \bar\alpha_n}
(-p_1^*,\cdots -p_n^*), \quad P_{\alpha\beta}(k)^* = P_{\bar\alpha \bar\beta} (-k^*)\, .
\ee
For fermions some more signs are needed that will be discussed separately. Now we can
proceed with our analysis of \S\ref{s2.3} as before. The main change is in the fact that
in relating $\langle b|T|a\rangle$ to $\langle a| T |b\rangle$ we not only need to change the
sign of all the external momenta, but also replace all the field labels $\alpha_i$ by
$\bar{\alpha_i}$.  In the second step of the analysis in \S\ref{s2.3}, where we relabel
the internal momenta by a change of sign, we also relabel the internal indices carried
by the vertices and propagators by their conjugates. In the third step the 
expression for $\langle b| T |a\rangle^*$ can be manipulated using 
\refb{erexx} to arrive at an expression that is a modification of
that of $\langle a|T|b\rangle$ 
by complex conjugation of 
each internal and external momenta and replacement of the factor of 
$i$ accompanying each loop integration by $-i$. In the fourth step we
relabel the  loop momentum integration variables by
their 
complex conjugates to arrive at an expression in which the integrand is related to
that in $\langle a|T|b\rangle$ by the replacement of the external momenta
by their complex conjugates, and the integration contour is related to the original one
by complex conjugation. Rest of the analysis remains unchanged.

Typically theories with higher spin fields have gauge symmetries, and as a result not
all states propagating in the propagator are physical.
In such cases Cutkosky rules do not by themselves imply unitarity -- we have to do 
additional work to show that the unphysical state contribution cancels. There are also
massless fields for which our analysis may break down. We shall return
to these points when we describe applications to string field theory.

\item {\bf Fermions:}
For fermionic fields there are a few additional signs that need to be 
taken care of.
First of all the vertices $V^{(n)}$ are no longer fully symmetric under permutations
of fields -- under the exchange of a pair of fermionic states 
they pick up minus signs.
Since
the complex conjugate of the product of grassmann 
variables involves reversing their order
in the product besides taking conjugates of each variable,
in the reality constraint \refb{erexx},
the order of the fermionic indices 
carried by the vertices and propagators on the two sides of the equation will have 
to be in opposite order leading to extra signs in our analysis.
Also the hermitian conjugate of a multi-particle state containing
fermions will involve the conjugate states arranged in opposite order. 
As a result in the analysis of \S\ref{s2.3}, the computation of $\langle b|T|a\rangle$ will now not
only involve reversing the signs of the external momenta and complex conjugating the labels
of external states, 
but also changing the order of the fermions in the external states.
After performing manipulations similar to that in \S\ref{s2.3} we arrive at
the result that the computation of $T^\dagger$ will involve evaluating
an integral whose integrand differs from that of the original integrand for $T$
by complex conjugation of external momenta and
\begin{enumerate}
\item \label{ia}
a reversal of the order of the fermionic labels in the external states,
\item \label{ib}
a reversal of the 
order of the fermionic
labels carried by the vertices, and
\item \label{ic}
a reversal of the order of the fermionic labels carried by the propagators.
\end{enumerate} 
Let us denote by $2n_e$ the total number of external fermions, by $2n_v$ the
total number of fermionic labels carried by the vertices and by $n_p$ the total
number of internal fermionic propagators. Then the net factor from the three effects
mentioned above is $(-1)^{n_p+n_v+n_e}$. Using the relation $n_v-n_e=n_p$ 
we see that this number is 1. 
Therefore we  get back the same integrand as that in the
computation of $T$ except for complex conjugation of the external momenta. 
As before the integration contour will be
given by the complex conjugate
of the integration contour for $T$.

In subsequent analysis,
another set of minus signs originate from the fact that each fermion loop 
is accompanied by a minus sign. Therefore if we have a cut diagram in which
$N$ fermion loops are cut, the diagram is accompanied by a factor of $(-1)^N$.
When we attempt to interpret this as a contribution to $T^\dagger T$ by inserting a
complete set of states $|\alpha\rangle\langle\alpha|$ between $T$ and $T^\dagger$,
then the order of the $2N$ fermions in $|\alpha\rangle$ and $\langle\alpha|$ must be
opposite. Reversing the order of the $2N$ fermions leads to another factor of $(-1)^N$
that cancels the $(-1)^N$ factor coming from the $N$ fermion loops.

\end{enumerate}

\subsection{String field theory} \label{s6.2}

We shall now describe the implication of our results for string field theory.
\begin{enumerate}
\item {\bf Exponential suppression of vertices at large momentum:}
The key feature of the toy model of \S\ref{s1} is the peculiar form 
of the interaction vertices in the momentum space, possessing an essential
singularity at infinity, diverging exponentially as $k^2\to -\infty$ and falling
off exponentially as $k^2\to \infty$ for any momentum $k$ carried by an
external line to the vertex. In string field theory this exponential factor comes from the
conformal transformation of the vertex operator.  In defining the off-shell vertex
we have to choose local coordinate system at the punctures and by scaling the
local coordinate at a puncture 
by some real number $\beta$, we can scale the off-shell vertex
by a factor of $\beta^h$ where $h$ is the $L_0+\bar L_0$ eigenvalue of the 
vertex operator. Since $L_0+\bar L_0$ has an additive contribution of $k^2/2$ 
besides the oscillator contribution, this introduces a factor of $\beta^{k^2/2}
=\exp[(k^2 \ln\beta) / 2]$, and this can be made small for large $k^2$ by taking $\beta$
to be small. In the string field theory literature this operation of scaling local coordinates
by $\beta$ is known as the
act of adding stubs of length $-\ln\beta$ to the vertices. Physically choosing
small $\beta$ i.e.\ long stubs amounts to ensuring that integration over 
most of the moduli space
of Riemann surfaces comes from the elementary vertices, and only small regions 
near the boundary of the moduli space come from Feynman diagrams with
internal propagators. 

\item {\bf Poles of the propagator:} We have assumed that the only poles of the
propagator occur at $k^2+m^2=0$ for different values of $m$. 
In string field theory this is automatic in the Siegel gauge.

\item {\bf Infinite number of states:}
Since string field theory has infinite number of fields, 
we also need to ensure
that the sum over fields that appear in the evaluation of the Feynman diagrams
converge.
The number of states below a certain mass $m$ grows as
$\exp[c_1 m]$ for some positive constant $c_1$, and the stubs suppress the
vertices
by a factor of $\exp[-c_2 m^2]$ for some positive constant $c_2$ that can be made 
arbitrarily large.
Therefore we expect the
sums to be convergent.

\item  {\bf Analyticity of the vertices at finite momentum:}
In our analysis we have assumed that the interaction vertices 
are analytic as function
of external momenta for finite momenta. In 
string field theory this is a consequence of the fact that the
$n$-point
interaction vertices are obtained as integrals over subspaces of moduli spaces of
genus $g$ Riemann surfaces with $n$-punctures for various values of $g$,
and that {\it these subspaces never include any degenerate Riemann surface.}

\item {\bf Reality of the action:}
In the derivation of the Cutkosky rules we needed to make use of the reality
of the action. Therefore to extend the proof to string field theory we need 
to prove the reality of the string field theory action. This has been proved for
the bosonic string theory\cite{9206084}. It is expected that a similar proof can be given
for superstring field theory with judicious choice of the locations of the
picture changing operators, but
this has not yet been worked out.

\item{\bf Decoupling of unphysical states:}
A more serious issue arises from the fact that among
the fields of string field theory there are many auxiliary fields, pure gauge fields
and ghost fields which do not correspond to physical particles.
In the Siegel gauge
in which the propagator is proportional to $(L_0+\bar L_0)^{-1} 
=2 (k^2 + C)^{-1}$ where $C/2$ gives the discrete
contribution to $L_0+\bar L_0$ from
the oscillators, all the fields contribute to the poles and hence will be summed
over as intermediate states in the Cutkosky rules. 
As a result, Cutkosky rules by themselves do not prove unitarity.
The complete proof will
involve showing that the contribution from all fields other than the physical 
fields vanish or cancel. 
In principle this should be possible with the help of Ward identities of the
kind described in \cite{1508.02481}, but the details need to be worked out.

\item {\bf Massless states:}
The spectrum of superstring theory contains massless states, and as a
result the S-matrix suffers from infrared divergences. 
Therefore unitarity may not hold in the usual sense. 
In our analysis this problem shows up in the breakdown of our implicit assumption
that at a pinch singularity only one of the two poles in a propagator blows up.
For example for $m=0$ 
in \refb{eqpos}, as $\vec \ell\to 0$ both poles $Q_1$ and $Q_2$ approach the
contour from two sides and pinch it at the origin even when all the external
energies are taken to be imaginary and the integration contour lies along the 
imaginary $\ell^0$ axis. Since we have worked with fixed values of the spatial
components of loop momenta, our analysis can still be used for generic values
of these spatial momenta, but will break down when one or more internal or
external massless particle carries zero spatial momenta. 
In sufficiently large dimensions ($>4$)
configurations with zero spatial momenta do not contribute to $T$ or
$T^\dagger T$ due to the vanishing of the integration measure dominating the
divergences from the propagators. In such cases infrared divergences are tame
and our result holds. 
In dimensions $\le 4$ we need to be more careful -- work with
cross section instead of S-matrix and sum over final states
and average over initial states\cite{kinoshita,lee,bloch,sterman}.
Since our proof of Cutkosky rules holds for fixed spatial components of loop
momenta, we expect that the method described in \cite{sterman} can be used
to prove finiteness of appropriate inclusive cross sections after averaging over
initial states, but the details need to be worked out.

\item  {\bf Vacuum shift:}
Like in ordinary quantum field theories, 
in string field theory the vacuum can get shifted
from the original classical vacuum by vacuum expectation values 
of certain fields. Since the  interaction vertices around
the shifted vacuum have the same analytic structure as in the original vacuum,
the Cutkosky rules will hold in the new vacuum as well. This of course
requires that the
fields acquiring vacuum expectation values satisfy appropriate reality condition
so that the action expanded around the new vacuum continues to be real.
\item {\bf Mass renormalization:} Massive particles in string theory 
undergo mass renormalization. 
The S-matrix has to be defined by taking into account these effects.
We expect that the proof of unitarity can be carried through even in the presence
of these effects following the same steps as in \S\ref{smass}. 
There is also the issue that most of the massive 
particles in string theory become unstable under quantum corrections and 
hence cease to be true candidates for asymptotic states. We expect that this
effect can also be taken into account following the same method as in a quantum
field theory\cite{veltman}.
\end{enumerate}

Therefore the two main technical problems that need to be solved before we
can declare superstring field theory amplitudes to be unitary are:
\begin{enumerate}
\item proving
reality of the
superstring field theory action, and 
\item showing that the contribution to the cut propagator
from  the unphysical and pure gauge states cancel, leaving behind only the contribution 
from physical states.
\end{enumerate}

\bigskip

\noindent {\bf Acknowledgement:}
We wish to thank Corinne de Lacroix, Eric D'Hoker, Harold Erbin,
George Sterman and 
Edward Witten for useful discussions.
The research of R.P.
was supported in part by Perimeter Institute for Theoretical Physics. Research at Perimeter 
Institute is supported by the Government of Canada through Industry Canada and by the Province of Ontario through the Ministry of Research and Innovation. 
This research  of A.S. was
supported in part by the 
DAE project 12-R\&D-HRI-5.02-0303 and J. C. Bose fellowship of 
the Department of Science and Technology, India.




\begin{thebibliography}{99}

\bibitem{1508.05387} 
  A.~Sen,
  ``BV Master Action for Heterotic and Type II String Field Theories,''
  arXiv:1508.05387 [hep-th].

\bibitem{wittenssft} 
  E.~Witten,
  ``Interacting Field Theory of Open Superstrings,''
  Nucl.\ Phys.\ B {\bf 276}, 291 (1986).
  
\bibitem{9202087} 
  R.~Saroja and A.~Sen,
  ``Picture changing operators in closed fermionic string field theory,''
  Phys.\ Lett.\ B {\bf 286}, 256 (1992)
  doi:10.1016/0370-2693(92)91772-2
  [hep-th/9202087].
  
  
  \bibitem{9503099}
  N.~Berkovits,
  ``SuperPoincare invariant superstring field theory,''
  Nucl.\ Phys.\ B {\bf 450} (1995) 90
   [Erratum-ibid.\ B {\bf 459} (1996) 439]
  [hep-th/9503099].

\bibitem{0109100}
  N.~Berkovits,
  ``The Ramond sector of open superstring field theory,''
  JHEP {\bf 0111} (2001) 047
  [hep-th/0109100].

\bibitem{0406212}
  Y.~Okawa and B.~Zwiebach,
  ``Heterotic string field theory,''
  JHEP {\bf 0407} (2004) 042
  [hep-th/0406212].

\bibitem{0409018}
  N.~Berkovits, Y.~Okawa and B.~Zwiebach,
  ``WZW-like action for heterotic string field theory,''
  JHEP {\bf 0411} (2004) 038
  [hep-th/0409018].

\bibitem{1312.2948}
  T.~Erler, S.~Konopka and I.~Sachs,
  ``Resolving Witten`s superstring field theory,''
  JHEP {\bf 1404} (2014) 150
  [arXiv:1312.2948 [hep-th]].


\bibitem{1312.7197}
  H.~Kunitomo,
  ``The Ramond Sector of Heterotic String Field Theory,''
  PTEP {\bf 2014} 4,  043B01
  [arXiv:1312.7197 [hep-th]].
  
  \bibitem{1403.0940}
  T.~Erler, S.~Konopka and I.~Sachs,
 ``NS-NS Sector of Closed Superstring Field Theory,''
  arXiv:1403.0940 [hep-th].

\bibitem{1407.8485} 
  H.~Matsunaga,
  ``Nonlinear gauge invariance and WZW-like action for NS-NS superstring field theory,''
  arXiv:1407.8485 [hep-th].
  
\bibitem{1412.5281} 
  H.~Kunitomo,1412.5281
  ``Symmetries and Feynman Rules for Ramond Sector in Open Superstring Field Theory,''
  arXiv: [hep-th].


\bibitem{1505.01659} 
  T.~Erler, Y.~Okawa and T.~Takezaki,
  ``$A_\infty$ structure from the Berkovits formulation of open superstring field theory,''
  arXiv:1505.01659 [hep-th].

\bibitem{1506.05774} 
  T.~Erler, S.~Konopka and I.~Sachs,
  ``Ramond Equations of Motion in Superstring Field Theory,''
  JHEP {\bf 1511}, 199 (2015)
  doi:10.1007/JHEP11(2015)199
  [arXiv:1506.05774 [hep-th]].

\bibitem{1506.06657} 
  K.~Goto and H.~Matsunaga,
  ``On-shell equivalence of two formulations for superstring field theory,''
  arXiv:1506.06657 [hep-th].

\bibitem{1507.08250} 
  S.~Konopka,
  ``The S-Matrix of superstring field theory,''
  JHEP {\bf 1511}, 187 (2015)
  doi:10.1007/JHEP11(2015)187
  [arXiv:1507.08250 [hep-th]].

\bibitem{1508.00366} 
  H.~Kunitomo and Y.~Okawa,
  ``Complete action for open superstring field theory,''
  doi:10.1093/ptep/ptv189
  arXiv:1508.00366 [hep-th].

\bibitem{1512.03379} 
  K.~Goto and H.~Matsunaga,
  ``$A_\infty / L_\infty$ structure and alternative action for WZW-like superstring field theory,''
  arXiv:1512.03379 [hep-th].

\bibitem{1602.02582} 
  T.~Erler, Y.~Okawa and T.~Takezaki,
  ``Complete Action for Open Superstring Field Theory with Cyclic $A_\infty$ Structure,''
  arXiv:1602.02582 [hep-th].
  
\bibitem{1602.02583} 
  S.~Konopka and I.~Sachs,
  ``Open Superstring Field Theory on the Restricted Hilbert Space,''
  arXiv:1602.02583 [hep-th].

\bibitem{1303.2323} 
  B.~Jurco and K.~Muenster,
  ``Type II Superstring Field Theory: Geometric Approach and Operadic Description,''
  JHEP {\bf 1304}, 126 (2013)
  doi:10.1007/JHEP04(2013)126
  [arXiv:1303.2323 [hep-th]].

\bibitem{Cutkosky} 
  R.~E.~Cutkosky,
  ``Singularities and discontinuities of Feynman amplitudes,''
  J.\ Math.\ Phys.\  {\bf 1}, 429 (1960).
  doi:10.1063/1.1703676

\bibitem{fowler}
M. Fowler, ``Introduction to Momentum Space Integration Techniques 
in Perturbation Theory'',
Journal of Mathematical Physics 3, 936 (1962); doi: 10.1063/1.1724310.

\bibitem{veltman} 
  M.~J.~G.~Veltman,
  ``Unitarity and causality in a renormalizable field theory with unstable particles,''
  Physica {\bf 29}, 186 (1963).
  doi:10.1016/S0031-8914(63)80277-3

\bibitem{diagrammar} 
  G.~'t Hooft and M.~J.~G.~Veltman,
  ``Diagrammar,''
  NATO Sci.\ Ser.\ B {\bf 4}, 177 (1974).

\bibitem{1512.01705} 
  S.~Bloch and D.~Kreimer,
  ``Cutkosky Rules and Outer Space,''
  arXiv:1512.01705 [hep-th].

\bibitem{1508.02481} 
  A.~Sen,
  ``Supersymmetry Restoration in Superstring Perturbation Theory,''
  arXiv:1508.02481 [hep-th].

\bibitem{dhoker} 
  K.~Aoki, E.~D'Hoker and D.~H.~Phong,
  ``Unitarity of Closed Superstring Perturbation Theory,''
  Nucl.\ Phys.\ B {\bf 342}, 149 (1990).
  doi:10.1016/0550-3213(90)90575-X

\bibitem{1209.5461} 
  E.~Witten,
  ``Superstring Perturbation Theory Revisited,''
  arXiv:1209.5461 [hep-th].

\bibitem{1304.7798} 
  R.~Donagi and E.~Witten,
  ``Supermoduli Space Is Not Projected,''
  arXiv:1304.7798 [hep-th].
  

\bibitem{1404.6257} 
  R.~Donagi and E.~Witten,
``Super Atiyah classes and obstructions to splitting of supermoduli space,''
  arXiv:1404.6257 [hep-th].

 \bibitem{1504.00609} 
  A.~Sen and E.~Witten,
  ``Filling The Gaps With PCO's,''
  arXiv:1504.00609 [hep-th].

\bibitem{berera} 
  A.~Berera,
  ``Unitary string amplitudes,''
  Nucl.\ Phys.\ B {\bf 411}, 157 (1994).

\bibitem{1307.5124}
E.~Witten,
``The Feynman $i \epsilon$ in String Theory,''
  arXiv:1307.5124 [hep-th].

\bibitem{sterman} 
  G.~F.~Sterman,
  ``An Introduction to quantum field theory,'' Cambridge University Press (1993).

\bibitem{9206084} 
  B.~Zwiebach,
  ``Closed string field theory: Quantum action and the B-V master equation,''
  Nucl.\ Phys.\ B {\bf 390}, 33 (1993)
  doi:10.1016/0550-3213(93)90388-6
  [hep-th/9206084].

\bibitem{kinoshita} 
  T.~Kinoshita,
  ``Mass singularities of Feynman amplitudes,''
  J.\ Math.\ Phys.\  {\bf 3}, 650 (1962).
  doi:10.1063/1.1724268

\bibitem{lee} 
  T.~D.~Lee and M.~Nauenberg,
  ``Degenerate Systems and Mass Singularities,''
  Phys.\ Rev.\  {\bf 133}, B1549 (1964).
  doi:10.1103/PhysRev.133.B1549

\bibitem{bloch} 
  F.~Bloch and A.~Nordsieck,
  ``Note on the Radiation Field of the electron,''
  Phys.\ Rev.\  {\bf 52}, 54 (1937).
  doi:10.1103/PhysRev.52.54

\end{thebibliography}
\end{document}